\documentclass[journal]{IEEEtran}
\usepackage[figuresright]{rotating}
\usepackage{amssymb}
\usepackage{amsthm}
\usepackage{multicol}
\usepackage{subfigure}
\usepackage{graphicx}
\usepackage{epstopdf}
\usepackage{fullpage}
\usepackage{latexsym,amsmath}
\usepackage{stmaryrd}
\usepackage{algorithm,algorithmic}
\usepackage{subfigure}
\usepackage{amsfonts}
\usepackage{xcolor,multirow}
\usepackage{cite}
\usepackage{bm}
\usepackage{url}
\usepackage{diagbox}
\usepackage{array}
\usepackage{booktabs}
\usepackage{footmisc} 
\usepackage{pifont}
\usepackage{tikz}
\newcommand*{\circled}[1]{\lower.7ex\hbox{\tikz\draw (0pt, 0pt)%
		circle (.5em) node {\makebox[1em][c]{\small #1}};}}

\newtheorem{theorem}{Theorem}
\newtheorem{definition}{Definition}

\ifCLASSINFOpdf
\else
\fi
\begin{document}
	
\title{Quaternion higher-order singular value decomposition and its applications in color image processing}

\author{Jifei~Miao and Kit~Ian~Kou

\thanks{The authors are with the Department of Mathematics, Faculty of Science
	and Technology, University of Macau, Macau 999078, China  (e-mail: jifmiao@163.com;
	kikou@umac.mo)}}

\markboth{Journal of \LaTeX\ Class Files,~Vol.~14, No.~8, August~2019}%
{Shell \MakeLowercase{\textit{et al.}}: Bare Demo of IEEEtran.cls for IEEE Journals}

\maketitle

\begin{abstract}
Higher-order singular value decomposition (HOSVD) is one of the most efficient tensor decomposition techniques. It has the salient ability to represent high-dimensional data and extract features. In more recent years, the quaternion has proven to be a very suitable tool for color pixel representation as it can well preserve cross-channel correlation  of color channels. Motivated by the advantages of the HOSVD and the quaternion tool, in this paper, we generalize the HOSVD to the quaternion domain and define quaternion-based HOSVD (QHOSVD). Due to the non-commutability of quaternion multiplication, QHOSVD is not a trivial extension of the HOSVD. They have similar but different calculation procedures. The defined QHOSVD can be widely used in various visual data processing with color pixels. In this paper, we present two applications of the defined QHOSVD in color image processing--- multi-focus color image fusion and color image denoising. The experimental results on the two applications respectively demonstrate the competitive performance
of the proposed methods over some existing ones.
\end{abstract}

\IEEEpeerreviewmaketitle

\section{Introduction}
\IEEEPARstart{H}igher-order singular value decomposition (HOSVD) is a natural extension of the singular value decomposition (SVD) in higher dimensional space \cite{DBLP:journals/siammax/LathauwerMV00}. Benefiting from the outstanding ability of  high-dimensional data representing and feature extracting, the HOSVD has been widely applied in many areas such as image fusion \cite{DBLP:journals/tip/LiangHLZ12}, face recognition \cite{2010Gurumoorthy}, texture synthesis \cite{2008HigherCostantini}, image denoising \cite{DBLP:journals/access/GaoGZCZ19}, \emph{etc.} On the other hand, quaternion, as an elegant color image representation tool, has attracted much attention in the
field of color image processing, recently.  For instance, it has achieved excellent results in color image filtering \cite{DBLP:journals/iet-ipr/ChenLSLS14}, color image edge detection \cite{Xiao2018Phase}, color image denoising \cite{DBLP:journals/tip/ChenXZ20,DBLP:journals/ijon/YuZY19}, color face recognition \cite{DBLP:journals/tip/ZouKW16,DBLP:journals/access/ZouKDZT19}, color image inpainting \cite{DBLP:journals/tsp/MiaoK20,DBLP:journals/nla/JiaNS19} and so on. However, most of the existing quaternion-based methods are limited to one-dimensional quaternion vector or two-dimensional quaternion matrix. Although they can perfectly preserve the connection between color channels, they cannot make good use of the spatial structure of higher dimensional data. Thus, in this paper, combining the advantages of HOSVD and quaternion, we aim to define quaternion-based HOSVD (QHOSVD) and use it for color image processing.

HOSVD was first defined by De Lathauwer et al. in \cite{DBLP:journals/siammax/LathauwerMV00}, which is a proper generalization of the SVD. The HOSVD  of  a tensor $\mathcal{T}\in\mathbb{R}^{N_{1}\times N_{2} \times\ldots \times N_{L}}$ is given by 
\begin{equation*}
	\label{hosvd}
	\mathcal{T}=\mathcal{S}\times_{1}\mathbf{U}_{1}\times_{2}\mathbf{U}_{2}\ldots\times_{L}\mathbf{U}_{L},
\end{equation*}
where $\mathbf{U}_{l}\in\mathbb{R}^{N_{l}\times N_{l}} (l=1,\ldots, L)$ are orthogonal matrices, $\mathcal{S}\in\mathbb{R}^{N_{1}\times N_{2} \times\ldots \times N_{L}}$ is called the core tensor. Following the calculation procedure in \cite{DBLP:journals/siammax/LathauwerMV00}, the core tensor $\mathcal{S}$ satisfies the all-orthogonality conditions\footnote{We will introduce the concept of all-orthogonality hereinafter. For more information about HOSVD, readers can refer to \cite{DBLP:journals/siammax/LathauwerMV00}.}. The concept of QHOSVD, in this paper, is an extension of the HOSVD on quaternions, but the limitation of non-commutativity of quaternion multiplication results in the definition and calculation procedure of QHOSVD cannot be the same as that of HOSVD. It should be noted that the definition of QHOSVD is application-oriented. The defined QHOSVD can be widely used in various visual data processing with color pixels. As examples, in this paper, we tend to use it to solve two common color image processing problems: multi-focus color image fusion and color image denoising.

Multi-focus image fusion is an effective technique to extend the depth-of-field of optical lenses by creating an all-in-focus image from a set of partially focused images of the same scene \cite{DBLP:journals/inffus/LiuWCLC20}. The existing multi-focus image fusion methods can be roughly  classified into three categories: Transform domain methods, \emph{e.g.}, \cite{DBLP:journals/cvgip/LiMM95, 2007Image, DBLP:journals/inffus/ZhouLW14, DBLP:journals/tip/LiKH13, DBLP:journals/tip/LiangHLZ12}, spatial domain methods, \emph{e.g.}, \cite{DBLP:journals/inffus/BaiZZX15, DBLP:journals/spic/QiuLZY19, DBLP:journals/jei/ZhanKLH19}, and deep learning
methods, \emph{e.g.}, \cite{DBLP:journals/inffus/LiuCPW17, DBLP:journals/access/LaiLGX19}. Most of the existing traditional methods are inherently designed for multi-focus grayscale image fusion. For color multi-focus images, they generally fuse three color channels independently. Quaternions have also been widely used in this field in the past few years. For example those in \cite{DBLP:journals/sigpro/LiuJWSD14,DBLP:journals/access/ChaiLZ17,DBLP:journals/jei/LiuJWSD13}, which mainly focus on grayscale images. In \cite{2013Multifocus} and \cite{2012Multifocus}, the authors respectively using quaternion wavelet
transform (QWT) and quaternion curvelet transform (QCT) to fuse multi-focus color images. However, the fixed base of QWT and QCT usually cannot obtain accurate features of color images. Motivated by the fact that HOSVD, an efficient data-driven decomposition technique, can obtain a dynamic adaptive base \cite{DBLP:journals/access/GaoGZCZ19}, and it has the salient ability for feature extraction \cite{DBLP:journals/jvcir/LuoZZW17}, we design a novel multi-focus color image fusion method based on the defined QHOSVD procedure. In consideration of the fact that image fusion mainly depends on local information of source images \cite{DBLP:journals/jvcir/LuoZZW17}, the source images are firstly divided into image patches using the sliding window technique. Then, since these partially focused color images refer to the same scene and are highly similar, we construct them (represented as quaternion matrices) into a third-order quaternion tensor \cite{DBLP:journals/pr/MiaoKL20} and adopt the proposed QHOSVD procedure to extract their features simultaneously. Then, The $L_{1}$-norm of the coefficient quaternion sub-tensor is employed as the activity measure and the maximum selection fusion rule is adopted to obtain the fused  coefficient quaternion sub-tensor. To construct the entire fused image, all the fused patches are pasted in their corresponding positions and the intensity value of an overlapped pixel is averaged by its accumulative times. 

Recently, quaternion has also been applied to color image denoising, and achieved promising results.
For example, the authors in \cite{DBLP:journals/ijon/YuZY19} and \cite{DBLP:journals/tip/ChenXZ20} extended the traditional low-rank matrix approximation 
methods to the quaternion domain and respectively proposed low-rank quaternion approximation (LRQA)  (including nuclear norm, Laplace function, Geman function, and weighted Schatten norm) and quaternion weighted nuclear norm minimization (QWNNM) algorithms. The nonlocal self-similarity (NSS) of the color image is exploited in these methods to promote the denoising performance. However, these methods will destroy the two-dimensional spatial structure of color image patches when they are expanded into one-dimensional column quaternion vectors. Aiming at this drawback of the existing quaternion matrix-based methods, we propose a novel color image denoising method based on the defined QHOSVD.  Selected similar color image patches (represented as quaternion matrices) are stacked into a third-order quaternion tensor, and then the proposed QHOSVD procedure is performed on it to obtain the singular value coefficients. Finally, manipulate these coefficients by noise level related hard thresholding, and invert the QHOSVD procedure to produce the denoised color image patches used to construct the final full denoised color image.

 The main contributions of this paper can be
summarized as follows.
\begin{itemize}
	\item We generalize the concept of the HOSVD to the quaternion domain and define QHOSVD. We also give the calculation procedure of the QHOSVD, which is different from that of HOSVD in \cite{DBLP:journals/siammax/LathauwerMV00}.
	\item Based on the defined QHOSVD, we propose a novel multi-focus color image fusion method and a novel color image denoising method.
	\item Both methods are used to deal with the corresponding color image processing tasks. The experimental results on both problems demonstrate their competitive performance.
\end{itemize}

The remainder of this paper is organized as follows. Section \ref{sec2} introduces some notations and preliminaries for quaternion algebra. Section \ref{Q_tensor} introduces the quaternion tensors and the related definitions, operations, and properties. Section \ref{sec:qhosvd} gives the definition of QHOSVD and the detailed calculation procedure.  Section \ref{app:qhosvd} presents two applications of the defined QHOSVD in color image processing, including multi-focus color image fusion and color image denoising. Section \ref{exresults} provides some experiments to illustrate the performance of the proposed corresponding color image processing methods. Finally, some conclusions are drawn in Section \ref{sec:Conc}.

\section{Notations and preliminaries}
\label{sec2}
In this section, we first summarize some main notations and
then introduce some basic knowledge of quaternion algebra.

\subsection{Notations}
In this paper, $\mathbb{R}$, $\mathbb{C}$, and $\mathbb{H}$ respectively denote the real space, complex space, and quaternion space. A scalar, a vector, a matrix, and a tensor are written as $a$, $\mathbf{a}$, $\mathbf{A}$, and $\mathcal{A}$ respectively. $\dot{a}$,  $\dot{\mathbf{a}}$, $\dot{\mathbf{A}}$, and $\dot{\mathcal{A}}$ 
respectively represent a quaternion scalar, a quaternion vector, a quaternion matrix, and a quaternion tensor. The $(n_{1},n_{2})$th and the $(n_{1}, n_{2},\ldots, n_{L})$th entries in $\dot{\mathbf{A}}\in\mathbb{H}^{N_{1}\times N_{2}}$ and $\dot{\mathcal{A}}\in\mathbb{H}^{N_{1}\times N_{2} \times\ldots \times N_{L}}$ are respectively denoted as $\dot{a}_{n_{1}n_{2}}$ and $\dot{a}_{n_{1}n_{2}\ldots n_{L}}$. The mode-$k$ unfolding of a quaternion tensor $\dot{\mathcal{A}}$ is denoted by ${\rm{Unfold}}_{k}(\dot{\mathcal{A}}):=\dot{\mathbf{A}}_{[k]}$. $\times_{k}$,  $\otimes$, and $\langle\cdot,\cdot\rangle$  respectively denote the $k$-mode product, Kronecker product, and inner product. $(\cdot)^{\ast}$, $(\cdot)^{T}$, and $(\cdot)^{H}$ denote the conjugation, transpose, and  conjugate transpose, respectively. $|\cdot|$, $\|\cdot\|_{F}$ and $\|\cdot\|_{L_{1}}$ are respectively the modulus, Frobenius norm, and $L_{1}$ norm.

\subsection{Basic knowledge of quaternion algebras}
Quaternion space $\mathbb{H}$ was first introduced by W. Hamilton \cite{articleHamilton84} in 1843, which is an extension of the complex space $\mathbb{C}$.  Some basic knowledge of quaternion algebras can be found in Appendix \ref{a_sec1}. In the following, we give some properties used in this paper about quaternion matrices.
\begin{theorem}\label{th2}
	Given two quaternion matrices $\dot{\mathbf{A}}\in\mathbb{H}^{M\times N}$ and $\dot{\mathbf{B}}\in\mathbb{H}^{N\times M}$. Then
	\begin{enumerate}
		\item [\circled{a}] $(\dot{\mathbf{A}}\dot{\mathbf{B}})^{H}=\dot{\mathbf{B}}^{H}\dot{\mathbf{A}}^{H}$ \cite{10029950538};
		\item [\circled{b}] $(\dot{\mathbf{A}}\dot{\mathbf{B}})^{\ast}\neq\dot{\mathbf{A}}^{\ast}\dot{\mathbf{B}}^{\ast}$ in general \cite{10029950538};
		\item [\circled{c}] $(\dot{\mathbf{A}}\dot{\mathbf{B}})^{T}\neq\dot{\mathbf{B}}^{T}\dot{\mathbf{A}}^{T}$ in general \cite{10029950538};
		\item [\circled{d}] If $\dot{\mathbf{A}}$ is unitary (i.e., $\dot{\mathbf{A}}^{H}\dot{\mathbf{A}}=\mathbf{I}$), $\dot{\mathbf{A}}^{\ast}$ and $\dot{\mathbf{A}}^{T}$ are generally no longer unitary.
	\end{enumerate}
\end{theorem}
One can  directly verify \circled{d} based on \circled{a} and \circled{c}.

\section{Quaternion tensors}
\label{Q_tensor}
Quaternion tensors are generalizations of quaternion vectors (that have one index) and quaternion matrices (that have two indices) to an arbitrary number of indices. The notation about quaternion tensors used here is very similar to that of the real-valued case in \cite{DBLP:journals/siamrev/KoldaB09}.
\begin{definition}(Quaternion tensor \cite{DBLP:journals/pr/MiaoKL20}) A multidimensional array or an $L$th-order tensor is called a quaternion tensor if its entries are quaternion numbers, i.e.,
	\begin{equation}
	\label{qtensor1}
	\begin{split}
	\dot{\mathcal{T}}&=(\dot{t}_{n_{1}n_{2}\ldots n_{L}})\in\mathbb{H}^{N_{1}\times N_{2} \times\ldots \times N_{L}}\\	
	&=\mathcal{T}_{0}+\mathcal{T}_{1}i+\mathcal{T}_{2}j+\mathcal{T}_{3}k,
	\end{split}
	\end{equation}	
	where $\mathcal{T}_{p}\in\mathbb{R}^{N_{1}\times N_{2} \times\ldots \times N_{L}}\: (p=0,1,2,3)$, $\dot{\mathcal{T}}$ is named a pure quaternion tensor when $\mathcal{T}_{0}$ is a zero tensor.
\end{definition}

\begin{definition}(Mode-$k$ unfolding \cite{DBLP:journals/pr/MiaoKL20}) For an $L$th-order quaternion tensor $\dot{\mathcal{T}}\in\mathbb{H}^{N_{1}\times N_{2} \times\ldots \times N_{L}}$, its mode-$k$ unfolding is denoted by $\dot{\mathbf{T}}_{[k]}$ and arranges the mode-$k$ fibers to be the columns of the quaternion matrix, i.e.,
	\begin{equation*}
	{\rm{Unfold}}_{k}(\dot{\mathcal{T}})=\dot{\mathbf{T}}_{[k]}\in\mathbb{H}^{N_{k}\times N_{1}\ldots N_{k-1}N_{k+1}\ldots N_{L}} 
	\end{equation*}
with entries
	\begin{equation*}
	\dot{\mathbf{T}}_{[k]}(n_{k}, n_{1}\ldots n_{k-1}n_{k+1}\ldots n_{L})=\dot{t}_{n_{1}n_{2}\ldots n_{L}},
	\end{equation*}
	where $\dot{t}_{n_{1}n_{2}\ldots n_{L}}$ is the $(n_{1},n_{2},\ldots,n_{L})$th entry of $\dot{\mathcal{T}}$.
\end{definition}

\begin{definition}
	\label{mode product}
	(The $k$-mode product) The $k$-mode product of a quaternion tensor $\dot{\mathcal{T}}\in\mathbb{H}^{N_{1}\times N_{2} \times\ldots \times N_{L}}$ with a matrix $\dot{\mathbf{U}}\in\mathbb{H}^{M\times N_{k}}$ is denoted by\footnote{It should be noted that different from the definition of the real-valued case in \cite{DBLP:journals/siamrev/KoldaB09}, the sequence of $\dot{u}_{mn_{k}}$ and $\dot{t}_{n_{1}n_{2}\ldots n_{L}}$ in (\ref{k_mode product}) is not commutative, which is determined by the non-commutativity of quaternion multiplication.}  $\dot{\mathcal{Y}}=\dot{\mathcal{T}}\times_{k}\dot{\mathbf{U}}\in\mathbb{H}^{N_{1}\times  \ldots \times N_{k-1} \times M \times N_{k+1}\times \ldots \times N_{L}}$ with entries
	\begin{equation}
	\label{k_mode product}
	\dot{y}_{n_{1}\ldots n_{k-1} m n_{k+1}\ldots  n_{L}}=\sum_{n_{k}=1}^{N_{k}}\dot{u}_{mn_{k}}\dot{t}_{n_{1}n_{2}\ldots n_{L}}.
	\end{equation}
\end{definition}
From \textbf{Definition \ref{mode product}}, we can find the following facts:
\begin{enumerate}
	\item [\ding{202}] The $k$-mode product can also be expressed in
	terms of unfolded quaternion tensors:
	\begin{equation*}
	\dot{\mathcal{Y}}=\dot{\mathcal{T}}\times_{k}\dot{\mathbf{U}}\Longleftrightarrow \dot{\mathbf{Y}}_{[k]}=\dot{\mathbf{U}}\dot{\mathbf{T}}_{[k]}.
	\end{equation*}
	\item [\ding{203}] If the modes are the same, then
	\begin{equation*}
	\dot{\mathcal{T}}\times_{k}\dot{\mathbf{U}}_{1}\times_{k}\dot{\mathbf{U}}_{2}=\dot{\mathcal{T}}\times_{k}(\dot{\mathbf{U}}_{2}\dot{\mathbf{U}}_{1}).
	\end{equation*} 
	\item [\ding{204}] For distinct modes in a series of multiplications, the sequence of the multiplication
	is vitally important, \emph{i.e.}, if $k_{1}\neq k_{2}$, generally 
	\begin{equation*}
	\dot{\mathcal{T}}\times_{k_{1}}\dot{\mathbf{U}}_{1}\times_{k_{2}}\dot{\mathbf{U}}_{2}\neq \dot{\mathcal{T}}\times_{k_{2}}\dot{\mathbf{U}}_{2}\times_{k_{1}}\dot{\mathbf{U}}_{1}.  
	\end{equation*}
	\item [\ding{205}] If $\dot{\mathcal{Y}}=\dot{\mathcal{T}}\times_{1}\dot{\mathbf{U}}_{1}\times_{2}\dot{\mathbf{U}}_{2}\ldots\times_{L}\dot{\mathbf{U}}_{L}$, then 
	\begin{equation*}
	\dot{\mathbf{Y}}_{[L]}=\dot{\mathbf{U}}_{L}\left(\big(\dot{\mathbf{U}}_{L-1}\otimes\ldots\otimes\dot{\mathbf{U}}_{1}\big)\dot{\mathbf{T}}_{[L]}^{T}\right)^{T}
	\end{equation*}
	but for $k\neq L$, generally
	\begin{equation*}
	\begin{split}
	\dot{\mathbf{Y}}_{[k]}&\neq\dot{\mathbf{U}}_{k}\left(\big(\dot{\mathbf{U}}_{L}\otimes\ldots \otimes\dot{\mathbf{U}}_{k+1}\otimes\dot{\mathbf{U}}_{k-1}\right.\\
	&\left.\otimes\ldots\otimes\dot{\mathbf{U}}_{1}\big)\dot{\mathbf{T}}_{[k]}^{T}\right)^{T}.
	\end{split}
	\end{equation*}
\end{enumerate}
Facts \ding{202} and \ding{203} are the same as their real-valued counterparts \cite{DBLP:journals/siamrev/KoldaB09}. However, due to the non-commutativity of quaternion multiplication, and the property of quaternion matrix (\emph{see} \textbf{Theorem \ref{th2}} \circled{c}), facts \ding{204} and \ding{205} are completely different from their real-valued counterparts.

\section{Quaternion higher order singular value decomposition}
\label{sec:qhosvd}
The HOSVD is one of the most efficient tensor decomposition techniques \cite{DBLP:journals/siammax/LathauwerMV00}. Motivated by the outstanding ability of the HOSVD to represent high-dimensional data and extract features, in this section, we try to generalize the HOSVD to the quaternion domain and define QHOSVD.

\begin{theorem} Every quaternion tensor $\dot{\mathcal{T}}\in\mathbb{H}^{N_{1}\times N_{2} \times\ldots \times N_{L}}$ can be written as the form
\begin{equation}
\label{qhosvd}
\dot{\mathcal{T}}=\dot{\mathcal{S}}\times_{1}\dot{\mathbf{U}}_{1}\times_{2}\dot{\mathbf{U}}_{2}\ldots\times_{L}\dot{\mathbf{U}}_{L},
\end{equation}	
where $\dot{\mathbf{U}}_{l}\in\mathbb{H}^{N_{l}\times N_{l}} (l=1,\ldots, L)$ are unitary quaternion matrices, $\dot{\mathcal{S}}\in\mathbb{H}^{N_{1}\times N_{2} \times\ldots \times N_{L}}$ is a quaternion tensor with the same size as $\dot{\mathcal{T}}$. The quaternion sub-tensors $\dot{\mathcal{S}}_{n_{l}=\alpha}$ obtained by fixing the $l$th index of $\dot{\mathcal{S}}$ to $\alpha$ have the property that
\begin{equation}\label{order0}
\|\dot{\mathcal{S}}_{n_{l}=1}\|_{F}\geq\|\dot{\mathcal{S}}_{n_{l}=2}\|_{F}\geq\ldots\geq\|\dot{\mathcal{S}}_{n_{l}=N_{l}}\|_{F}\geq 0.
\end{equation}
\end{theorem}
\textit{proof.}
From fact \ding{205}, equation (\ref{qhosvd}) can be expressed in a quaternion matrix format as
\begin{equation}
\label{ul}
\dot{\mathbf{T}}_{[L]}=\dot{\mathbf{U}}_{L}\left(\big(\dot{\mathbf{U}}_{L-1}\otimes\ldots\otimes\dot{\mathbf{U}}_{1}\big)\dot{\mathbf{S}}_{[L]}^{T}\right)^{T}.
\end{equation}
Now consider the particular case where $\dot{\mathbf{U}}_{L}$ is obtained from the QSVD of $\dot{\mathbf{T}}_{[L]}$ as
\begin{equation}\label{ulsvd}
\dot{\mathbf{T}}_{[L]}=\dot{\mathbf{U}}_{L}\mathbf{\Sigma}_{L}\dot{\mathbf{V}}_{L}^{H}.
\end{equation}
 Equation (\ref{qhosvd}) can then be expressed in a quaternion matrix format as 
\begin{equation}
\label{ul-1}
\begin{split}
{\rm{Unfold}}_{L-1}(\dot{\mathcal{T}}\times_{L}\dot{\mathbf{U}}_{L}^{H})=&\dot{\mathbf{U}}_{L-1}\left(\big(\dot{\mathbf{U}}_{L-2}\otimes\ldots\right.\\
&\left.\otimes\dot{\mathbf{U}}_{1}\big)\dot{\mathbf{S}}_{[L-1]}^{T}\right)^{T}.
\end{split}
\end{equation}
Then, the particular case where $\dot{\mathbf{U}}_{L-1}$ is obtained from the QSVD of ${\rm{Unfold}}_{L-1}(\dot{\mathcal{T}}\times_{L}\dot{\mathbf{U}}_{L}^{H})$ as
\begin{equation}\label{ul-1svd}
{\rm{Unfold}}_{L-1}(\dot{\mathcal{T}}\times_{L}\dot{\mathbf{U}}_{L}^{H})=\dot{\mathbf{U}}_{L-1}\mathbf{\Sigma}_{L-1}\dot{\mathbf{V}}_{L-1}^{H}.
\end{equation} 
 Continue the above procedure in sequence, and equation (\ref{qhosvd}) finally can be expressed in a quaternion matrix format as 
\begin{equation}\label{u1}
\begin{split}
{\rm{Unfold}}_{1}(\dot{\mathcal{T}}\times_{L}\!\dot{\mathbf{U}}_{L}^{H}\times_{L-1}\!\dot{\mathbf{U}}_{L-1}^{H}\ldots\times_{2}\!\dot{\mathbf{U}}_{2}^{H})\!=\!\dot{\mathbf{U}}_{1}\dot{\mathbf{S}}_{[1]}.
\end{split}
\end{equation}
The particular case where $\dot{\mathbf{U}}_{1}$ is obtained from the QSVD of ${\rm{Unfold}}_{1}(\dot{\mathcal{T}}\times_{L}\dot{\mathbf{U}}_{L}^{H}\times_{L-1}\dot{\mathbf{U}}_{L-1}^{H}\ldots\!\times_{2}\!\dot{\mathbf{U}}_{2}^{H})$ as
\begin{equation}\label{u1svd}
{\rm{Unfold}}_{1}(\dot{\mathcal{T}}\times_{L}\dot{\mathbf{U}}_{L}^{H}\times_{L-1}\dot{\mathbf{U}}_{L-1}^{H}\ldots\times_{2}\dot{\mathbf{U}}_{2}^{H})\!=\!\dot{\mathbf{U}}_{1}\mathbf{\Sigma}_{1}\dot{\mathbf{V}}_{1}^{H}.
\end{equation}
Afterwards, the quaternion tensor $\dot{\mathcal{S}}$ can be obtained by 
\begin{equation}\label{s}
\dot{\mathcal{S}}=\dot{\mathcal{T}}\times_{L}\dot{\mathbf{U}}_{L}^{H}\times_{L-1}\dot{\mathbf{U}}_{L-1}^{H}\ldots\times_{1}\dot{\mathbf{U}}_{1}^{H}.
\end{equation}
Then, taking $\dot{\mathcal{S}}_{n_{L}}$ as an example, we show the property (\ref{order0}). Comparing (\ref{ul}) and (\ref{ulsvd}), we can find that 
\begin{equation}\label{sl}
\dot{\mathbf{S}}_{[L]}=\mathbf{\Sigma}_{L}\left(\dot{\mathbf{V}}_{L}^{T}\big(\dot{\mathbf{U}}_{L-1}\otimes\ldots\otimes\dot{\mathbf{U}}_{1}\big)\right)^{\ast},
\end{equation}
in which $\mathbf{\Sigma}_{L}={\rm{diag}}(\sigma_{L}^{(1)},\sigma_{L}^{(2)},\ldots,\sigma_{L}^{(N_{L})})$, where 
\begin{equation}\label{order}
\sigma_{L}^{(1)}\geq\sigma_{L}^{(2)}\geq\ldots\geq\sigma_{L}^{(N_{L})}\geq 0.
\end{equation}
We call $r_{L}$ the highest index for which $\sigma_{L}^{r_{L}}>0$, which combining (\ref{order}) means\footnote{Note that $ \dot{\mathbf{V}}_{L}$ and  $(\dot{\mathbf{U}}_{L-1}\otimes\ldots\otimes\dot{\mathbf{U}}_{1})$ are unitary, thus the Frobenius norm of each row vector of $\left(\dot{\mathbf{V}}_{L}^{T} \big(\dot{\mathbf{U}}_{L-1}\otimes\ldots\otimes\dot{\mathbf{U}}_{1}\big)\right)^{\ast}$ is $1$.} 
\begin{equation}\label{order2}
\|\dot{\mathcal{S}}_{n_{L}=1}\|_{F}\geq\|\dot{\mathcal{S}}_{n_{L}=2}\|_{F}\geq\ldots\geq\|\dot{\mathcal{S}}_{n_{L}=r_{L}}\|_{F},
\end{equation}
and
\begin{equation}\label{order3}
\|\dot{\mathcal{S}}_{n_{L}=r_{L}+1}\|_{F}=\ldots=\|\dot{\mathcal{S}}_{n_{L}=N_{L}}\|_{F}=0.
\end{equation}
Consequently, for any quaternion tensor $\dot{\mathcal{T}}\in\mathbb{H}^{N_{1}\times N_{2} \times\ldots \times N_{L}}$, we can always find unitary quaternion matrices $\dot{\mathbf{U}}_{l}\in\mathbb{H}^{N_{l}\times N_{l}} (l=1,\ldots, L)$ and a corresponding quaternion tensor $\dot{\mathcal{S}}$ with the property (\ref{order0}) such that equation (\ref{qhosvd}) holds. 
The proof is completed.

\begin{figure}[htbp]
	\centering
	\includegraphics[width=7.2cm,height=3.5cm]{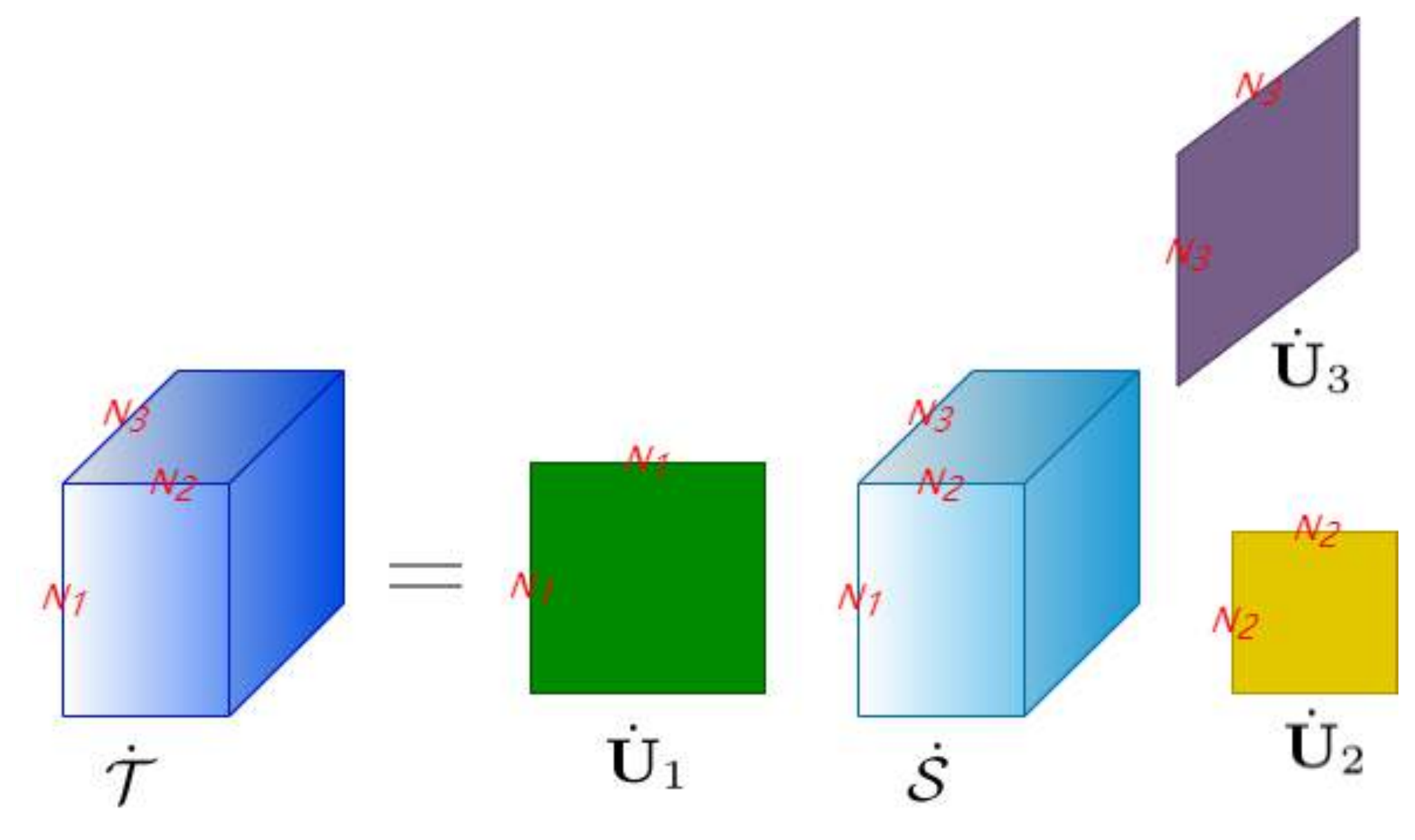}
	\caption{Visualization of the QHOSVD for a third-order quaternion tensor.}
	\label{vis_QHOSVD}
\end{figure}

Based on \textbf{Theorem 1}, we can define the following QHOSVD.
\begin{definition}(QHOSVD) For an $L$th-order quaternion tensor $\dot{\mathcal{T}}\in\mathbb{H}^{N_{1}\times N_{2} \times\ldots \times N_{L}}$, its QHOSVD is defined by
\begin{equation}
\label{qhosvd_d}
\dot{\mathcal{T}}=\dot{\mathcal{S}}\times_{1}\dot{\mathbf{U}}_{1}\times_{2}\dot{\mathbf{U}}_{2}\ldots\times_{L}\dot{\mathbf{U}}_{L},
\end{equation}	
where $\dot{\mathbf{U}}_{l}\in\mathbb{H}^{N_{l}\times N_{l}} (l=1,\ldots, L)$ are unitary quaternion matrices, $\dot{\mathcal{S}}\in\mathbb{H}^{N_{1}\times N_{2} \times\ldots \times N_{L}}$ is a quaternion tensor with the property
\begin{equation*}
\|\dot{\mathcal{S}}_{n_{l}=1}\|_{F}\geq\|\dot{\mathcal{S}}_{n_{l}=2}\|_{F}\geq\ldots\geq\|\dot{\mathcal{S}}_{n_{l}=N_{l}}\|_{F}\geq 0
\end{equation*}
for all possible values of $l$. The Frobenius-norms $\|\dot{\mathcal{S}}_{n_{l}=n}\|_{F}$,  labeled by $\sigma_{n}^{(l)}$, are $l$th-mode singular values of $\dot{\mathcal{T}}$ and the vector $\dot{\mathbf{U}}_{l}(:,n)$ is an $n$th $l$-mode singular vector. The decomposition is visualized for third-order quaternion tensors in Fig. \ref{vis_QHOSVD}.
\end{definition}

\textbf{Remark 1.} The proof procedure of \textbf{Theorem 1} actually indicates how the QHOSVD of a given quaternion tensor $\dot{\mathcal{T}}$ can be computed, which is summarized in Table \ref{tab_qhosvd}.
\begin{table}[htbp]
	\caption{The calculation procedure of QHOSVD}
	\hrule
	\label{tab_qhosvd}
	\begin{algorithmic}[1]
		\REQUIRE Quaternion tensor $\dot{\mathcal{T}}\in\mathbb{H}^{N_{1}\times N_{2} \times\ldots \times N_{L}}$.
		\STATE $\dot{\mathbf{U}}_{L}\longleftarrow  \text{left singular vectors of}\   {\rm{Unfold}}_{L}(\dot{\mathcal{T}})$.
		\FOR {$l=(L-1):-1:1$}
		\STATE  $\dot{\mathbf{U}}_{l}\longleftarrow  \text{left singular vectors of}\   {\rm{Unfold}}_{l}(\dot{\mathcal{T}}\times_{L}\dot{\mathbf{U}}_{L}\ldots\times_{l+1}\dot{\mathbf{U}}_{l+1})$.
		\ENDFOR
		\STATE  $\dot{\mathcal{S}}\longleftarrow \dot{\mathcal{T}}\times_{L}\dot{\mathbf{U}}_{L}^{H}\times_{L-1}\dot{\mathbf{U}}_{L-1}^{H}\ldots\times_{1}\dot{\mathbf{U}}_{1}^{H}$.
		\ENSURE $\dot{\mathcal{S}}$, $\{\dot{\mathbf{U}}_{1},\dot{\mathbf{U}}_{2},\ldots,\dot{\mathbf{U}}_{L}\}$.
	\end{algorithmic}
	\hrule
\end{table}

\textbf{Remark 2.} The fact \ding{205} determines that the calculation procedure of QHOSVD is different from that of HOSVD proposed in \cite{DBLP:journals/siammax/LathauwerMV00}. Another difference from HOSVD is that our definition of QHOSVD does not require $\dot{\mathcal{S}}$ to be all-orthogonality\footnote{All-orthogonality for real-valued tensor $\mathcal{S}\in\mathbb{R}^{N_{1}\times N_{2} \times\ldots \times N_{L}}$  \cite{DBLP:journals/siammax/LathauwerMV00}: two sub-tensors $\mathcal{S}_{n_{l}=\alpha}$ and $\mathcal{S}_{n_{l}=\beta}$ for all possible values of $l$, $\alpha$ and $\beta$ satisfy $\langle\mathcal{S}_{n_{l}=\alpha},\mathcal{S}_{n_{l}=\beta}\rangle=0$ when $\alpha\neq\beta$.}. The first reason is that it is quite difficult for the quaternion tensor $\dot{\mathcal{S}}$ to satisfy the same all-orthogonality as that of the real-valued tensor in \cite{DBLP:journals/siammax/LathauwerMV00}. The orthogonality is satisfied only when  $l=1$ for the obtained $\dot{\mathcal{S}}$. Since comparing (\ref{u1}) and (\ref{u1svd}), we can find that $\dot{\mathbf{S}}_{[1]}=\mathbf{\Sigma}_{1}\dot{\mathbf{V}}_{1}^{H}$, where $\dot{\mathbf{V}}_{1}^{H}$ is unitary, which means $\langle\dot{\mathcal{S}}_{n_{1}=\alpha},\dot{\mathcal{S}}_{n_{1}=\beta}\rangle=0$. However, for other values of $l$, the orthogonality is generally not satisfied.
 From equation (\ref{sl}), for example, we can find that   $\left(\dot{\mathbf{V}}_{L}^{T}\big(\dot{\mathbf{U}}_{L-1}\otimes\ldots\otimes\dot{\mathbf{U}}_{1}\big)\right)^{\ast}$ is generally no longer a unitary matrix since $\dot{\mathbf{V}}_{L}^{T}$ is generally  no longer unitary (\emph{see} \textbf{Theorem \ref{th2}} \circled{d}), which means that generally $\langle\dot{\mathcal{S}}_{n_{L}=\alpha},\dot{\mathcal{S}}_{n_{L}=\beta}\rangle\neq0$, the same is true for all the other values of $l$ except $l=1$. The second reason is that the definition of QHOSVD in this paper is application-oriented. In practice, it is irrelevant whether $\dot{\mathcal{S}}$ is all-orthogonality or not in the considered applications.

\section{Applications of QHOSVD in color image processing}
\label{app:qhosvd}
\begin{figure}[htbp]
	\centering
	\includegraphics[width=8cm,height=2.5cm]{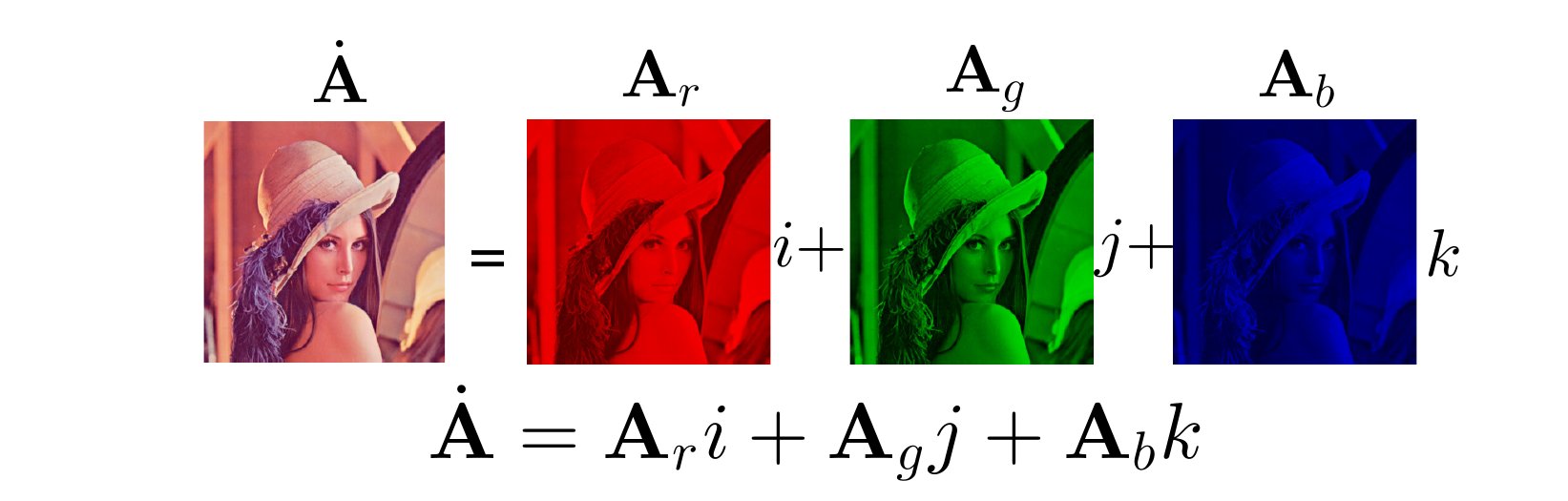}
	\caption{Quaternion matrix represented color image.}
	\label{rgb}
\end{figure}

In this section, based on the defined QHOSVD, we develop a multi-focus color image fusion method and a color image denoising method. In these methods, each color pixel with R, G, B channels is encoded as a pure quaternion unit. That is
\begin{equation}
\dot{a}=a_{r}i+a_{g}j+a_{b}k,
\end{equation}
where $\dot{a}$ denotes a color pixel, the coefficients $a_{r}$, $a_{g}$, $a_{b}$ of the imaginary parts are respectively the pixel values in R, G, B channels. Naturally, a given color image with the spatial resolution of $N_{1}\times N_{2}$ pixels can be represented by a pure quaternion matrix $\dot{\mathbf{A}}=(\dot{a}_{n_{1}n_{2}})\in\mathbb{H}^{N_{1}\times N_{2}}$, $1\leq n_{1}\leq N_{1}$, $1\leq n_{2}\leq N_{2}$ as follows:
\begin{equation}
	\label{quaternion_rc}
\dot{\mathbf{A}}=\mathbf{A}_{r}i+\mathbf{A}_{g}j+\mathbf{A}_{b}k,
\end{equation}
where $\mathbf{A}_{r}, \mathbf{A}_{g}, \mathbf{A}_{b}\in\mathbb{R}^{N_{1}\times N_{2}}$ containing respectively R, G, B pixel values. Fig. \ref{rgb} shows an example of using a quaternion matrix to represent a color image.

From (\ref{quaternion_rc}), we can see that, based on the quaternion representation, when we process the color image $\dot{\mathbf{A}}$, the three RGB channels can be handled holistically. Thus, the  high correlation among RGB  channels can be fully utilized. Different from the real third-order tensor representation, the quaternion matrix is still a two-dimensional array, which has no dimension increase and is easy to calculate and process.

\subsection{Multi-focus color image fusion using QHOSVD}
Multi-focus color image fusion is to create an all-in-focus color image from a set of partially focused color images of the same scene. Since these partially focused color images refer to the same scene and are highly similar, we construct them (represented as quaternion matrices) into a third-order quaternion tensor and adopt the proposed QHOSVD technique to extract their features simultaneously.

\begin{figure*}[htbp]
	\centering
	\includegraphics[width=15cm,height=7cm]{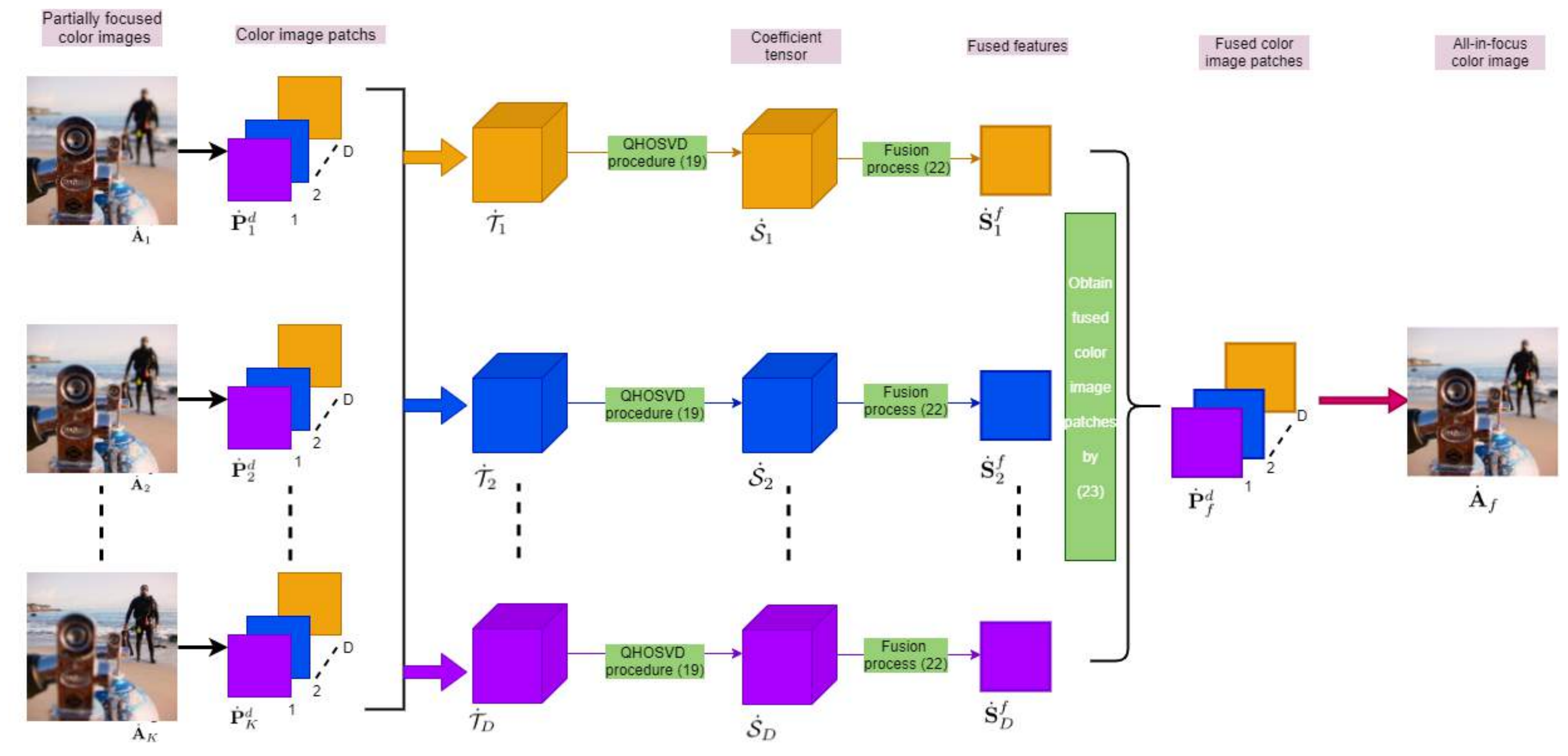}
	\caption{The flowchart of multi-focus color image fusion by QHOSVD.}
	\label{Flow_diagram1}
\end{figure*}

The whole framework of the proposed multi-focus color image fusion method is shown in Fig. \ref{Flow_diagram1}. We, in the following, show the detailed procedure of the proposed method step by step.

\textbf{Step 1}: All the partially focused color images are represented by pure quaternion matrices $\dot{\mathbf{A}}_{1}, \dot{\mathbf{A}}_{2}, \ldots, \dot{\mathbf{A}}_{K}\in\mathbb{H}^{N_{1}\times N_{2}}$. Then, the sliding window technique is applied to divide each
source color image $\dot{\mathbf{A}}_{k}, (k=1, \ldots, K)$ into $D$ overlapping patches $\dot{\mathbf{P}}_{k}^{d}\in\mathbb{H}^{M_{1}\times M_{2}}, (d=1, \ldots, D)$. Afterwards, for each $d$, all of the patches from $K$ source color images are stacked together to form a third-order quaternion tensor $\dot{\mathcal{T}}_{d}\in\mathbb{H}^{M_{1}\times M_{2}\times K}, (d=1, \ldots, D)$.

\textbf{Step 2}: Using QHOSVD procedure on $\dot{\mathcal{T}}_{d}$ to find two unitary matrices $\dot{\mathbf{U}}_{1}^{d}\in\mathbb{H}^{M_{1}\times M_{1}}$ and $\dot{\mathbf{U}}_{2}^{d}\in\mathbb{H}^{M_{2}\times M_{2}}$, and the corresponding third-order quaternion tensor $\dot{\mathcal{S}}_{d}\in\mathbb{H}^{M_{1}\times M_{2}\times K}$ such that
\begin{equation}
\label{qhosvd_two}
\dot{\mathcal{T}}_{d}=\dot{\mathcal{S}}_{d}\times_{1}\dot{\mathbf{U}}_{1}^{d}\times_{2}\dot{\mathbf{U}}_{2}^{d}.
\end{equation}
\textbf{Remark 3.} It can be found that in (\ref{qhosvd_two}) we do not perform a complete QHOSVD procedure (\emph{i.e.}, there is no $\dot{\mathbf{U}}_{3}^{d}$ in (\ref{qhosvd_two})), which is different from the traditional HOSVD-based methods \cite{DBLP:journals/tip/LiangHLZ12,DBLP:journals/jvcir/LuoZZW17}. The reasons are listed below.
\begin{enumerate}
	\item[(a)]
 In (\ref{qhosvd_two}), $\dot{\mathcal{S}}_{d}$ is the coefficient tensor, and generally can be seen as the feature of $\dot{\mathcal{T}}_{d}$. Our goal is to find the features corresponding to each source color image patch $\dot{\mathcal{T}}_{d}(:,:,k), (k=1, \ldots, K)$. After a simple but important derivation, we can find that
	\begin{equation}
	\label{feature}
		\begin{split}
	\dot{\mathcal{T}}_{d}(:,:,k)&=\left(\dot{\mathbf{U}}_{2}^{d}\,\big(\dot{\mathbf{U}}_{1}^{d}\,\dot{\mathcal{S}}_{d}(:,:,k)\big)^{T}\right)^{T},\\& \text{for} \quad k=1, \ldots, K.
		\end{split}
	\end{equation}
From (\ref{feature}), $\dot{\mathcal{S}}_{d}(:,:,k)$ can be seen as the feature of source color image patch $\dot{\mathcal{T}}_{d}(:,:,k)$, for $k=1, \ldots, K$.  Because for all $\dot{\mathcal{S}}_{d}(:,:,k) \ (k=1, \ldots, K)$, they share the common $\dot{\mathbf{U}}_{1}$ and $\dot{\mathbf{U}}_{2}$, the features are extracted simultaneously.
\item[(b)] If we perform a complete QHOSVD procedure on $\dot{\mathcal{T}}_{d}$, \emph{i.e.},
\begin{equation}
\label{qhosvd_three}
\dot{\mathcal{T}}_{d}=\tilde{\dot{\mathcal{S}}}_{d}\times_{1}\dot{\mathbf{U}}_{1}^{d}\times_{2}\dot{\mathbf{U}}_{2}^{d}\times_{3}\dot{\mathbf{U}}_{3}^{d},
\end{equation}
 it will be pretty difficult to find the corresponding features of each source color image patch $\dot{\mathcal{T}}_{d}(:,:,k), (k=1, \ldots, K)$. Since, based on the fact \ding{204}, generally $\tilde{\dot{\mathcal{S}}}_{d}\times_{3}\dot{\mathbf{U}}_{3}^{d}\neq\dot{\mathcal{S}}_{d}$, which is different from real-valued HOSVD procedure \cite{DBLP:journals/siammax/LathauwerMV00}.
 \item[(c)] The traditional HOSVD-based methods \cite{DBLP:journals/tip/LiangHLZ12,DBLP:journals/jvcir/LuoZZW17}  are to make a complete HOSVD  procedure first, and then multiply the third mode of the obtained feature tensor by the corresponding third singular vector matrix, to obtain a new feature tensor. Even if we could do the same process, it would be redundant and not conducive to saving calculation costs.
\end{enumerate}

\textbf{Step 3}: To fuse features $\dot{\mathcal{S}}_{d}(:,:,k), (k=1, \ldots, K)$ by the following way:
\begin{equation}\label{fuse_features}
\dot{\mathbf{S}}_{d}^{f}=\dot{\mathcal{S}}_{d}(:,:,\xi),
\end{equation}
in which the index $\xi$ satisfies
\begin{equation*}
\|\dot{\mathcal{S}}_{d}(:,:,\xi)\|_{L_{1}}=\mathop{{\rm{max}}}\limits_{k\in(1,2,\ldots,K)}\{\|\dot{\mathcal{S}}_{d}(:,:,k)\|_{L_{1}}\}.
\end{equation*} 
Or $\dot{\mathbf{S}}_{d}^{f}=\big(\dot{\mathcal{S}}_{d}(:,:,1)+\ldots+\dot{\mathcal{S}}_{d}(:,:,K)\big)/K$, if $\|\dot{\mathcal{S}}_{d}(:,:,1)\|_{L_{1}}=\ldots=\|\dot{\mathcal{S}}_{d}(:,:,K)\|_{L_{1}}$.

\textbf{Step 4}:
The fused color image patch $\dot{\mathbf{P}}_{f}^{d}$ is obtained by:
\begin{equation}\label{fuse_images}
\dot{\mathbf{P}}_{f}^{d}=\left(\dot{\mathbf{U}}_{2}^{d}\,\big(\dot{\mathbf{U}}_{1}^{d}\,\dot{\mathbf{S}}_{d}^{f}\big)^{T}\right)^{T}.
\end{equation}

\textbf{Step 5}: Take all the fused color image patches $\dot{\mathbf{P}}_{f}^{d}, (d=1, \ldots, D)$ back to their original corresponding
positions. The final fused color image $\dot{\mathbf{A}}_{f}$ is reconstructed by averaging the overlapping fused color image patches. 

\subsection{Color image denoising using QHOSVD}
Color image denoising is to recover the clear image $\dot{\mathbf{X}}\in\mathbb{H}^{M\times N}$ from its noisy observation
\begin{equation}
\label{noise_eq1}
\dot{\mathbf{Y}}=\dot{\mathbf{X}}+\dot{\mathbf{G}},
\end{equation}
where $\dot{\mathbf{G}}$ is assumed to be Gaussian noise. The recent quaternion matrix low-rank approximation-based color image denoising methods proposed in \cite{DBLP:journals/tip/ChenXZ20,DBLP:journals/ijon/YuZY19} are also based on the above observation model (\ref{noise_eq1}).
And the NNS which has significantly boosted the performance of image denoising is adopted in these works. However, these methods may destroy the two-dimensional structure information of color image patches when they are expanded into one-dimensional quaternion vectors. To avoid pulling color image patches into one-dimensional quaternion vectors, similar image patches are stacked into a third-order quaternion tensor, and then the proposed QHOSVD can be performed on it. The detailed procedure of color image denoising is listed below.

Firstly, using the sliding window technique, the noisy color image represented by pure quaternion matrix $\dot{\mathbf{Y}}$ is divided into overlapped patches $\dot{\mathbf{P}}_{i}\in\mathbb{H}^{w\times w}$. The $j$th reference patch $\dot{\mathbf{P}}_{ref}^{j}\in\mathbb{H}^{w\times w}$ is taken as the center of the searching window with size $W\times W$. For each reference patch $\dot{\mathbf{P}}_{ref}^{j}$, search for $K$ similar patches (including the reference patch itself) within its local searching window. The similarity is determined by the following distance:
\begin{equation*}
d(\dot{\mathbf{P}}_{ref}^{j},\dot{\mathbf{P}}_{i})=\|\dot{\mathbf{P}}_{ref}^{j}-\dot{\mathbf{P}}_{i}\|_{F}^{2}.
\end{equation*}
At last, for each reference patch $\dot{\mathbf{P}}_{ref}^{j}$, its $K$ similar patches are stacked together to form a third-order quaternion tensor $\dot{\mathcal{Y}}_{j}\in\mathbb{H}^{w\times w \times K}$. It should be noted that this step is different from that in \cite{DBLP:journals/tip/ChenXZ20} and \cite{DBLP:journals/ijon/YuZY19}, in which the $K$ similar patches are first expanded into one-dimensional quaternion vectors which then construct a quaternion matrix with size $w^2\times K$. Such operations may destroy the two-dimensional spatial structure of color image patches and may lead to the curse of dimensionality.

Following (\ref{noise_eq1}), each third-order quaternion tensor $\dot{\mathcal{Y}}_{j}$ has the following form:
\begin{equation}\label{noise_eq2}
\dot{\mathcal{Y}}_{j}=\dot{\mathcal{X}}_{j}+\dot{\mathcal{G}}_{j},
\end{equation}
where, $\dot{\mathcal{X}}_{j}\in\mathbb{H}^{w\times w \times K}$ is the quaternion tensor stacked by clear color image patches, which is what we need to estimate, and $\dot{\mathcal{G}}_{j}\in\mathbb{H}^{w\times w \times K}$ is the corresponding noise part. Then, we perform the QHOSVD procedure on $\dot{\mathcal{Y}}_{j}$, \emph{i.e.},
\begin{equation}\label{noise_eq3}
(\dot{\mathcal{S}}_{j}, \{\dot{\mathbf{U}}_{1}^{j},\dot{\mathbf{U}}_{2}^{j},\dot{\mathbf{U}}_{3}^{j}\})=QHOSVD(\dot{\mathcal{Y}}_{j}).
\end{equation}
The third-mode (similar to other modes) singular values of $\dot{\mathcal{Y}}_{j}$ are
\begin{equation}\label{noise_eq4}
\sigma_{k}^{(3)}=\|\dot{\mathcal{S}}_{j}(:,:,k)\|_{F}, \quad k=1,2,\ldots, K.
\end{equation}
From the formula (\ref{noise_eq4}), we can see that the coefficients in $\dot{\mathcal{S}}_{j}$ are actually to re-decompose the corresponding singular values so that the noise energy originally concentrated on the singular values is mainly transferred to the smaller coefficients in $\dot{\mathcal{S}}_{j}$ while larger coefficients are less affected by noise \cite{DBLP:journals/access/GaoGZCZ19}. Thus, to perform denoising, the smaller coefficients are truncated by adopting the following method of hard threshold shrinkage:
\begin{equation}\label{noise_eq5}
\hat{\dot{\mathcal{S}}}_{j}={\rm{Hard}}(\dot{\mathcal{S}}_{j},\tau),
\end{equation}
where $\rm{Hard}(\ast,\tau)$ is defined as
\begin{align}
\rm{Hard}(\ast,\tau)=\left\{
\begin{array}{lc}
\ast, \quad &\text{if} \ |\ast|\geq \tau,\\
0, &\text{otherwise},
\end{array}
\right.
\end{align}
and $\tau$ is chosen to be $\eta\sigma\sqrt{2log(\omega^{2}K)}$ (in which, $\sigma$ stands for the standard deviation of the noise,  $\eta$ is a scale that controls the smoothing degree in the denoising stage, $\sigma\sqrt{2log(\omega^{2}K)}$ is the optimal threshold from a statistical risk viewpoint \cite{1994Ideal}).

Afterwards, the denoised quaternion tensor $\hat{\dot{\mathcal{X}}}_{j}$ (\emph{i.e.}, the estimated $\dot{\mathcal{X}}_{j}$) can be obtained by inverse QHOSVD:
\begin{equation}\label{noise_eq6}
\hat{\dot{\mathcal{X}}}_{j}=\hat{\dot{\mathcal{S}}}_{j}\times_{1}\dot{\mathbf{U}}_{1}^{j}\times_{2}\dot{\mathbf{U}}_{2}^{j}\times_{3}\dot{\mathbf{U}}_{3}^{j}.
\end{equation}
That means the denoised $K$ similar patches corresponding to the reference patch $\dot{\mathbf{P}}_{ref}^{j}$ are obtained. Then, put all the denoised patches back to the original position of the color image. The final denoised color image $\hat{\dot{\mathbf{X}}}$ (\emph{i.e.}, the estimated $\dot{\mathbf{X}}$) is reconstructed by averaging all the overlapping color image patches together. And the iterative regularization scheme \cite{DBLP:journals/ijcv/GuXMZFZ17,DBLP:journals/ijon/YuZY19,DBLP:journals/tip/ChenXZ20}, \emph{etc.}, is adopted
\begin{equation*}
\dot{\mathbf{Y}}^{(\gamma)}=\hat{\dot{\mathbf{X}}}^{(\gamma-1)}+\delta(\dot{\mathbf{Y}}-\hat{\dot{\mathbf{X}}}^{(\gamma-1)}),
\end{equation*}
where $\gamma$ and $\delta$ are respectively the iteration number and the relaxation parameter.

\begin{figure}[htbp]
	\centering
	\includegraphics[width=7.5cm,height=4cm]{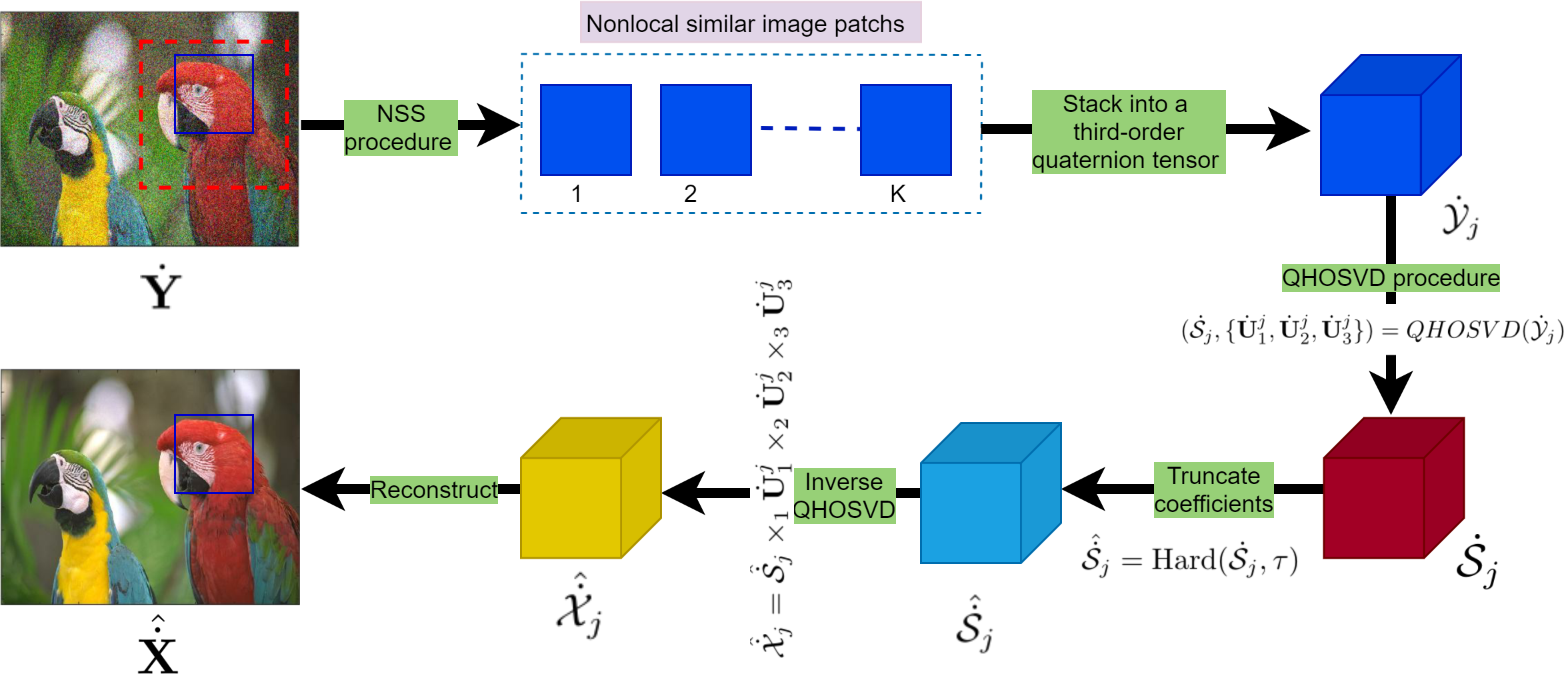}
	\caption{The flowchart of color image denoising by QHOSVD.}
	\label{Flow_diagram2}
\end{figure}

The  flowchart  of  QHOSVD  for  color image denoising is shown in Fig. \ref{Flow_diagram2},  The whole denoising algorithm can be summarized in Table \ref{tab_algorithm0}.

\begin{table}[htbp]
	\caption{Color image denoising by QHOSVD.}
	\hrule
	\label{tab_algorithm0}
	\begin{algorithmic}[1]
		\REQUIRE The observed noisy color image represented by a quaternion matrix $\dot{\mathbf{Y}}$, regularization parameter $\delta$, search window size $W\times W$, patch size $\omega\times \omega$, similar patch number $K$, iteration number $\varUpsilon$.
		\STATE \textbf{Initialize} $\hat{\dot{\mathbf{X}}}^{(0)}=\dot{\mathbf{Y}}$, $\dot{\mathbf{Y}}^{(0)}=\dot{\mathbf{Y}}$.
		\FOR{$\gamma=1 : \varUpsilon$}
		\STATE Iterative regularization $\dot{\mathbf{Y}}^{(\gamma)}=\hat{\dot{\mathbf{X}}}^{(\gamma-1)}+\delta(\dot{\mathbf{Y}}-\hat{\dot{\mathbf{X}}}^{(\gamma-1)})$
		\FOR{each  reference patch $\dot{\mathbf{P}}_{ref}^{j}$ in $\dot{\mathbf{Y}}^{(i)}$}
		\STATE Find $K$ similar patches, and stack into a third-order quaternion tensor $\dot{\mathcal{Y}}_{j}\in\mathbb{H}^{w\times w \times K}$.
		\STATE  Perform the QHOSVD on $\dot{\mathcal{Y}}_{j}$: $(\dot{\mathcal{S}}_{j}, \{\dot{\mathbf{U}}_{1}^{j},\dot{\mathbf{U}}_{2}^{j},\dot{\mathbf{U}}_{3}^{j}\})=QHOSVD(\dot{\mathcal{Y}}_{j})$.
		\STATE Truncate the coefficients $\dot{\mathcal{S}}_{j}$: $\hat{\dot{\mathcal{S}}}_{j}={\rm{Hard}}(\dot{\mathcal{S}}_{j},\tau)$.
		\STATE Perform the inverse QHOSVD on $\hat{\dot{\mathcal{S}}}_{j}$: $\hat{\dot{\mathcal{X}}}_{j}=\hat{\dot{\mathcal{S}}}_{j}\times_{1}\dot{\mathbf{U}}_{1}^{j}\times_{2}\dot{\mathbf{U}}_{2}^{j}\times_{3}\dot{\mathbf{U}}_{3}^{j}$.
		\ENDFOR 
		\STATE Compute the average of repeated estimates for every patch, and aggregate all patches together, get the clean image $\hat{\dot{\mathbf{X}}}^{(\gamma)}$.
		\ENDFOR  
		\ENSURE  Denoised color image $\hat{\dot{\mathbf{X}}}^{(\varUpsilon)}$.
	\end{algorithmic}
	\hrule
\end{table}


\section{Experimental results}
\label{exresults}
In this section, several experiments are conducted to evaluate the effectiveness of the  proposed QHOSVD-based multi-focus color image fusion method and color image denoising method. All the experiments  are run in MATLAB $2014b$ (except for one of the multi-focus color image fusion methods: MADCNN, which is performed in Python) under Windows $10$ on a personal computer with a $1.60$GHz CPU and $8$GB memory.

\subsection{Multi-focus color image fusion}
\begin{figure}[htbp]
	\centering
	\includegraphics[width=7.5cm,height=2.5cm]{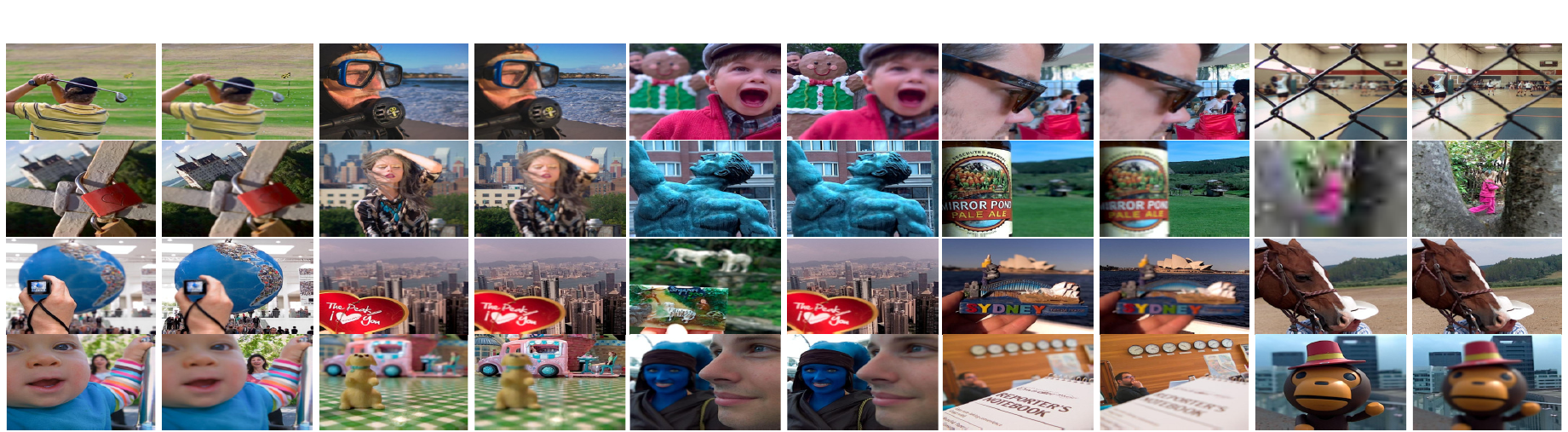}
	\caption{The used multi-focus color images in our experiments.}
	\label{fig1}
\end{figure}

\textbf{Dataset:} We conduct the experiments on a popular multi-focus image dataset, Lytro \cite{DBLP:journals/inffus/NejatiSS15}, which is publicly available online\footnote{\url{https://mansournejati.ece.iut.ac.ir/content/lytro-multi-focus-dataset}.}. The $20$ pairs (with two images) of color multi-focus images of size $520\times 520$ pixels are adopted from the dataset. They are shown in Fig. \ref{fig1}.

\textbf{Objective metrics:} In our experiments, six metrics that are widely used in multi-focus image fusion are adopted for evaluation and are listed below.  
\begin{itemize}
\item Information theory-based: Normalized mutual information $\rm{Q_{MI}}$ \cite{2008Comments} and Nonlinear correlation information entropy $\rm{Q_{NCIE}}$ \cite{Wang2005A}, which measure the amount of mutual information between the fused image and the source images.
\item Image feature-based: Multiscale scheme based $\rm{Q_{M}}$ \cite{2008A}, which measures the extent of edge information injected into the fused image from the source
images. 
\item Image structural similarity-based: $\rm{Q_{Y}}$ \cite{2008A}, which measures the amount of structural information preserved in the fused image.
\item Human perception-based: $\rm{Q_{CB}}$ \cite{DBLP:journals/ivc/ChenB09} and $\rm{Q_{CV}}$ \cite{DBLP:journals/inffus/ChenV07}, which address the major features in the human visual system.
\end{itemize}
For all metrics except $\rm{Q_{CV}}$, a larger value indicates a better fusion performance \cite{DBLP:journals/pami/LiuBXZLW12,DBLP:journals/inffus/LiuWCLC20}.	

\textbf{Methods for comparison:} We compare our QHOSVD-based method with several representative methods  including traditional ones and latest ones. Specifically, they are transform domain methods:
DWT \cite{DBLP:journals/cvgip/LiMM95}, NSCT \cite{2007Image},  MWGF \cite{DBLP:journals/inffus/ZhouLW14}, GFF \cite{DBLP:journals/tip/LiKH13}, QCT \cite{2012Multifocus},  HOSVD \cite{DBLP:journals/tip/LiangHLZ12}; spatial domain methods: QUADTREE \cite{DBLP:journals/inffus/BaiZZX15}, GFDF \cite{DBLP:journals/spic/QiuLZY19}, MISF \cite{DBLP:journals/jei/ZhanKLH19}; deep learning methods:
CNN \cite{DBLP:journals/inffus/LiuCPW17}, MADCNN \cite{DBLP:journals/access/LaiLGX19}.

\textbf{Parameter setting:}
For our method, the size of each patch is empirically set as $M_{1}=M_{2}=25$, and the patches are overlapped with $6$ pixels. All compared methods are from the source codes, and the parameter settings are based on the suggestions in the original papers.


\begin{figure*}[htbp]
	\centering
	\vspace{-0.05in}
	\subfigure[]{
		\includegraphics[width=7cm,height=3.5cm]{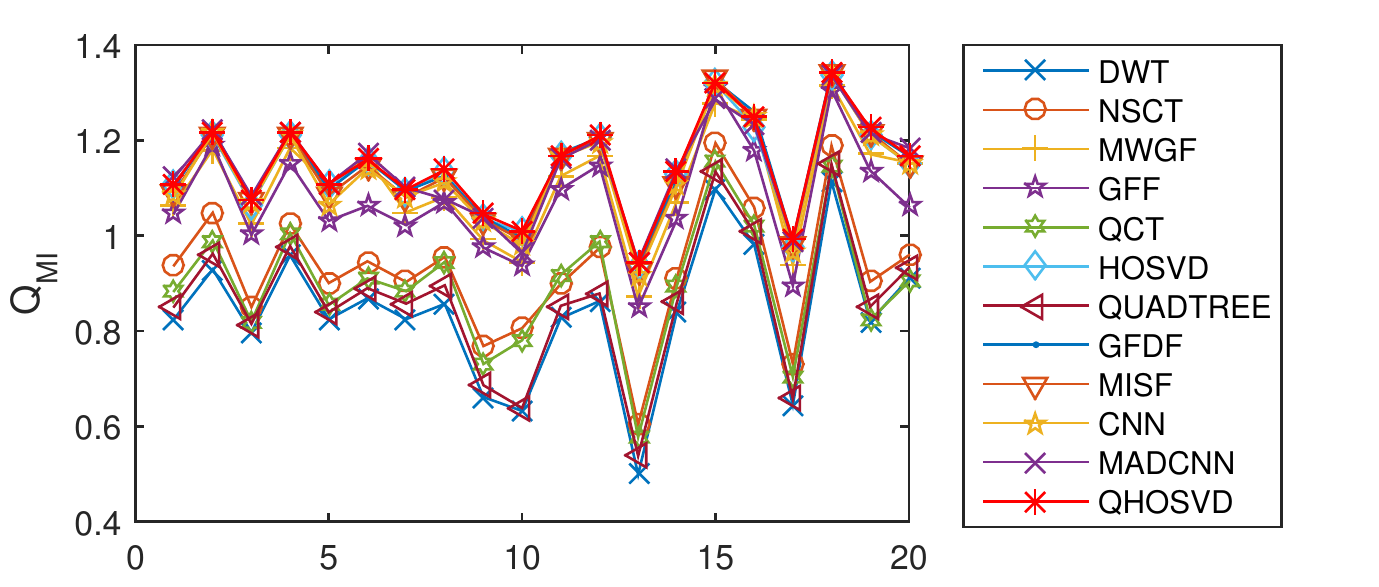}
	}
	\vspace{-0.05in}
	\subfigure[]{
		\includegraphics[width=7cm,height=3.5cm]{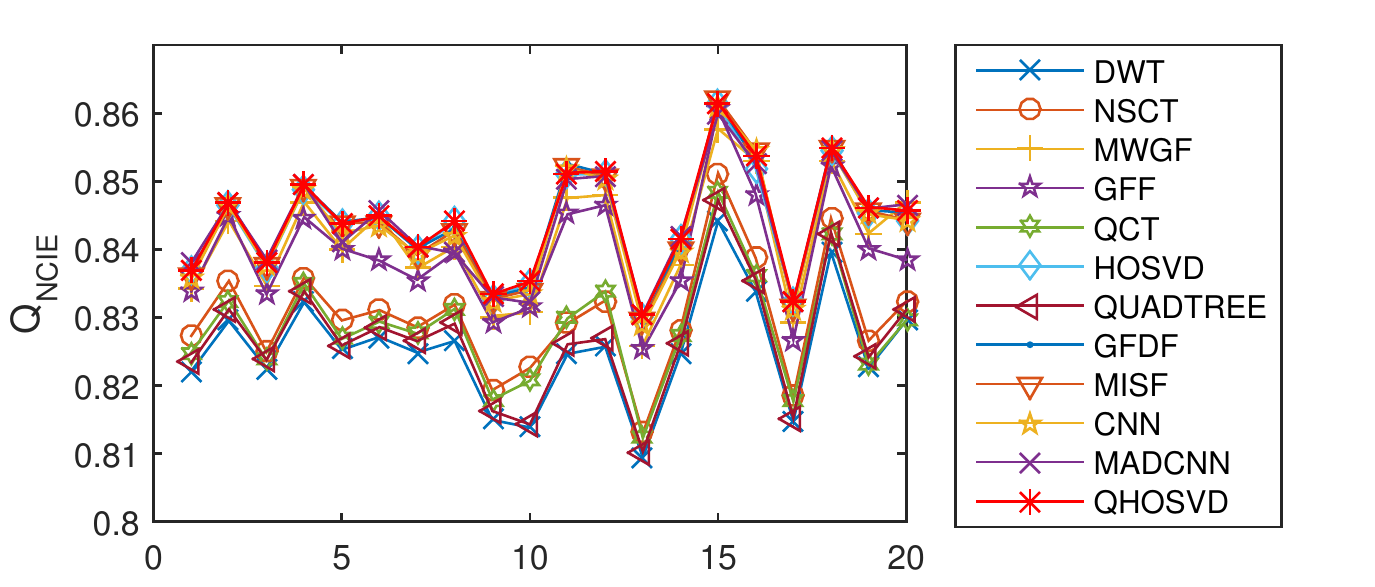}
	}
	\vspace{-0.05in}
	\subfigure[]{
		\includegraphics[width=7cm,height=3.5cm]{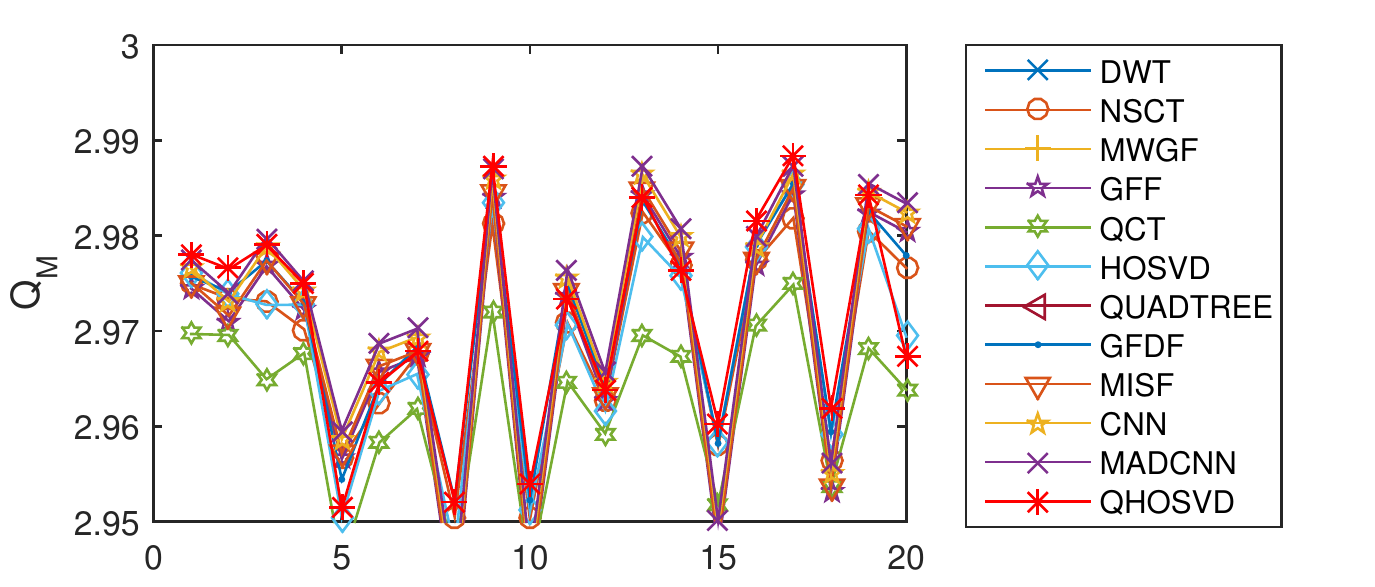}
	}
	\vspace{-0.05in}
	\subfigure[]{
		\includegraphics[width=7cm,height=3.5cm]{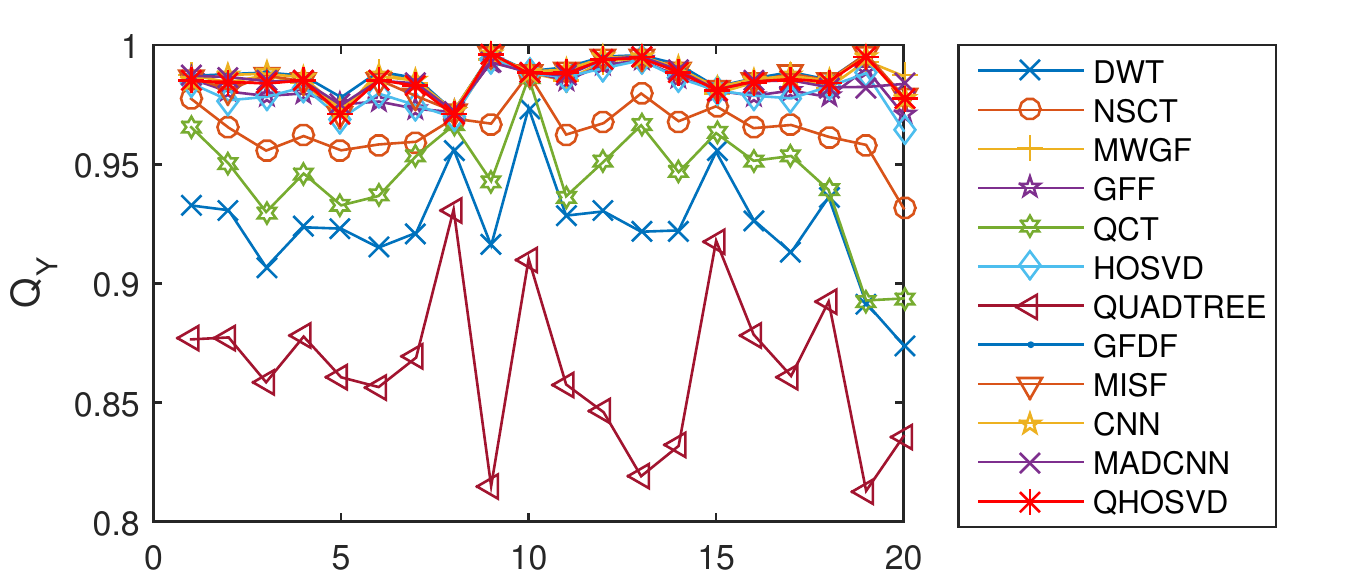}
	}\\
	\vspace{-0.05in}
	\subfigure[]{
		\includegraphics[width=7cm,height=3.5cm]{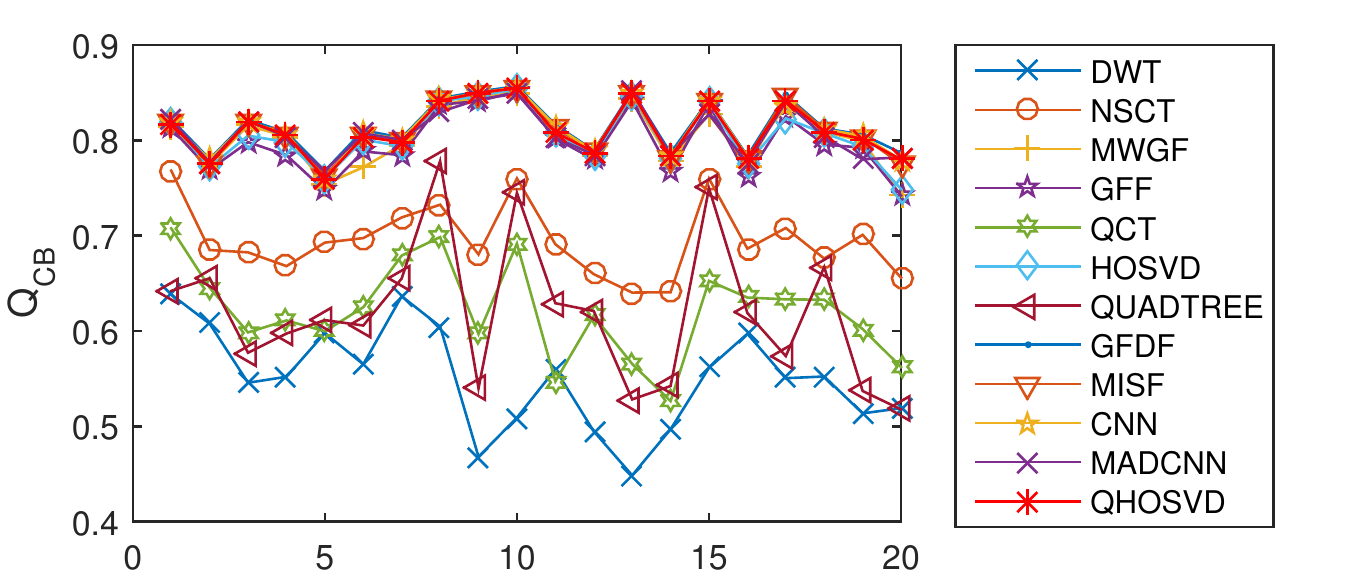}
	}
	\vspace{-0.05in}
	\subfigure[]{
		\includegraphics[width=7cm,height=3.5cm]{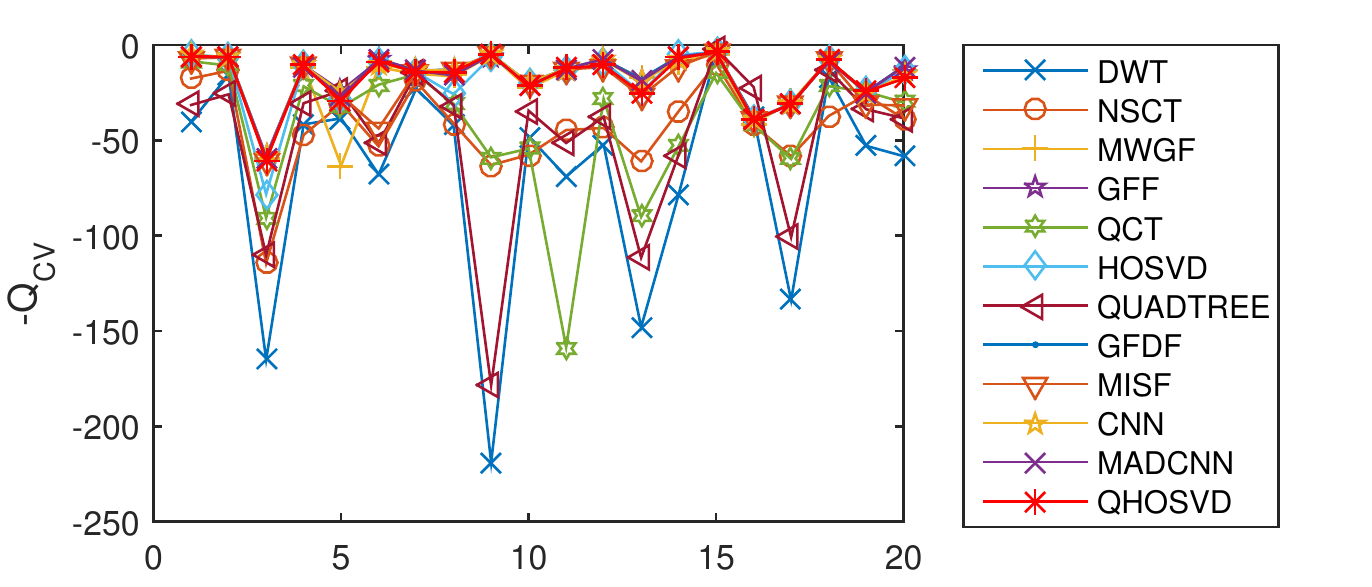}
	}	
	\caption{Objective performance of different fusion methods on the $20$ pairs of color multi-focus images (\textbf{the figure is viewed better in zoomed PDF}).}
	\label{fig01}
\end{figure*}

\begin{table}[htbp]  \caption{The average scores of different fusion methods (\textbf{bold} fonts denote the best performance; \underline{underline} ones represent the second-best results).}	
	\label{table_objective}
	\centering
	\resizebox{7.5cm}{2.5cm}{
		\begin{tabular}{|c|cccccc|c|}  	
			\hline	  		
			\diagbox{\scriptsize{Methods}}{\scriptsize{Metrics}} &$\rm{Q_{MI}}$ &$\rm{Q_{NCIE}}$&$\rm{Q_{M}}$  &$\rm{Q_{Y}}$&$\rm{Q_{CB}}$&$\rm{Q_{CV}}$&\textit{Time(s)}\\  \toprule		
			\hline  		
			\scriptsize{DWT} \cite{DBLP:journals/cvgip/LiMM95} &0.8389&0.8254&2.9091&0.9249&0.6232&67.5382 &0.2890\\  		
			\scriptsize{NSCT} \cite{2007Image}&0.9284&0.8301&2.9692&0.9646&0.6953&22.5183&14.1024\\      		
			\scriptsize{MWGF} \cite{DBLP:journals/inffus/ZhouLW14} &1.1076&0.8422&2.8991&0.9863&0.8036&19.3433&7.4339\\  		
			\scriptsize{GFF} \cite{DBLP:journals/tip/LiKH13} &1.0755 &0.8395&2.9688&0.9820&0.7981&\textbf{16.4563} &1.1895\\  	
			\scriptsize{QCT} \cite{2012Multifocus} &0.8970&0.8285&2.9622&0.9453&0.6214&43.3072&97.1247\\  
			\scriptsize{HOSVD} \cite{DBLP:journals/tip/LiangHLZ12}&1.1299&0.8431&2.9690&0.9813&0.8050&18.6918 &5.0940\\  	
			\scriptsize{QUADTREE} \cite{DBLP:journals/inffus/BaiZZX15}&0.8639&0.8269&2.8385&0.8643&0.6197&49.9656& 1.3634\\  
			\scriptsize{GFDF} \cite{DBLP:journals/spic/QiuLZY19}&1.1431 &\underline{0.8440}&2.9712&0.9870&\textbf{0.8145}&16.8288&0.3519  \\  
			\scriptsize{MISF} \cite{DBLP:journals/jei/ZhanKLH19}&1.1359&0.8435&2.9693&0.9860  &0.8114&20.7648  & 0.4688\\  
			\scriptsize{CNN} \cite{DBLP:journals/inffus/LiuCPW17}&1.1305 &0.8432&2.9710&\underline{0.9871}  &\underline{0.8118}&16.7943 &209.6695 \\  
			\scriptsize{MADCNN} \cite{DBLP:journals/access/LaiLGX19}&\underline{1.1421} &0.8439&\textbf{2.9717}&0.9858  &0.8078&\underline{16.7676}  &$\setminus$ \\  
			\scriptsize{\textbf{QHOSVD}} &\textbf{1.1465} &\textbf{0.8441}&\underline{2.9715}&\textbf{0.9874} &0.8093&17.6980 &7.8943\\  
		\hline  		
	\end{tabular}}	
\end{table}

\begin{figure}[htbp]
	\centering
	\hspace{-0.18in}
	\subfigure[\tiny{DWT} \cite{DBLP:journals/cvgip/LiMM95}]{
		\includegraphics[width=2.7cm,height=2cm]{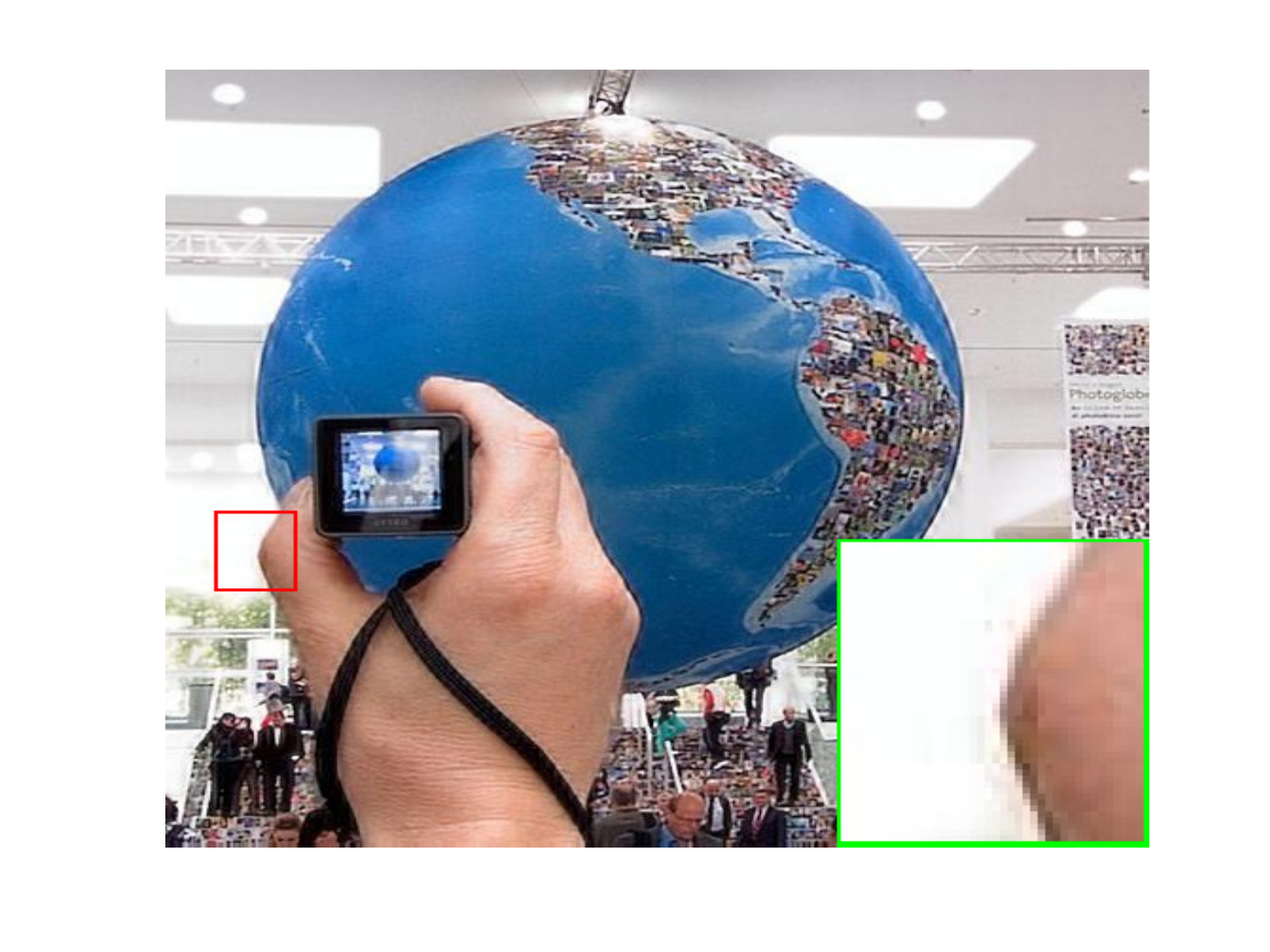}
	}
	\hspace{-0.18in}
	\subfigure[\tiny{NSCT} \cite{2007Image}]{
		\includegraphics[width=2.7cm,height=2cm]{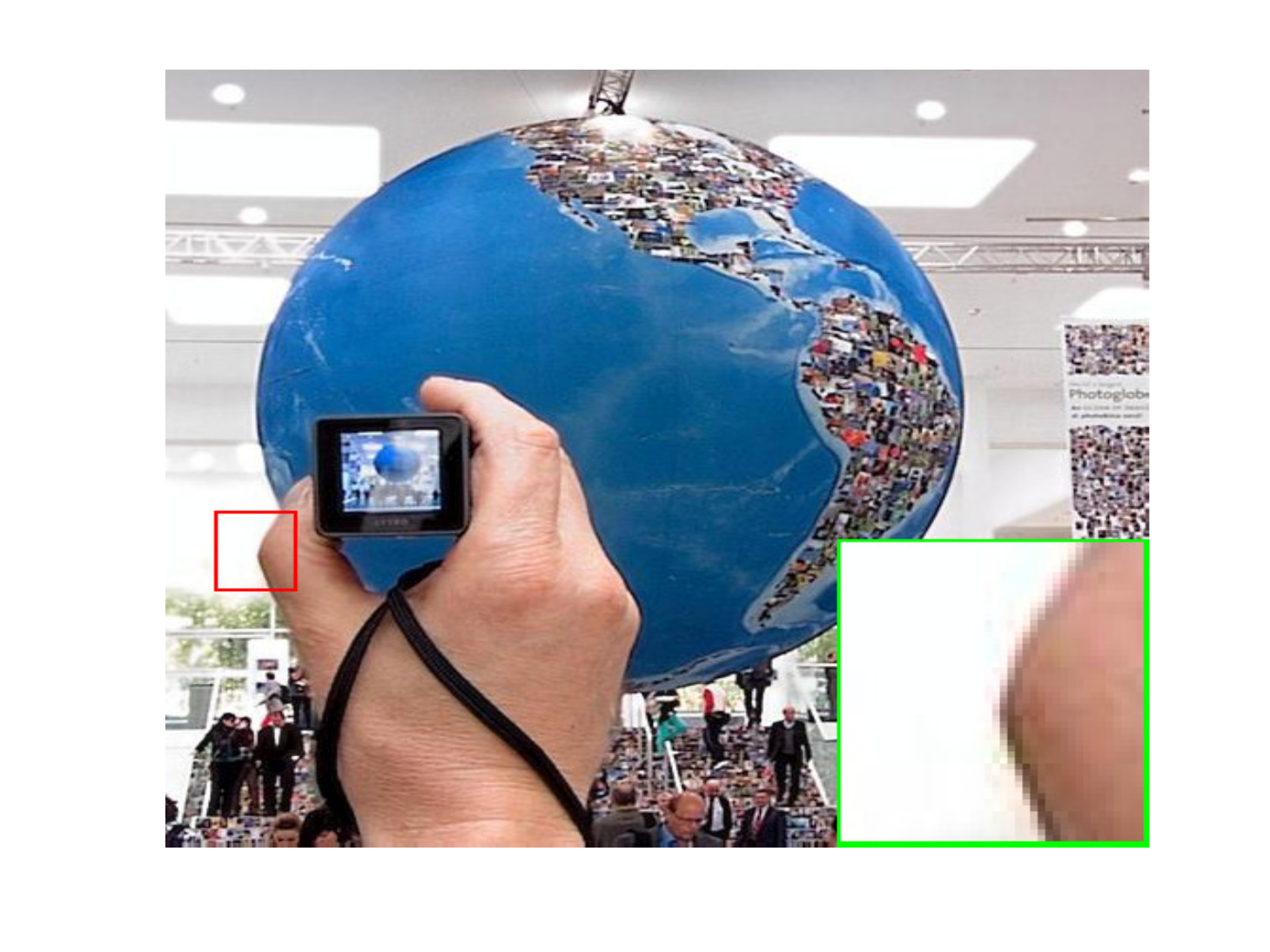}
	}
	\hspace{-0.18in}
	\subfigure[\tiny{MWGF} \cite{DBLP:journals/inffus/ZhouLW14}]{
		\includegraphics[width=2.7cm,height=2cm]{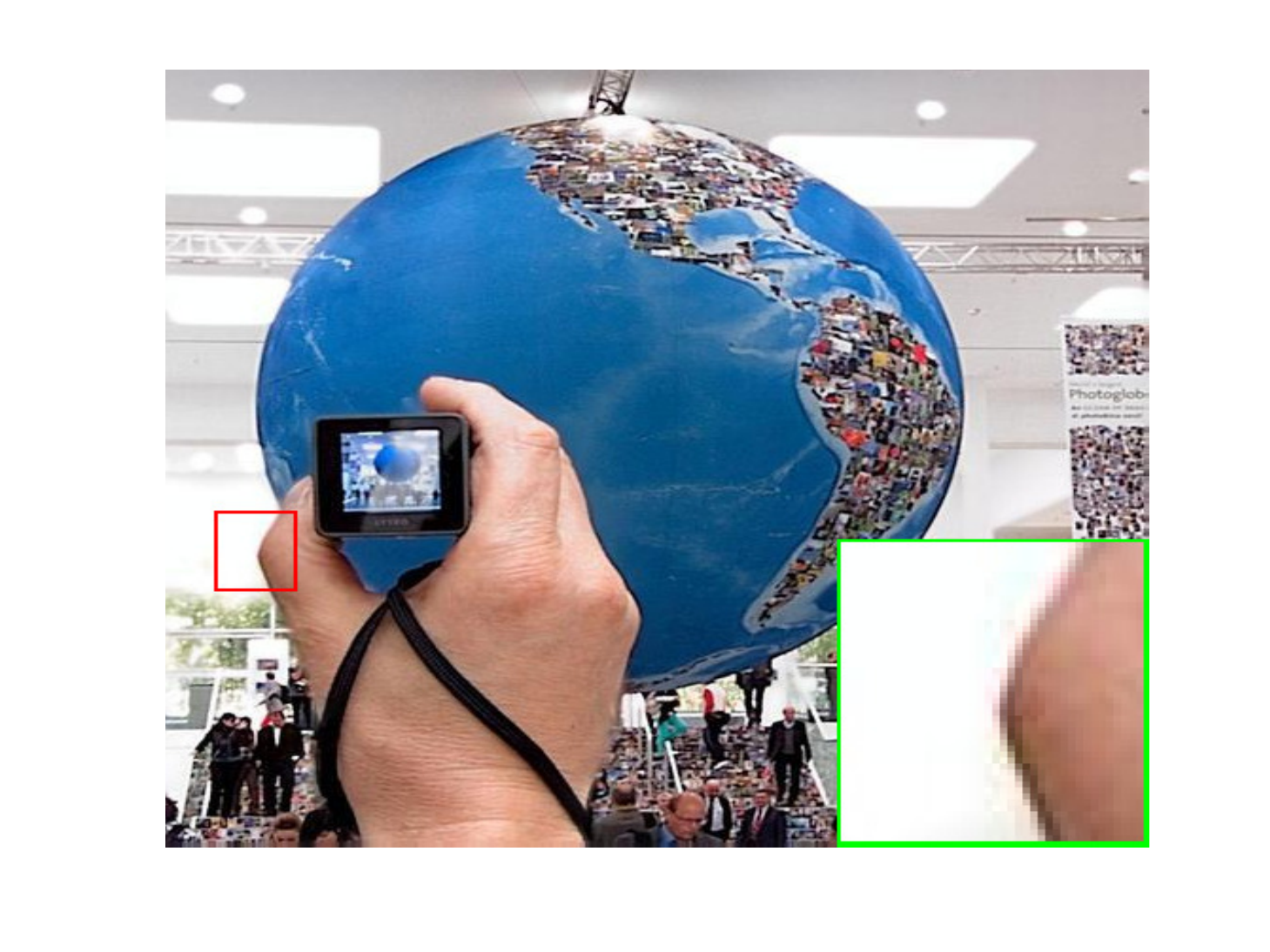}
	}\\
	\hspace{-0.18in}
	\subfigure[\tiny{GFF} \cite{DBLP:journals/tip/LiKH13}]{
		\includegraphics[width=2.7cm,height=2cm]{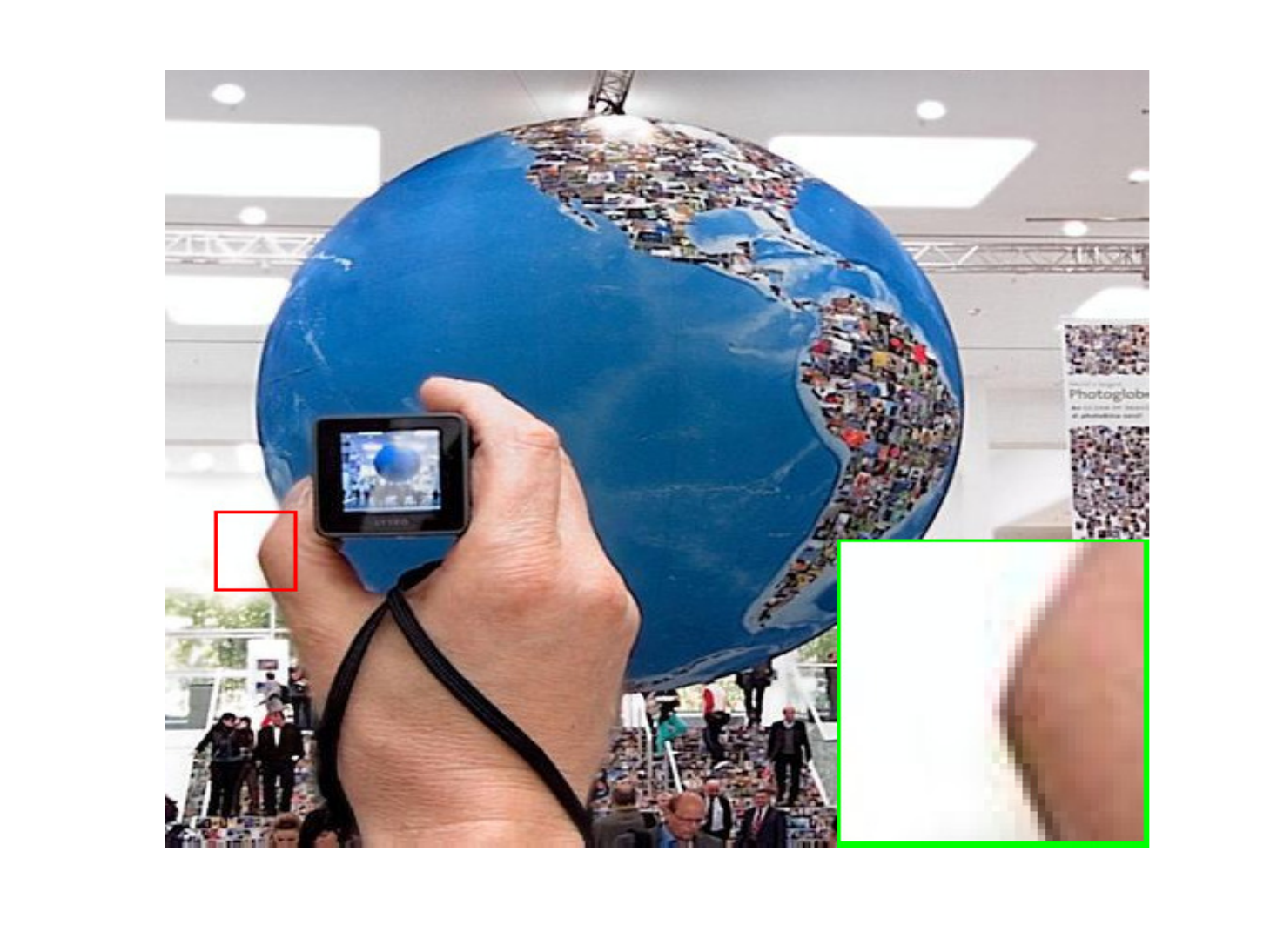}
	}
	\hspace{-0.18in}
	\subfigure[\tiny{QCT} \cite{2012Multifocus}]{
		\includegraphics[width=2.7cm,height=2cm]{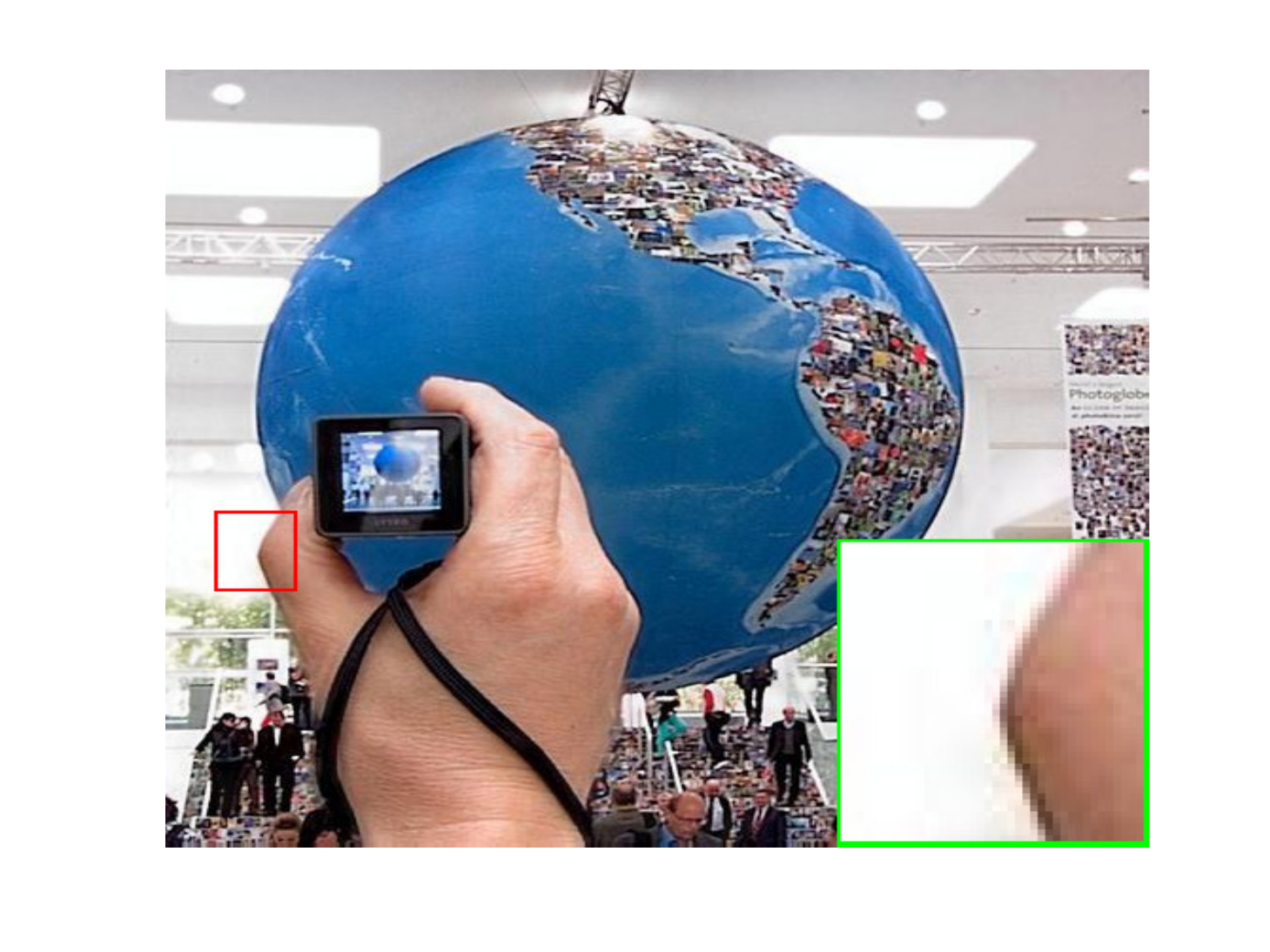}
	}
	\hspace{-0.18in}
	\subfigure[\tiny{HOSVD} \cite{DBLP:journals/tip/LiangHLZ12}]{
		\includegraphics[width=2.7cm,height=2cm]{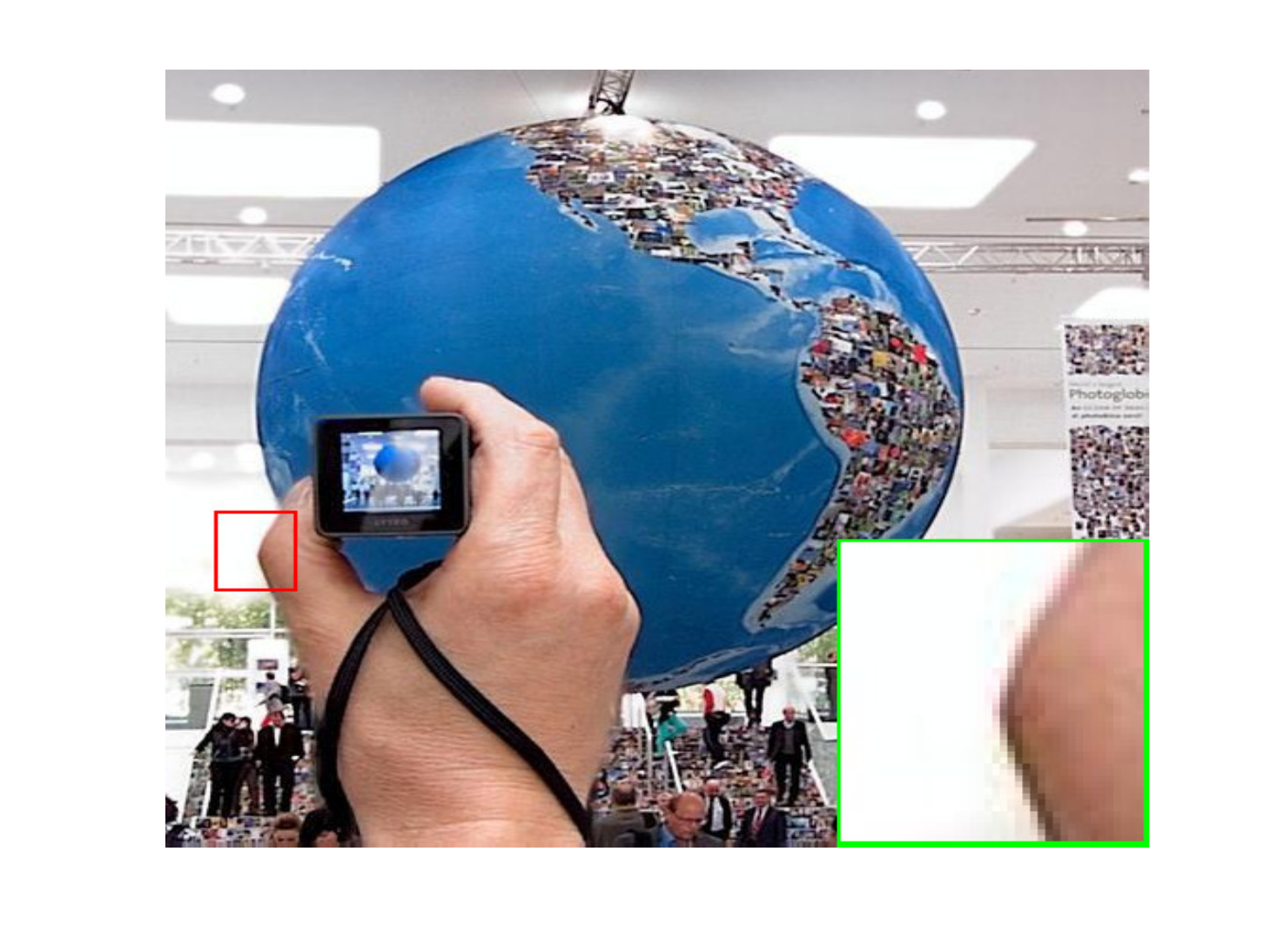}
	}\\
	\hspace{-0.18in}
	\subfigure[\tiny{QUADTREE} \cite{DBLP:journals/inffus/BaiZZX15}]{
		\includegraphics[width=2.7cm,height=2cm]{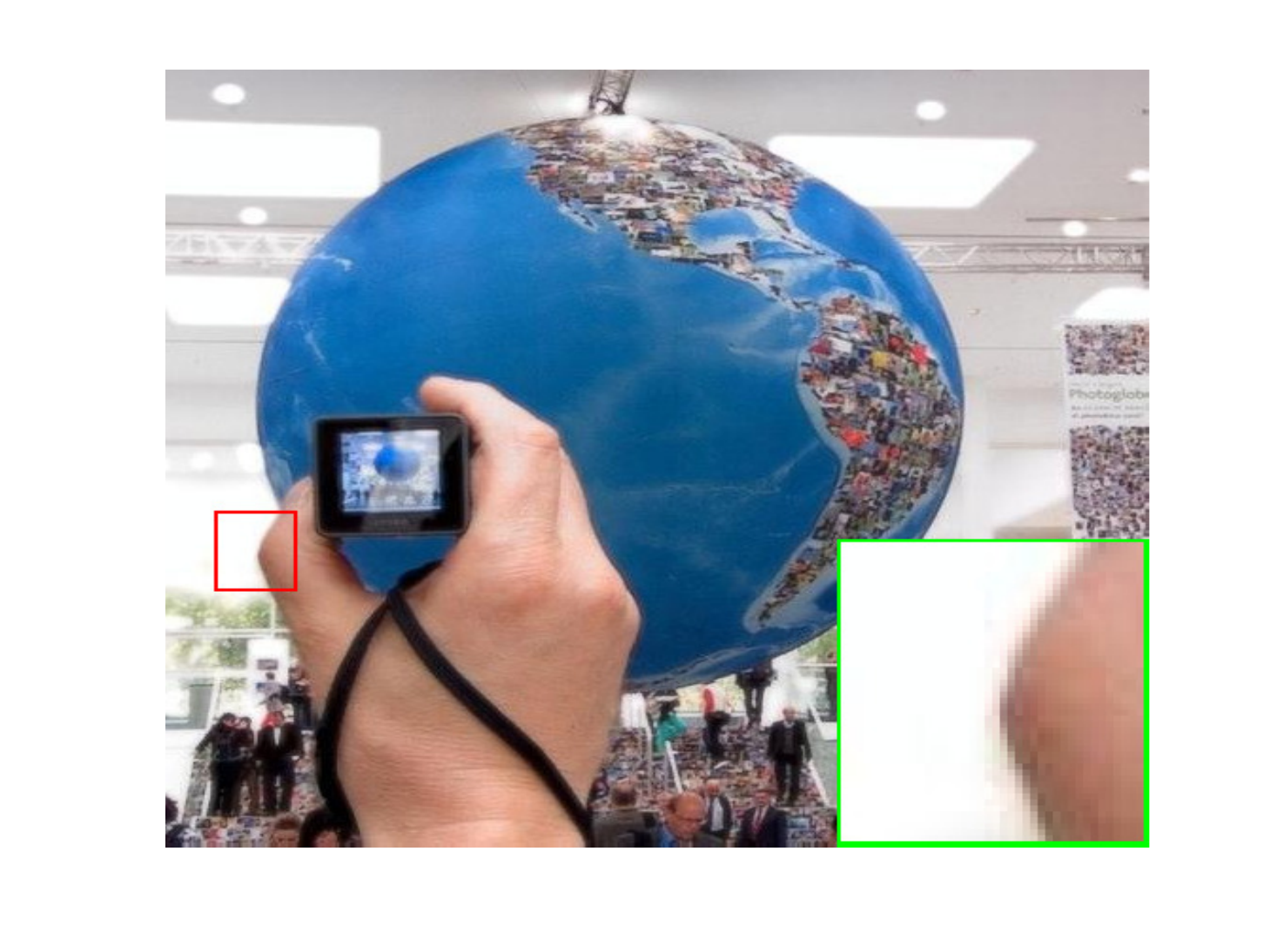}
	}
	\hspace{-0.18in}
	\subfigure[\tiny{GFDF} \cite{DBLP:journals/spic/QiuLZY19}]{
		\includegraphics[width=2.7cm,height=2cm]{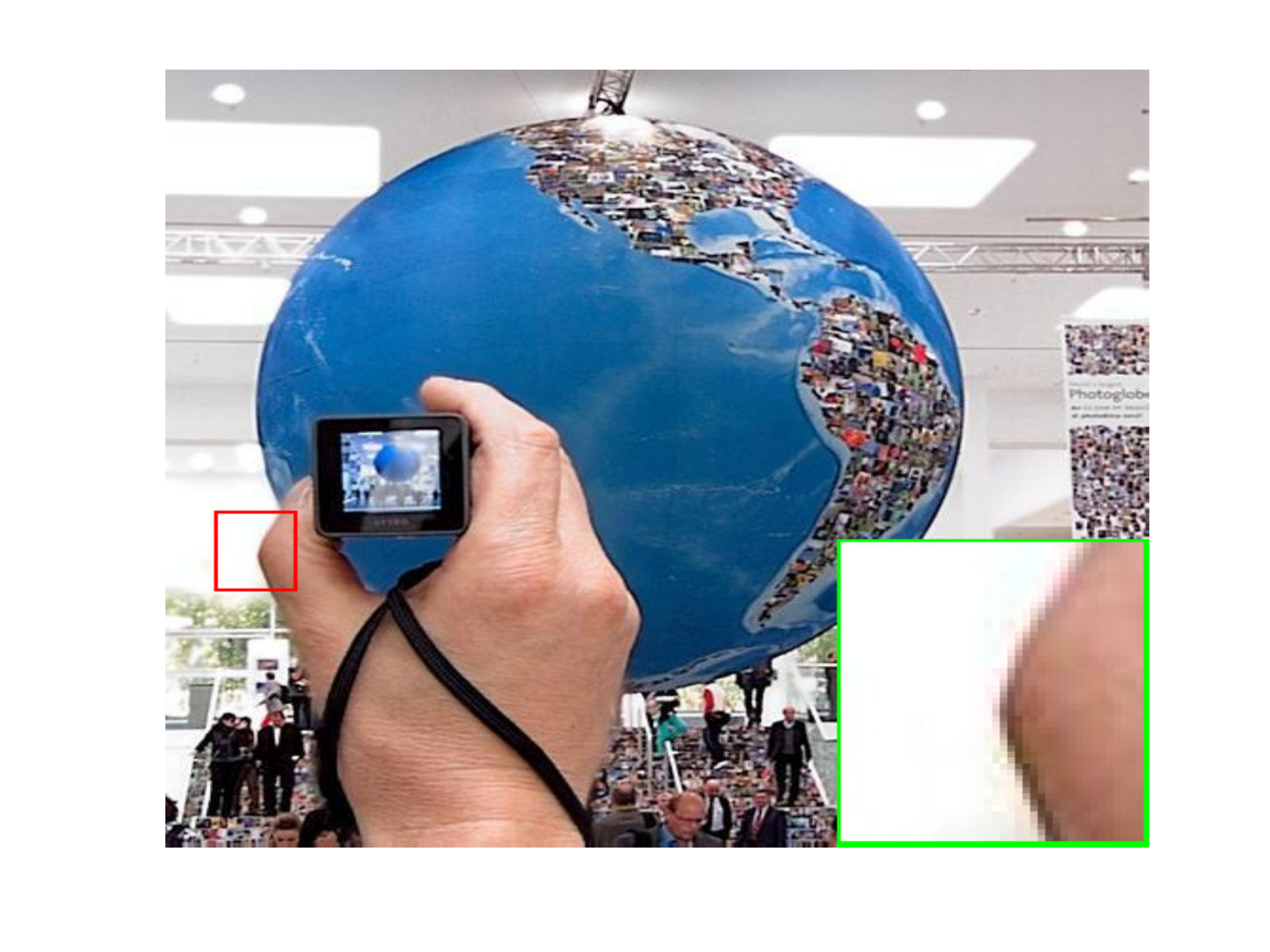}
	}
	\hspace{-0.18in}
	\subfigure[\tiny{MISF} \cite{DBLP:journals/jei/ZhanKLH19}]{
		\includegraphics[width=2.7cm,height=2cm]{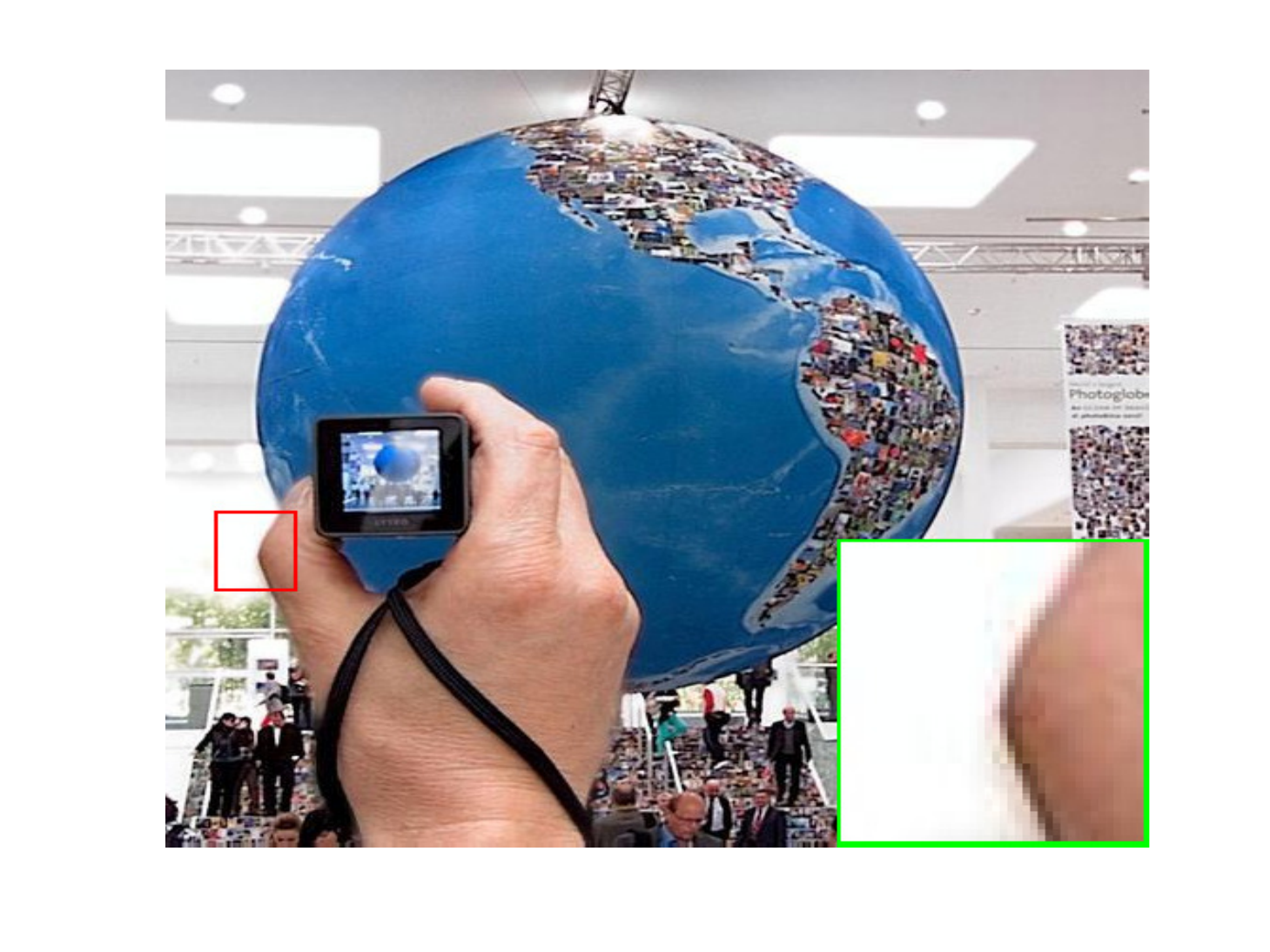}
	}\\
	\hspace{-0.18in}
	\subfigure[\tiny{CNN} \cite{DBLP:journals/inffus/LiuCPW17}]{
		\includegraphics[width=2.7cm,height=2cm]{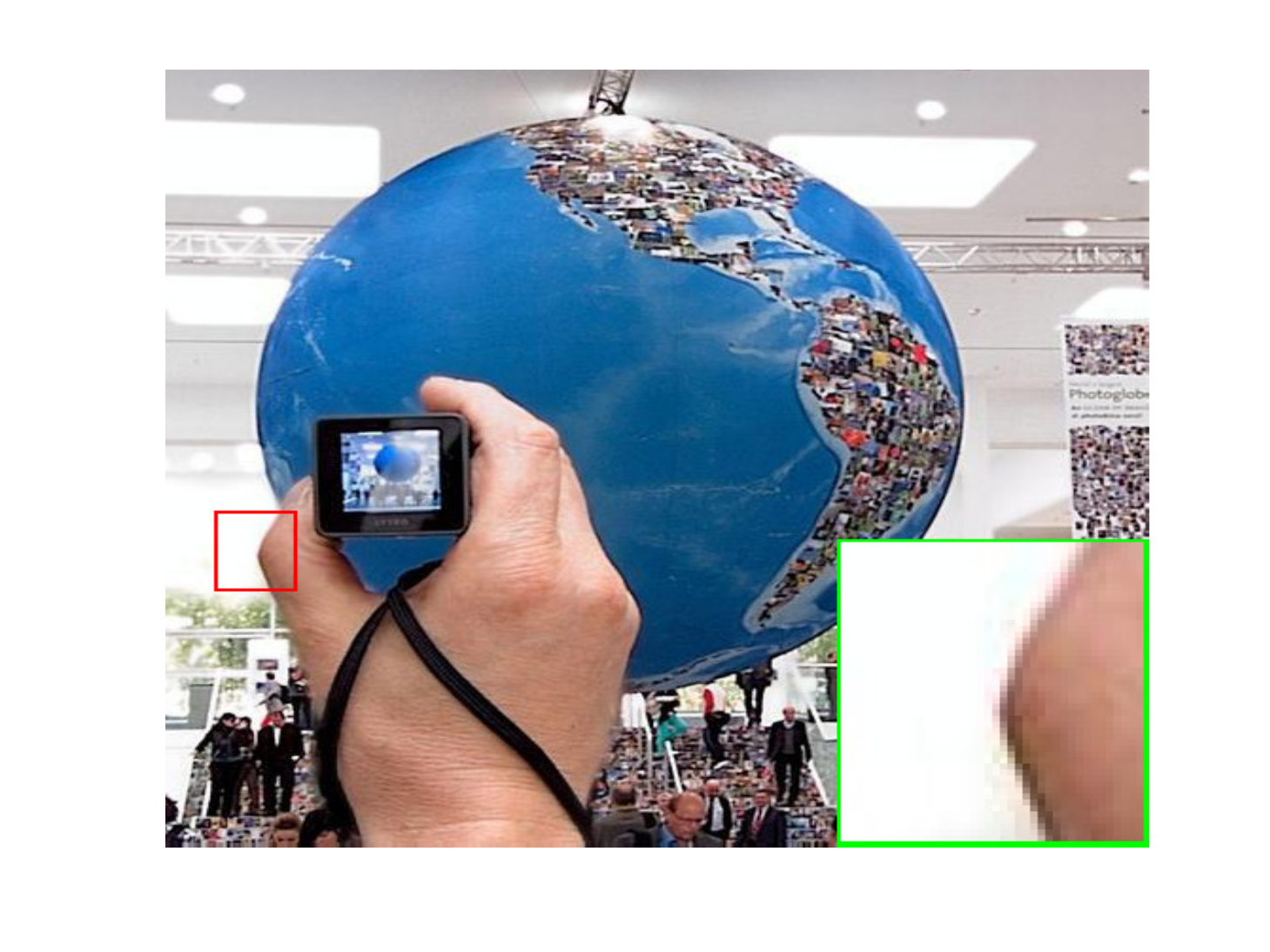}
	}
	\hspace{-0.18in}
	\subfigure[\tiny{MADCNN} \cite{DBLP:journals/access/LaiLGX19}]{
		\includegraphics[width=2.7cm,height=2cm]{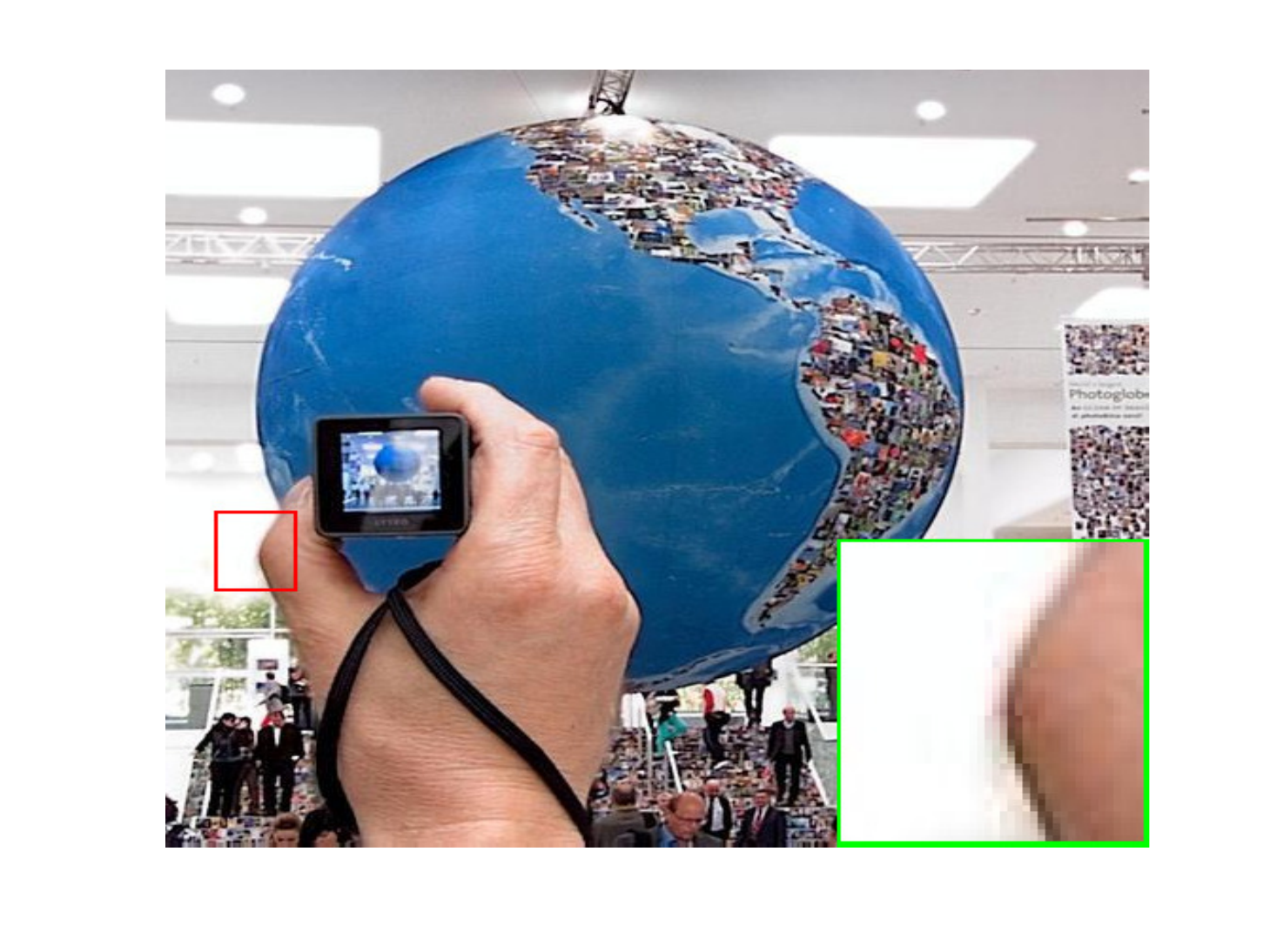}
	}
	\hspace{-0.18in}
	\subfigure[\tiny{QHOSVD}]{
		\includegraphics[width=2.7cm,height=2cm]{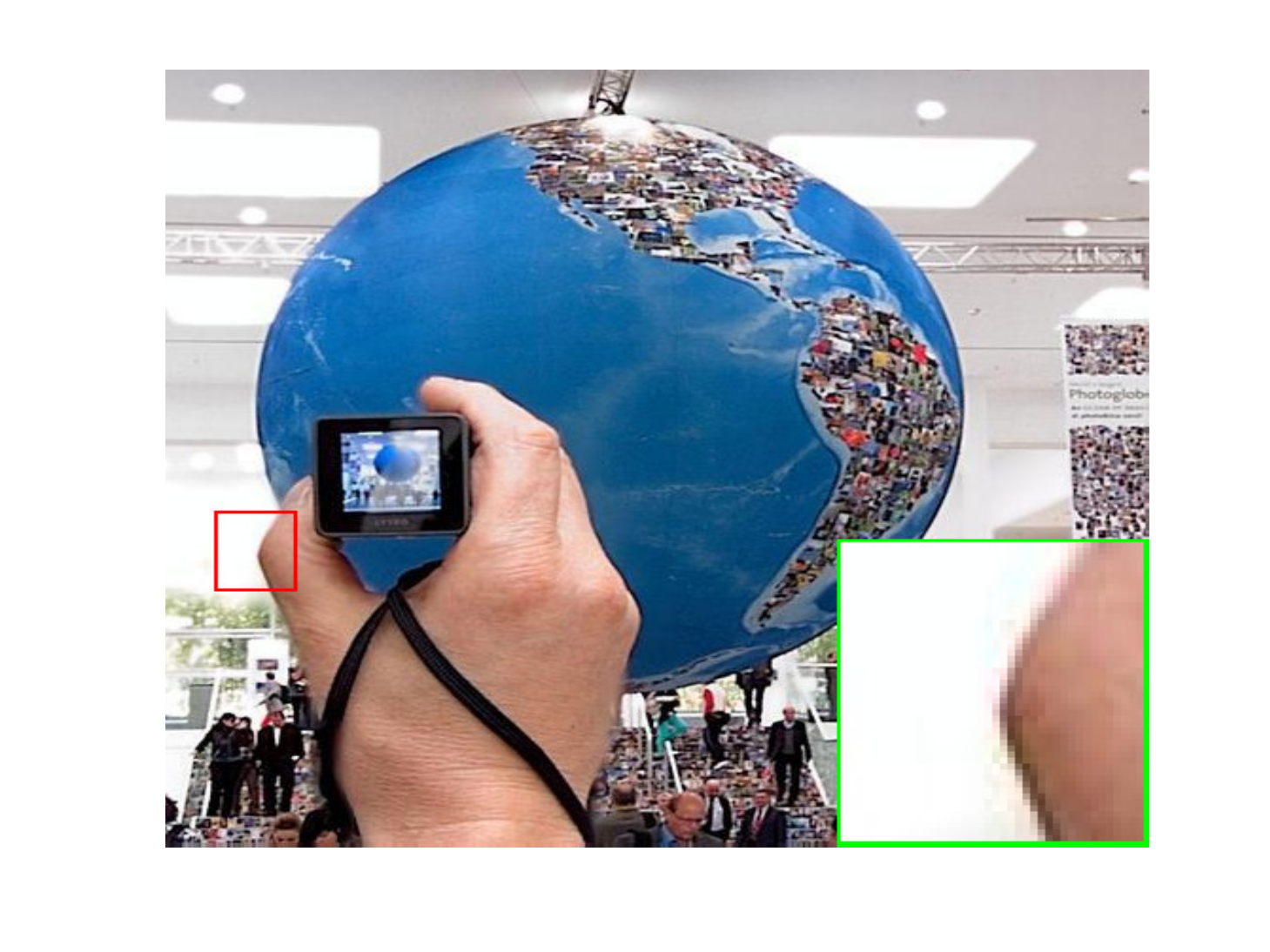}
	}
	\caption{The fusion results on the $11$th pair with detail magnified for all methods.}
	\label{fig02}
\end{figure}

\textbf{Results and discussions:} Fig. \ref{fig01} provides insights about the objective performance of different fusion methods on the $20$ pairs of color multi-focus images. TABLE \ref{table_objective} lists the objective performance (average score of each metric) and average running time of different fusion methods. Fig. \ref{fig02} visually shows one of the fusion results of different methods. From all the experimental results about multi-focus color image fusion, we can observe and summarize the following points.
\begin{itemize}
\item The defined QHOSVD can indeed be applied to multi-focus color image fusion problems.
\item Compared with the traditional fusion methods (\emph{e.g.}, DWT, NSCT, MWGF, GFF, QCT, HOSVD, and QUADTREE ), the proposed   QHOSVD-based method has great advantages. Compared with the latest fusion methods (\emph{e.g.}, GFDF and MISF) and even methods based on deep learning (\emph{e.g.}, CNN and MADCNN), our method still has a small advantage on $\rm{Q_{MI}}$, $\rm{Q_{NCIE}}$, and $\rm{Q_{Y}}$. Although our method is not competitive on $\rm{Q_{CB}}$ and $\rm{Q_{CV}}$, they are not far from the best value.
\item  Except for CNN and QCT, most of the methods only require a few seconds or less one second to accomplish the fusion task. We do not list the running time of MADCNN because its source code is a Python version, we perform it in Python. Therefore, it is not comparable.
\end{itemize}

\subsection{Color image denoising}
\begin{figure}[htbp]
	\centering
	\includegraphics[width=7.3cm,height=2cm]{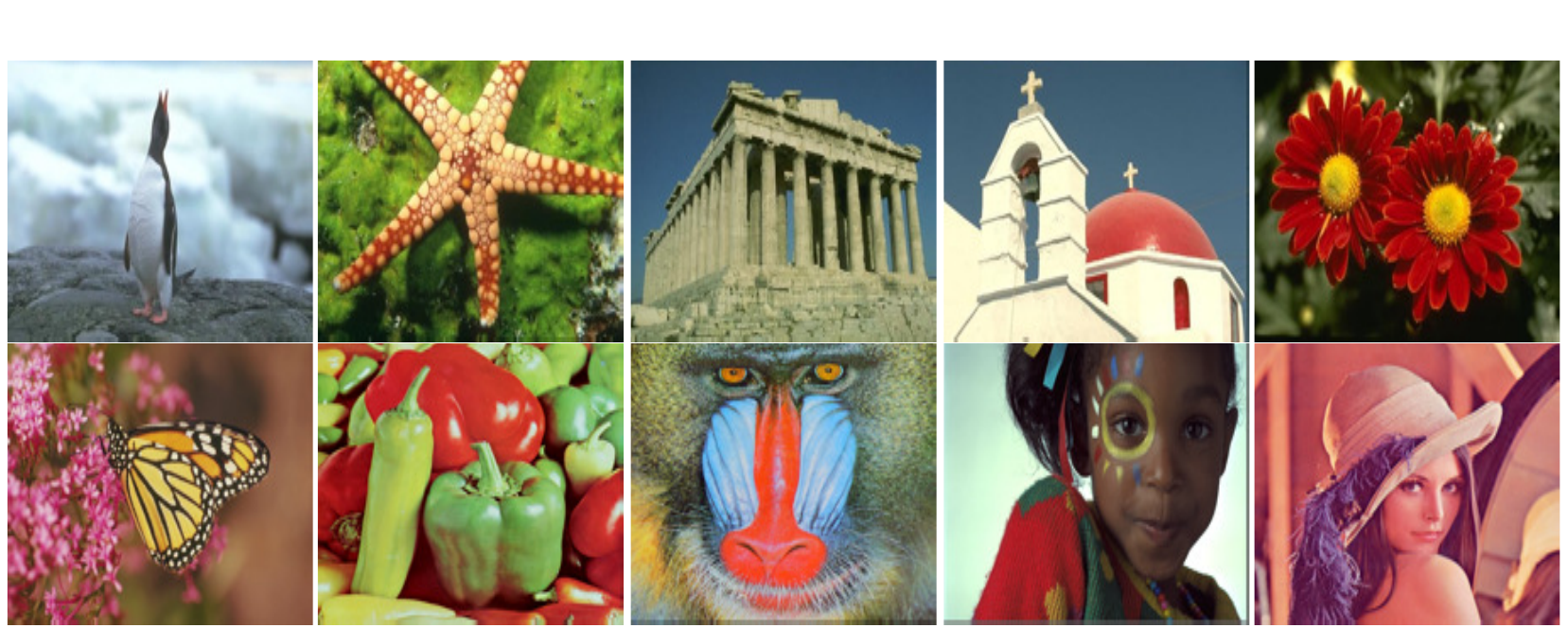}
	\caption{The $10$ color images for color image denoising in our experiments (Image(1)-Image(10) from left to right, top to bottom).}
	\label{fig2}
\end{figure}

\textbf{Dataset:}  We conduct the experiments on $10$ color images with size $121\times 181$ shown in Fig. \ref{fig2}. The $5$ images in the first row are randomly selected from the Berkeley Segmentation Dataset (BSD)\footnote{\url{https://www2.eecs.berkeley.edu/Research/Projects/CS/vision/
		bsds/}.}. The second row is the most widely used $5$ images in digital image processing.

\textbf{Objective metrics:} To evaluate the performance of proposed methods, except visual quality, we employ two widely used quantitative quality indexes, including the peak signal-to-noise ratio (PSNR) and the structure similarity (SSIM) \cite{DBLP:journals/tip/WangBSS04}. For both metrics, a larger value indicates a better denoising performance.

\textbf{Methods for comparison:} We compare our QHOSVD-based method with four state-of-the-art denoising methods including QNNM \cite{DBLP:journals/tip/ChenXZ20} (a quaternion nuclear norm minimization algorithm labeled as LRQA-1 in \cite{DBLP:journals/tip/ChenXZ20}), QWNNM \cite{DBLP:journals/ijon/YuZY19} (a quaternion weighted nuclear norm minimization algorithm), LRQA-WSNN \cite{DBLP:journals/tip/ChenXZ20} (a quaternion weighted Schatten norm minimization algorithm labeled as LRQA-4 in \cite{DBLP:journals/tip/ChenXZ20}), and HOSVD-based method \cite{DBLP:journals/access/GaoGZCZ19}.

\textbf{Parameter setting:} For the NSS procedure, the same parameters are set for all methods. Specifically, when $\delta=10, 20, 30, 50$, the patch size is set to $6\times 6$, $6\times 6$, $7\times7$, and  $8\times 8$, respectively. The number of similar patches $K$ is set to $70$, $70$, $90$, and $120$, respectively. The iteration $\varUpsilon$ is set to $8$, $8$, $14$, and $20$, respectively. For all noise levels, the iterative relaxation parameter $\delta$ is fixed to $0.1$, and the searching window  size is fixed to $30\times 30$. For our QHOSVD-based method, $\eta$ is empirically set to $0.70$, $0.55$, $0.45$, and $0.35$, respectively. In addition, all compared methods are from the source codes and the parameter settings are based
on the suggestions in the original papers. 

\begin{table}[htbp]
	\caption{Objective metrics (PSNR/SSIM) of different methods on the ten color images (\textbf{Bold} fonts denote the best performance; \underline{underline} ones represent the second-best results).}
	\centering
	\resizebox{7.8cm}{7cm}{
		\begin{tabular}{|c|ccccc|}		
			\hline
			Methods:& QNNM \cite{DBLP:journals/tip/ChenXZ20} & QWNNM \cite{DBLP:journals/ijon/YuZY19} & LRQA-WSNN \cite{DBLP:journals/tip/ChenXZ20} &HOSVD  \cite{DBLP:journals/access/GaoGZCZ19} &\textbf{QHOSVD} \\ \toprule
			\hline
			Images:  &\multicolumn{5}{c|}{$\sigma=10$}\\
			\hline
			Image(1)&35.234/0.952&\underline{35.881}/\underline{0.966}&35.800/\underline{0.966}&35.496/0.961&\textbf{35.895}/\textbf{0.967}\\
			Image(2)&31.113/0.975&31.356/\underline{0.978}&\underline{31.371}/\textbf{0.979}&31.240/0.977&\textbf{31.373}/\underline{0.978}\\
			Image(3)&32.624/0.975&\underline{33.138}/\underline{0.980}&33.123/\underline{0.980}&32.682/0.977&\textbf{33.337}/\textbf{0.982}\\
			Image(4)&36.117/0.992&37.330/\underline{0.994}&\underline{37.358}/0.993&36.550/0.993&\textbf{37.472}/\textbf{0.995}\\  
			Image(5)&32.915/0.981&33.767/\underline{0.983}&\underline{33.803}/\underline{0.983}&33.134/0.980&\textbf{33.814}/\textbf{0.984}\\
			Image(6)&31.735/0.987&\underline{32.873}/\textbf{0.991}&32.682/\underline{0.990}&32.388/0.989&\textbf{33.024}/\textbf{0.991}\\
			Image(7)&32.937/0.990&34.069/\underline{0.992}&\textbf{34.116}/\underline{0.992}&33.805/\underline{0.992}&\underline{34.094}/\textbf{0.993}\\
			Image(8)&30.743/0.948&30.866/0.951&\underline{30.881}/\underline{0.952}&30.603/0.945&\textbf{31.044}/\textbf{0.954}\\
			Image(9)&34.512/0.969&\underline{35.444}/\underline{0.976}&35.366/0.974&35.044/0.975&\textbf{35.454}/\textbf{0.977}\\
			Image(10)&33.613/0.990&34.443/\underline{0.991}&\underline{34.530}/\underline{0.991}&33.900/0.990&\textbf{34.604}/\textbf{0.992}\\
			\hline
			Aver. &33.154/0.976&\underline{33.917}/\underline{0.980}&33.903/\underline{0.980}&33.484/0.977&\textbf{34.011}/\textbf{0.981}\\ \toprule
			\hline
			Images  &\multicolumn{5}{c|}{$\sigma=20$}\\
			\hline
			Image(1)&31.904/0.907&\underline{32.312}/\underline{0.926}&32.218/0.925&32.014/0.923&\textbf{32.326}/\textbf{0.927}\\
			Image(2)&27.311/0.931&27.585/0.952&\underline{27.801}/\textbf{0.954}&27.592/\underline{0.953}&\textbf{27.914}/\textbf{0.954}\\
			Image(3)&29.099/0.930&29.630/0.952&\underline{29.715}/\underline{0.954}&29.042/0.949&\textbf{29.794}/\textbf{0.958}\\
			Image(4)&31.926/0.978&\textbf{33.693}/\textbf{0.988}&\underline{33.682}/\underline{0.987}&32.389/0.983&33.624/\textbf{0.988}\\  
			Image(5)&29.293/0.949&29.984/\underline{0.964}&\underline{30.012}/\underline{0.964}&29.671/0.963&\textbf{30.023}/\textbf{0.966}\\
			Image(6)&27.267/0.947&28.668/\underline{0.975}&\underline{28.709}/\underline{0.975}&28.217/0.902&\textbf{28.753}/\textbf{0.976}\\
			Image(7)&29.062/0.968&\underline{30.144}/\underline{0.982}&30.115/\underline{0.982}&29.891/0.981&\textbf{30.151}/\textbf{0.983}\\
			Image(8)&27.049/0.815&27.449/\underline{0.886}&\underline{27.513}/0.876&27.171/0.883 &\textbf{27.574}/\textbf{0.888}\\
			Image(9)&30.917/0.931&31.819/0.952&\underline{31.844}/\underline{0.953}&31.242/0.948&\textbf{31.884}/\textbf{0.954}\\
			Image(10)&30.034/0.969&30.846/\underline{0.982}&\underline{30.901}/0.981&30.093/0.978&\textbf{30.914}/\textbf{0.983}\\
			\hline
			Aver. &29.386/0.933&30.213/\underline{0.956}&\underline{30.251}/0.955&29.732/0.946&\textbf{30.296}/\textbf{0.958}\\	
			\hline	
			Images  &\multicolumn{5}{c|}{$\sigma=30$}\\
			\hline
			Image(1)&28.747/0.864&\underline{30.351}/\underline{0.896}&30.181/0.895&30.064/0.892&\textbf{30.534}/\textbf{0.899}\\
			Image(2)&25.201/0.908&25.633/0.933&\underline{25.793}/\underline{0.935}&25.504/0.930&\textbf{25.900}/\textbf{0.936}\\
			Image(3)&26.645/0.903&27.567/0.929&\underline{27.638}/\underline{0.930}&27.073/0.924&\textbf{27.644}/\textbf{0.931}\\
			Image(4)&27.939/0.957&\underline{30.929}/\underline{0.977}&30.918/0.976&29.486/0.969&\textbf{30.944}/\textbf{0.979}\\  
			Image(5)&26.626/0.921&27.818/0.941&\underline{27.820}/\underline{0.942}&27.714/0.940&\textbf{27.848}/\textbf{0.943}\\
			Image(6)&24.816/0.912&26.129/\underline{0.957}&\underline{26.202}/\underline{0.957}&25.339/0.950&\textbf{26.217}/\textbf{0.961}\\
			Image(7)&26.302/0.952&\underline{27.772}/\underline{0.970}&27.737/0.967&27.540/0.969&\textbf{27.803}/\textbf{0.971}\\
			Image(8)&25.335/0.778&25.694/0.822&\underline{25.726}/\underline{0.834}&25.481/\underline{0.834}&\textbf{25.774}/\textbf{0.845}\\
		    Image(9)&27.651/0.897&29.447/0.922&\underline{29.529}/\underline{0.924}&28.979/0.919&\textbf{29.615}/\textbf{0.931}\\
	     	Image(10)&27.158/0.957& 28.785/\underline{0.971}&\underline{28.790}/\underline{0.971}&27.982/0.965&\textbf{28.798}/\textbf{0.974}\\
			\hline
			Aver. &26.739/0.905&28.013/0.932&\underline{28.033}/\underline{0.933}&27.516/0.921&\textbf{28.189}/\textbf{0.937}\\
			\hline
			Images  &\multicolumn{5}{c|}{$\sigma=50$}\\
			\hline
			Image(1)&27.198/0.815&23.262/0.714&27.959/0.855&\underline{28.110}/\underline{0.856}&\textbf{28.180}/\textbf{0.861}\\
			Image(2)&22.753/0.835&23.498/\underline{0.895}&\underline{23.525}/0.889&23.319/\underline{0.895}&\textbf{23.534}/\textbf{0.898}\\
			Image(3)&24.702/0.827&25.229/0.886&\underline{25.321}/\underline{0.890}&24.945/0.883&\textbf{25.394}/\textbf{0.894}\\
			Image(4)&25.751/0.891&\underline{27.251}/\underline{0.948}&27.201/0.944&26.277/0.937&\textbf{27.291}/\textbf{0.951}\\  
			Image(5)&24.438/0.840&25.255/0.900&\underline{25.301}/0.899&25.221/\underline{0.901}&\textbf{25.318}/\textbf{0.906}\\
			Image(6)&21.707/0.823&\underline{23.208}/0.913&23.199/\underline{0.915}&22.495/0.904&\textbf{23.234}/\textbf{0.919}\\
			Image(7)&23.776/0.893&\underline{24.946}/\underline{0.946}&24.915/\underline{0.946}&24.830/0.945&\textbf{25.001}/\textbf{0.948}\\
			Image(8)&23.452/0.677&23.841/\underline{0.773}&\underline{23.857}/0.768&23.742/0.771&\textbf{23.904}/\textbf{0.786}\\
			Image(9)&25.659/0.806&26.768/\underline{0.881}&\underline{26.782}/0.880& 26.410/0.873&\textbf{26.984}/\textbf{0.892}\\
			Image(10)&25.251/0.921&\underline{26.218}/\underline{0.951}&26.206/0.950&25.647/0.945&\textbf{26.244}/\textbf{0.952}\\
			\hline
			Aver. &24.510/0.833&24.948/0.881&\underline{25.427}/\underline{0.894}&25.100/0.891&\textbf{25.508}/\textbf{0.901}\\
			\hline
	\end{tabular}}
	\label{Index4SR_2}
\end{table}

\begin{figure}[htbp]
	\centering
	\hspace{-0.3in}
	\subfigure[\tiny{Original}]{
		\includegraphics[width=2.7cm,height=2.5cm]{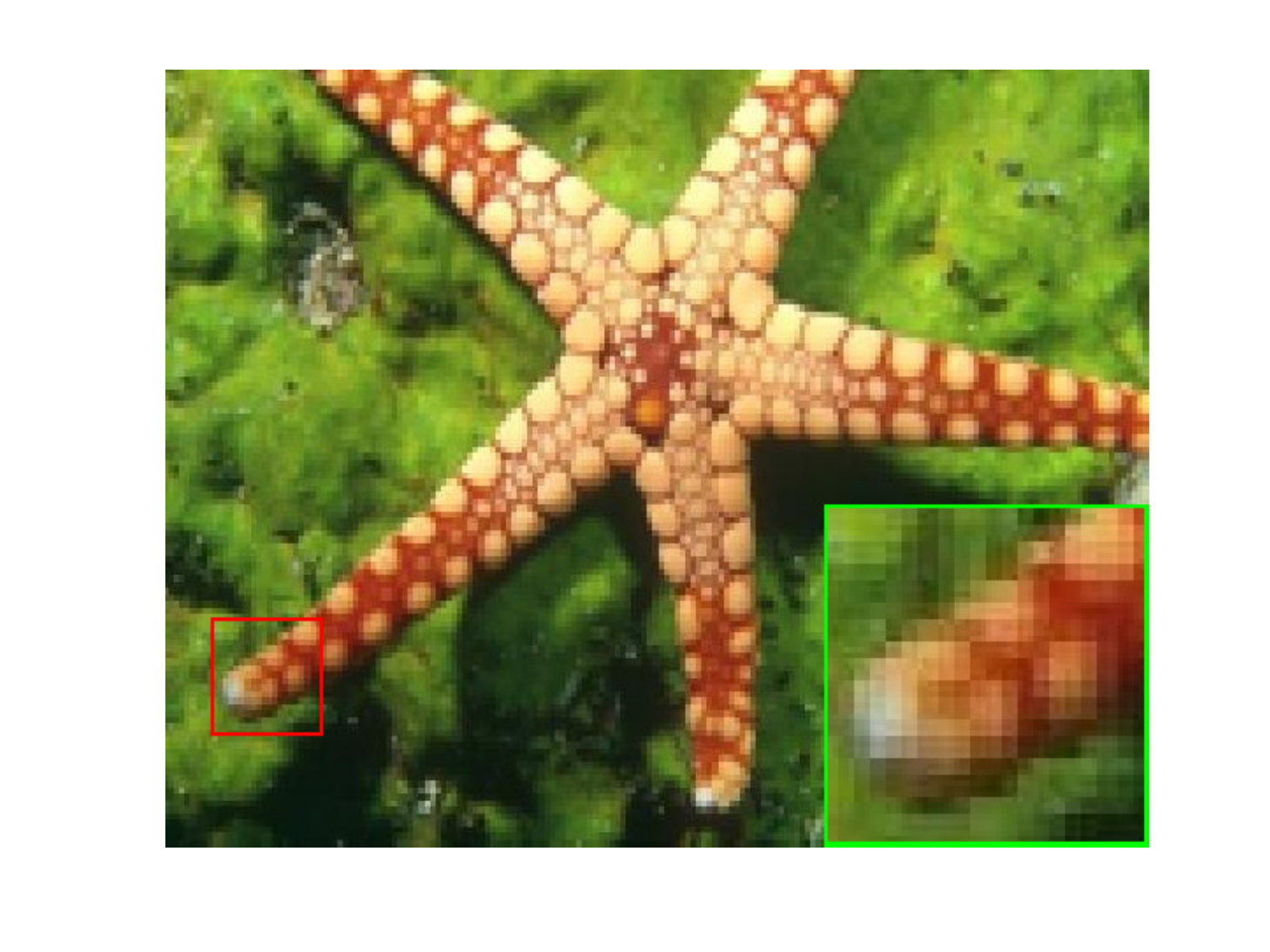}
	}\\
	\hspace{-0.3in}
	\subfigure[\tiny{Observed}]{
		\includegraphics[width=2.7cm,height=2.5cm]{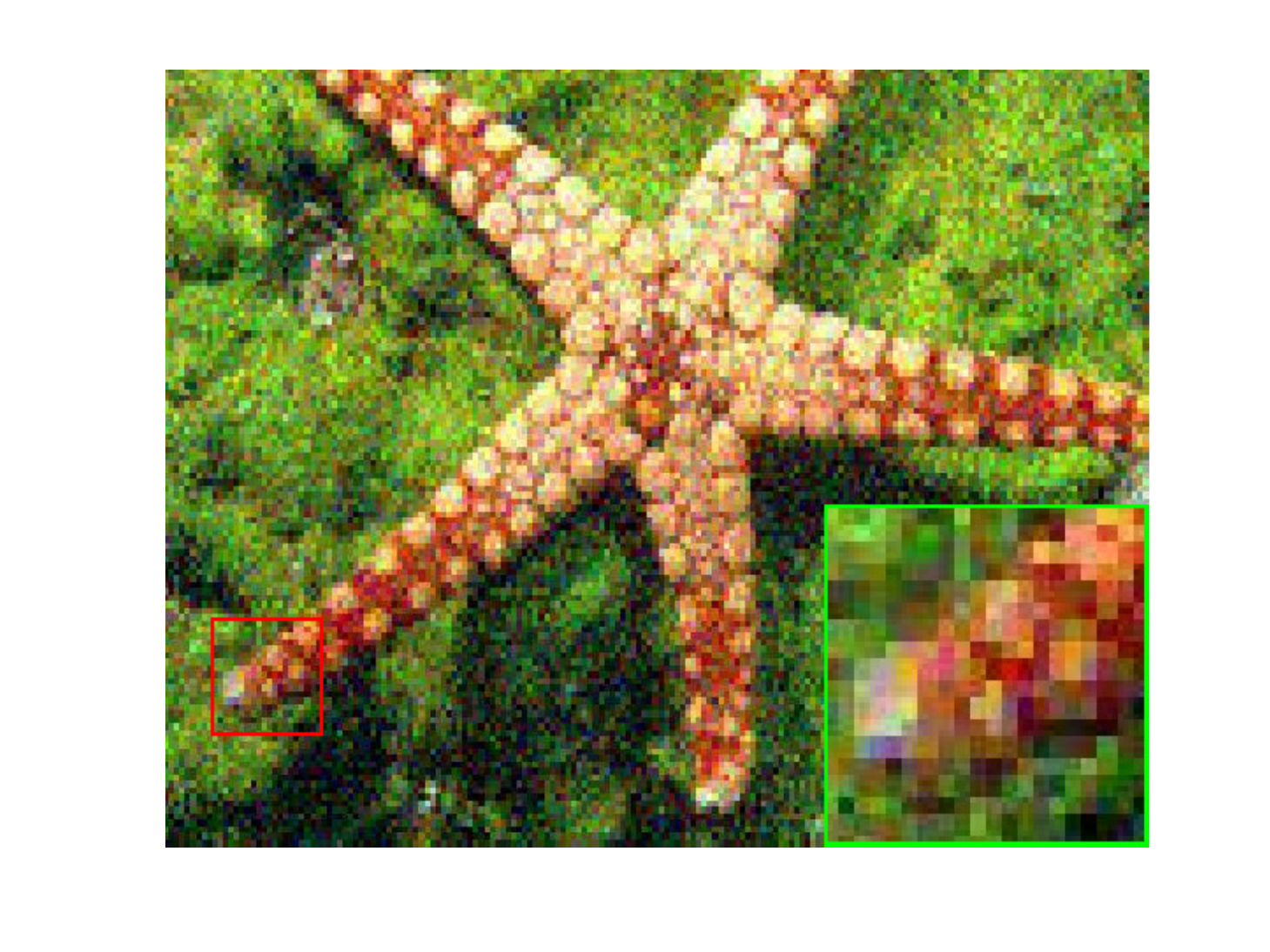}
	}
	\hspace{-0.3in}
	\subfigure[\tiny{QNNM} \cite{DBLP:journals/tip/ChenXZ20}]{
		\includegraphics[width=2.7cm,height=2.5cm]{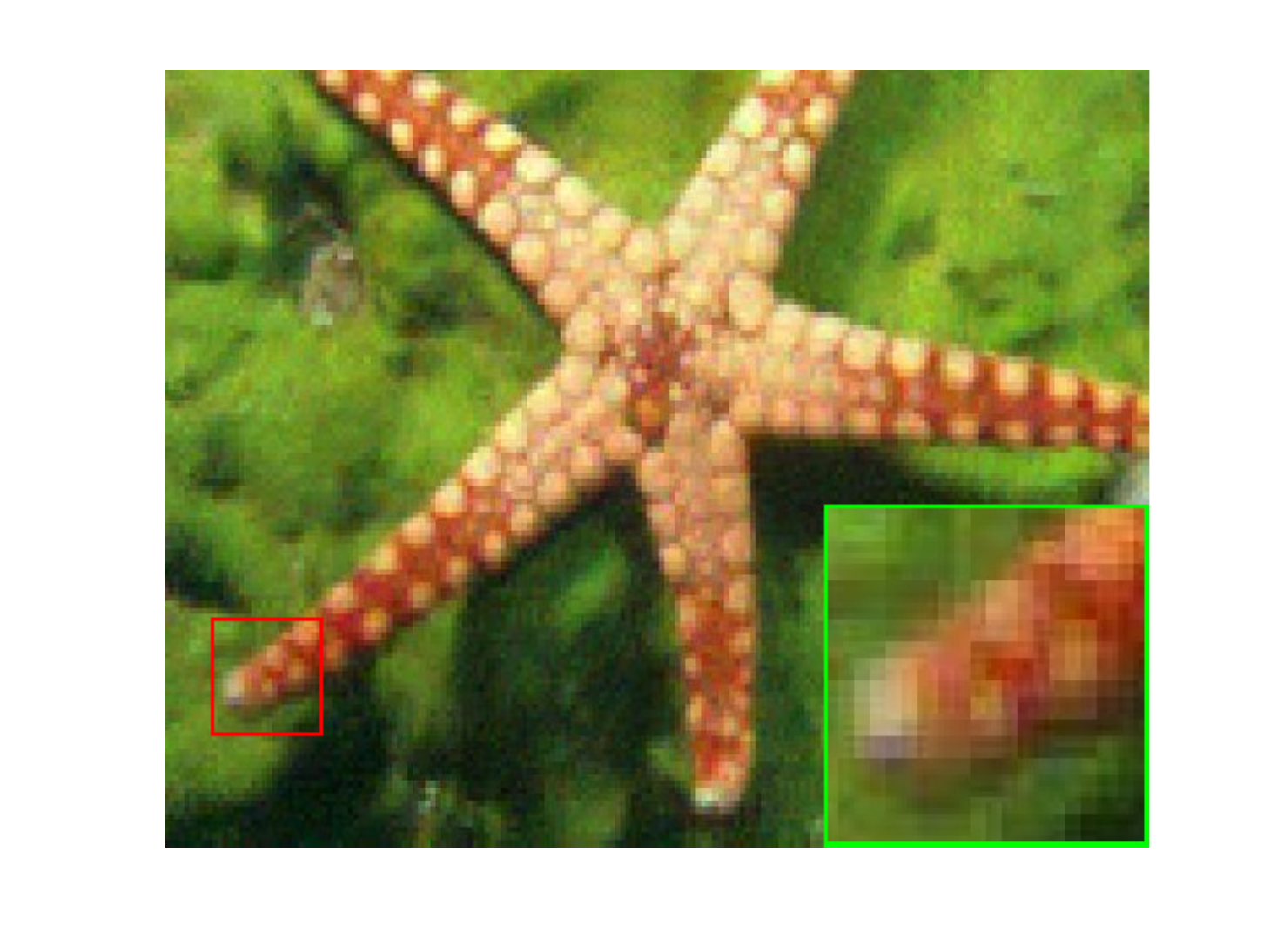}
	}
	\hspace{-0.3in}
	\subfigure[\tiny{QWNNM} \cite{DBLP:journals/ijon/YuZY19}]{
		\includegraphics[width=2.7cm,height=2.5cm]{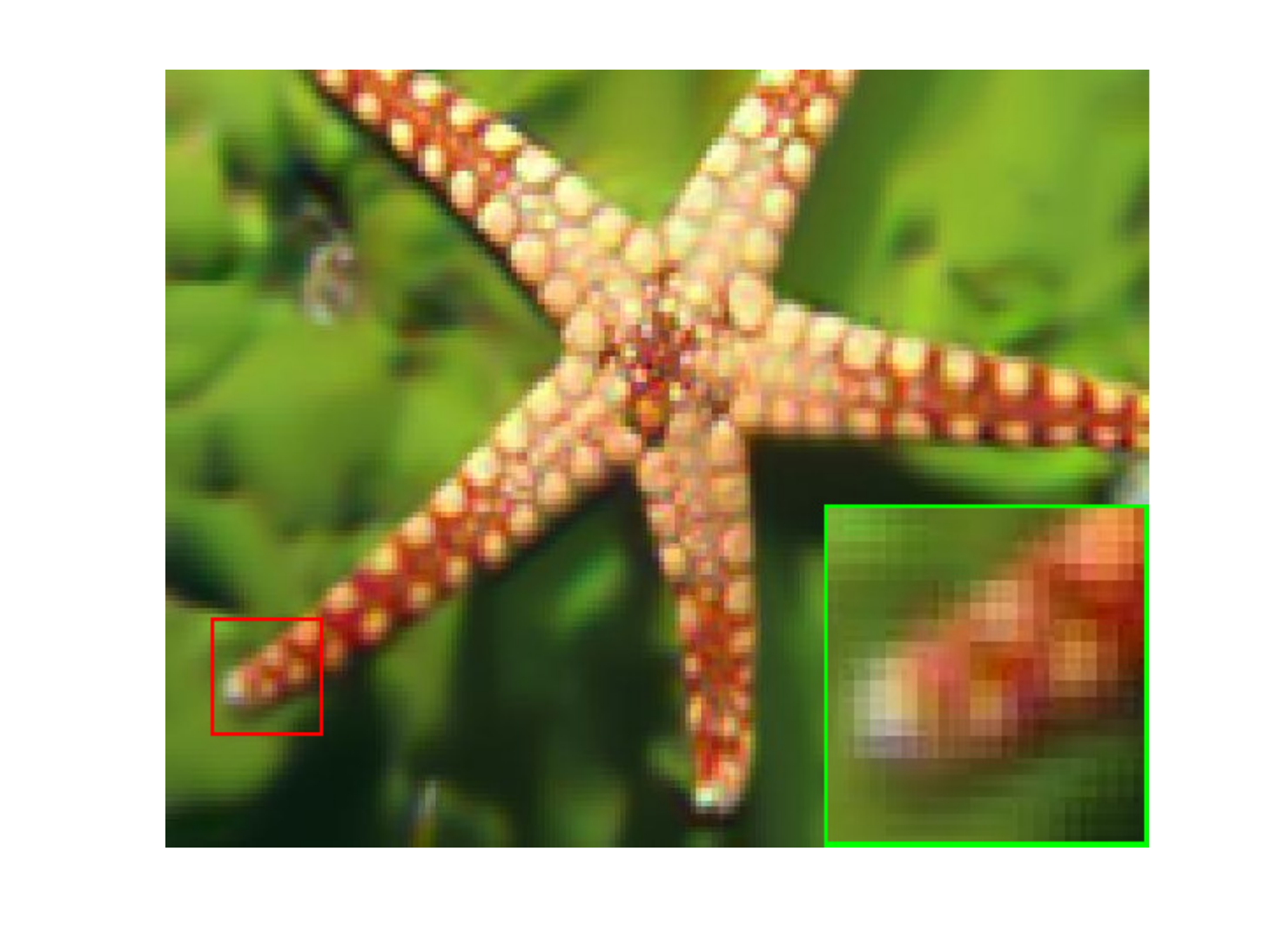}
	}\\
	\hspace{-0.3in}
	\subfigure[\tiny{LRQA-WSNN} \cite{DBLP:journals/tip/ChenXZ20}]{
		\includegraphics[width=2.7cm,height=2.5cm]{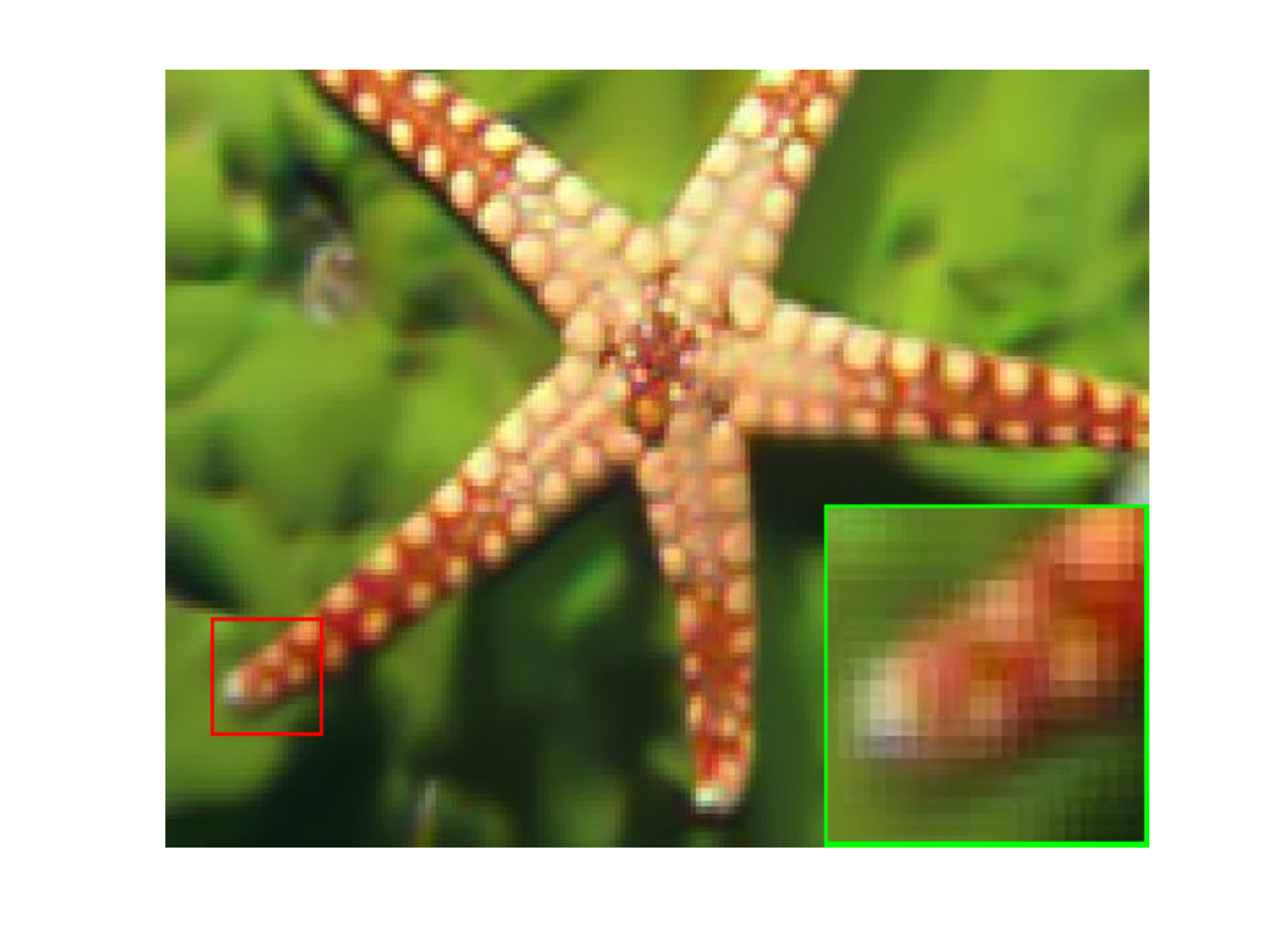}
	}
	\hspace{-0.3in}
	\subfigure[\tiny{HOSVD}  \cite{DBLP:journals/access/GaoGZCZ19}]{
		\includegraphics[width=2.7cm,height=2.5cm]{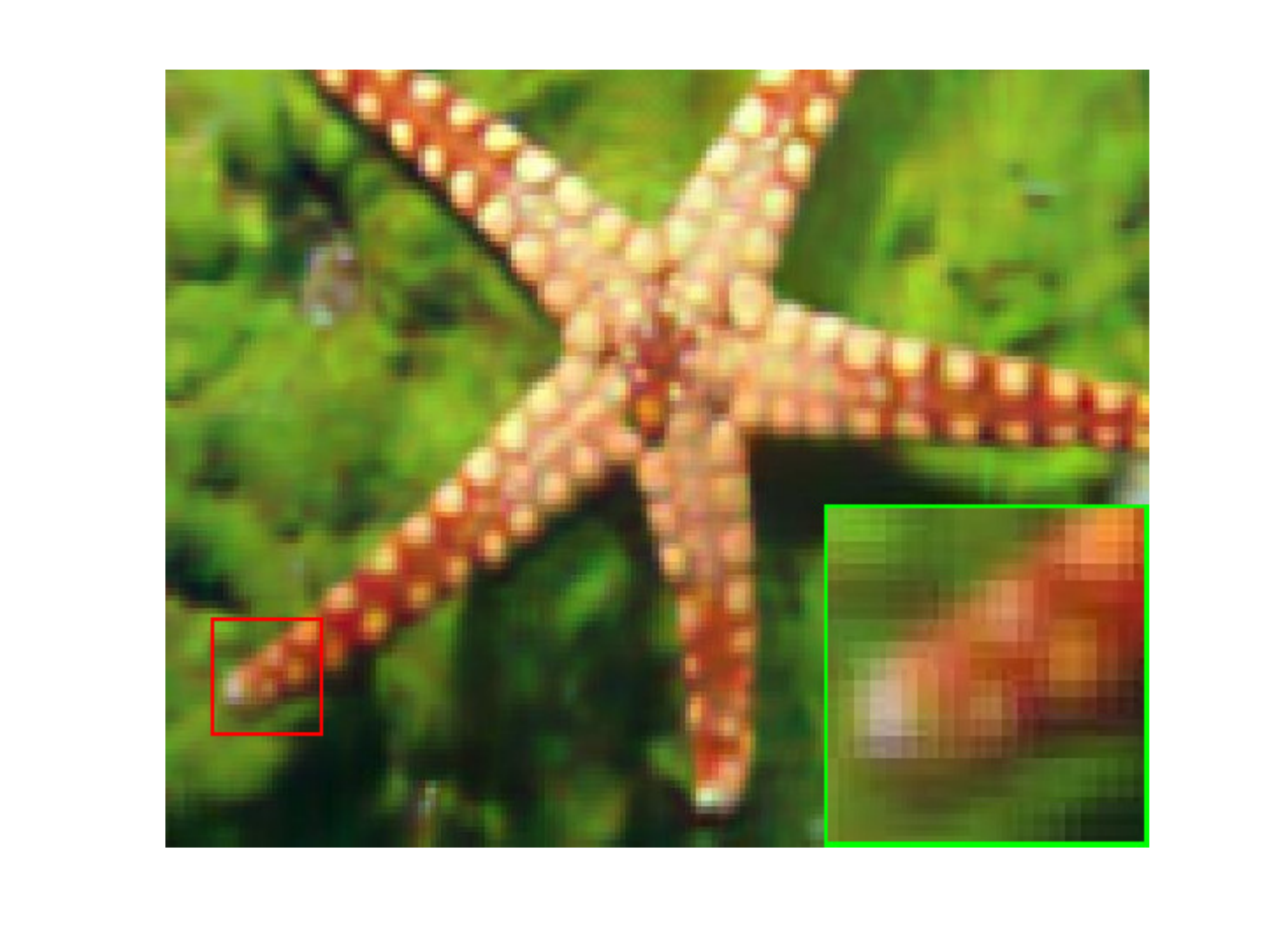}
	}
	\hspace{-0.3in}
	\subfigure[\tiny{QHOSVD}]{
		\includegraphics[width=2.7cm,height=2.5cm]{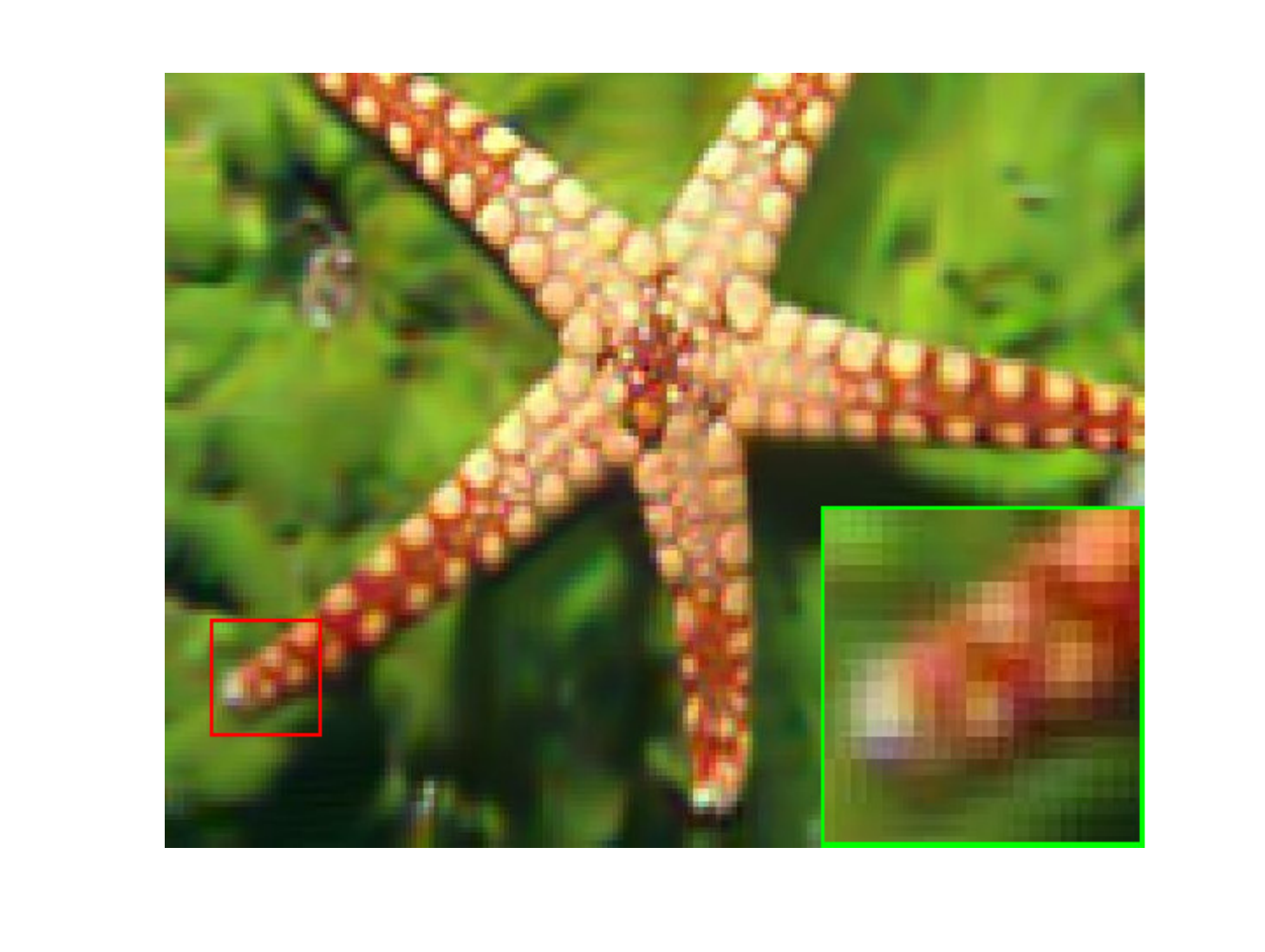}
	}	
	\caption{Color image denoising results on the Image(2) ($\sigma=20$).}
	\label{fig03}
\end{figure}

\begin{figure}[htbp]
	\centering
	\hspace{-0.3in}
	\subfigure[\tiny{Original}]{
		\includegraphics[width=2.7cm,height=2.5cm]{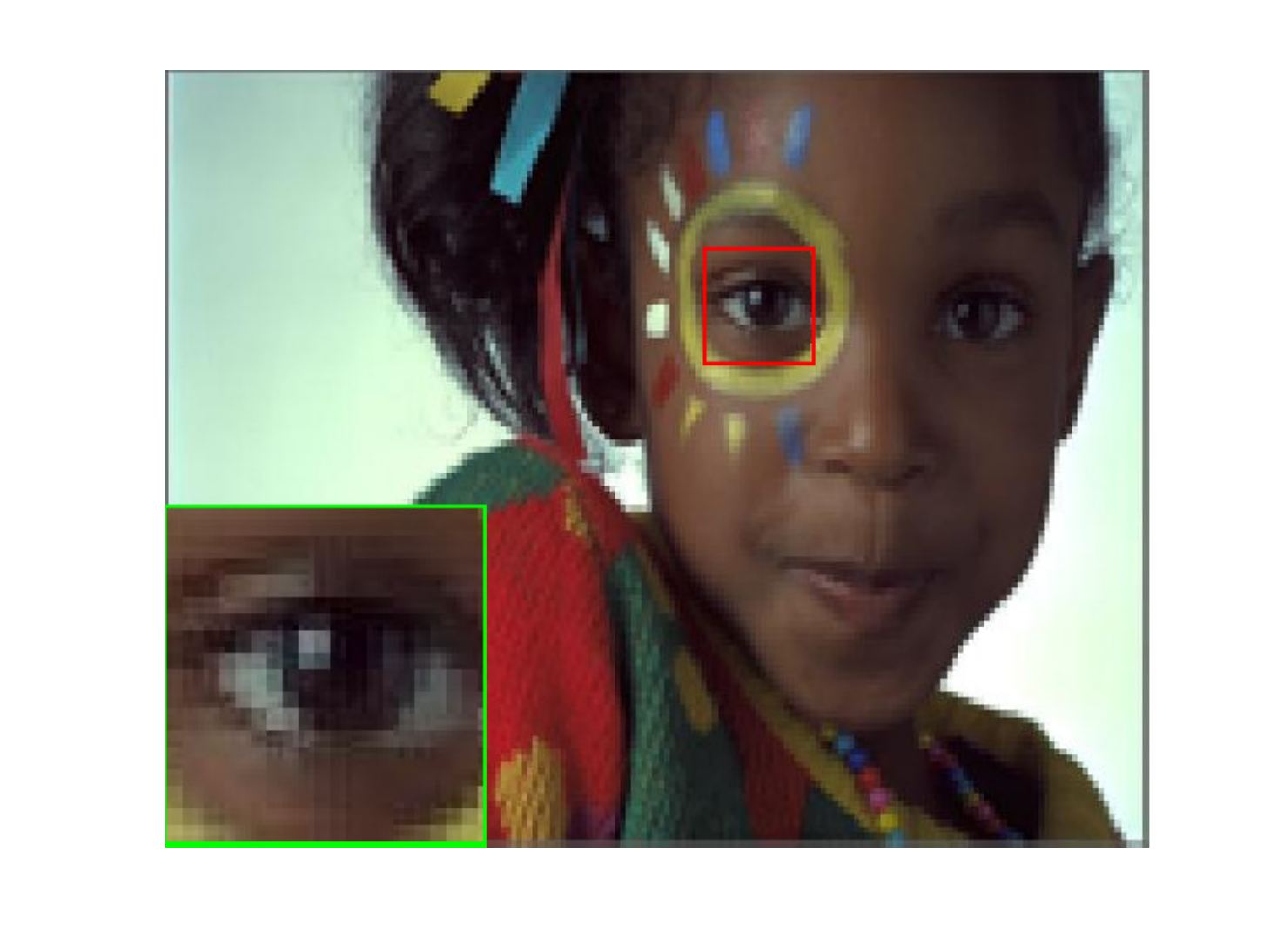}
	}\\
	\hspace{-0.3in}
	\subfigure[\tiny{Observed}]{
		\includegraphics[width=2.7cm,height=2.5cm]{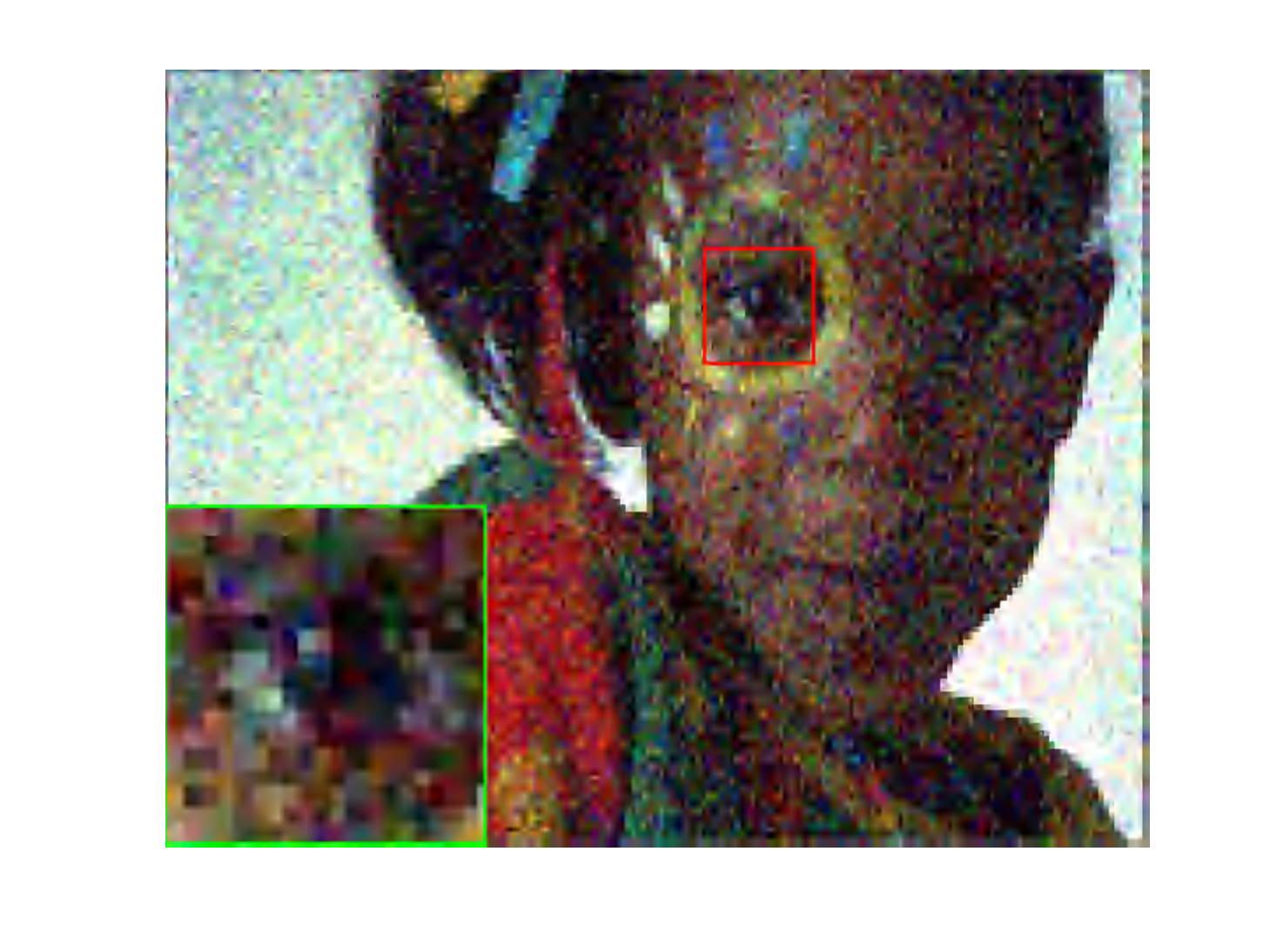}
	}
	\hspace{-0.3in}
	\subfigure[\tiny{QNNM} \cite{DBLP:journals/tip/ChenXZ20}]{
		\includegraphics[width=2.7cm,height=2.5cm]{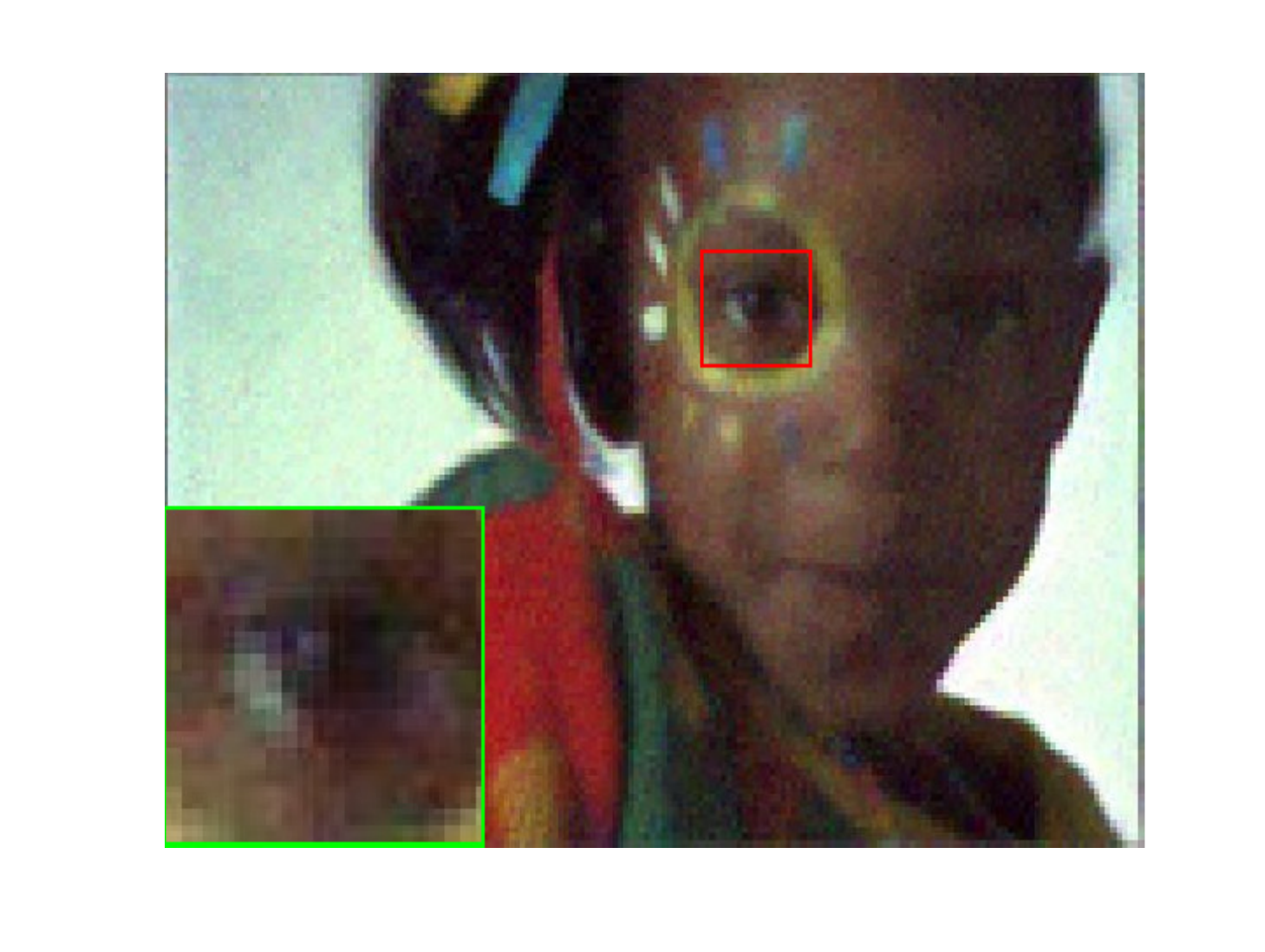}
	}
	\hspace{-0.3in}
	\subfigure[\tiny{QWNNM} \cite{DBLP:journals/ijon/YuZY19}]{
		\includegraphics[width=2.7cm,height=2.5cm]{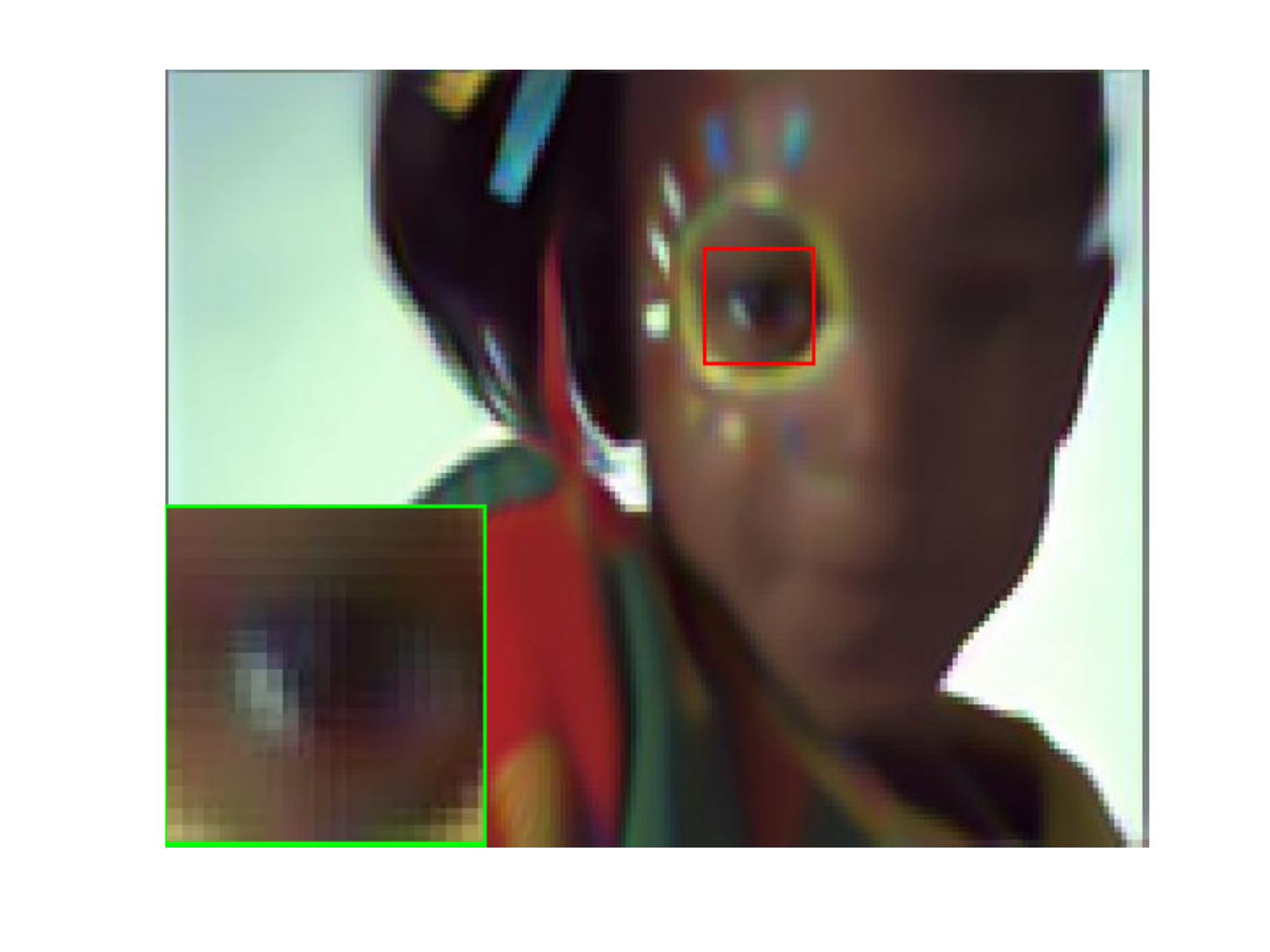}
	}\\
	\hspace{-0.3in}
	\subfigure[\tiny{LRQA-WSNN} \cite{DBLP:journals/tip/ChenXZ20}]{
		\includegraphics[width=2.7cm,height=2.5cm]{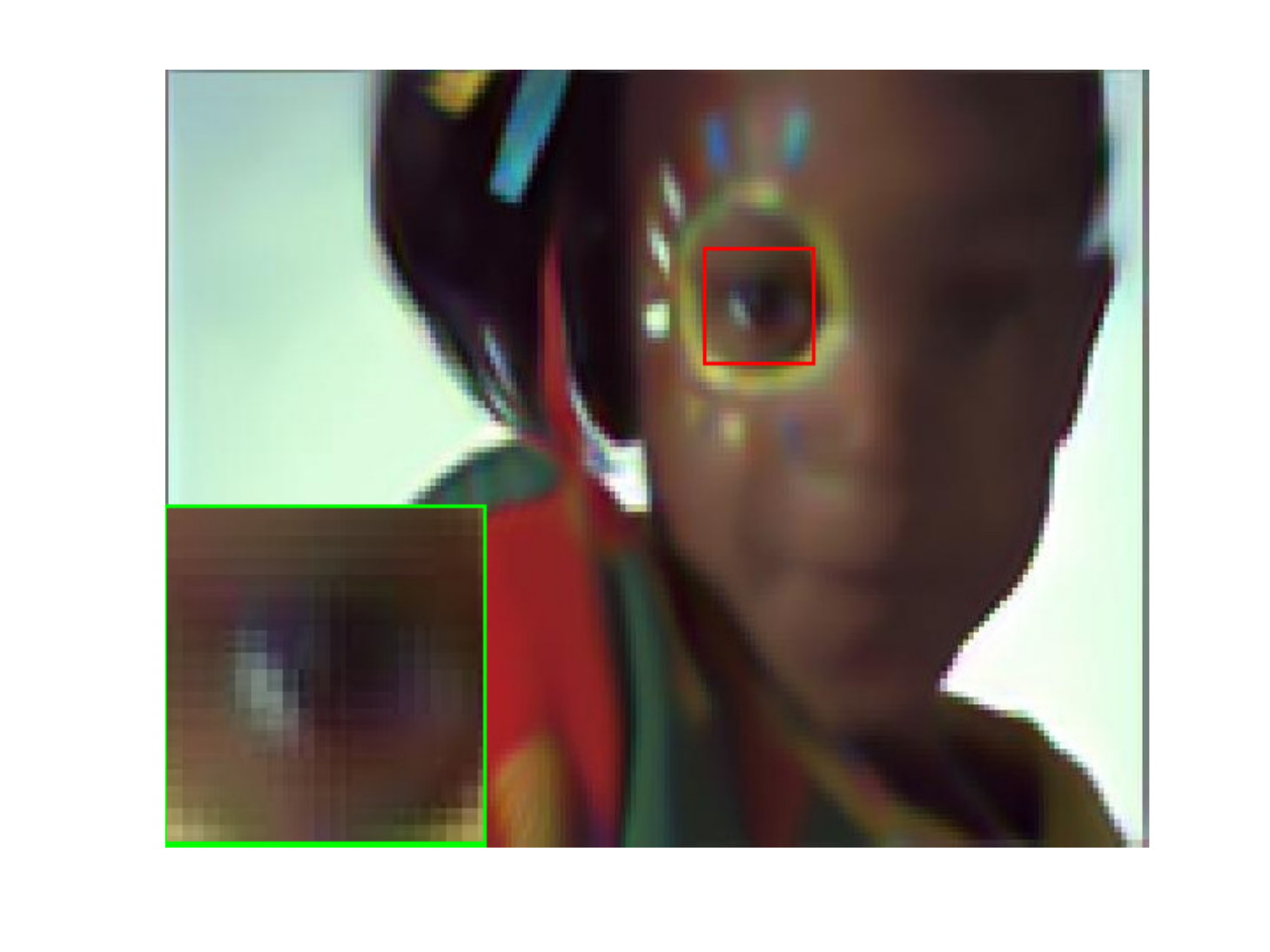}
	}
	\hspace{-0.3in}
	\subfigure[\tiny{HOSVD}  \cite{DBLP:journals/access/GaoGZCZ19}]{
		\includegraphics[width=2.7cm,height=2.5cm]{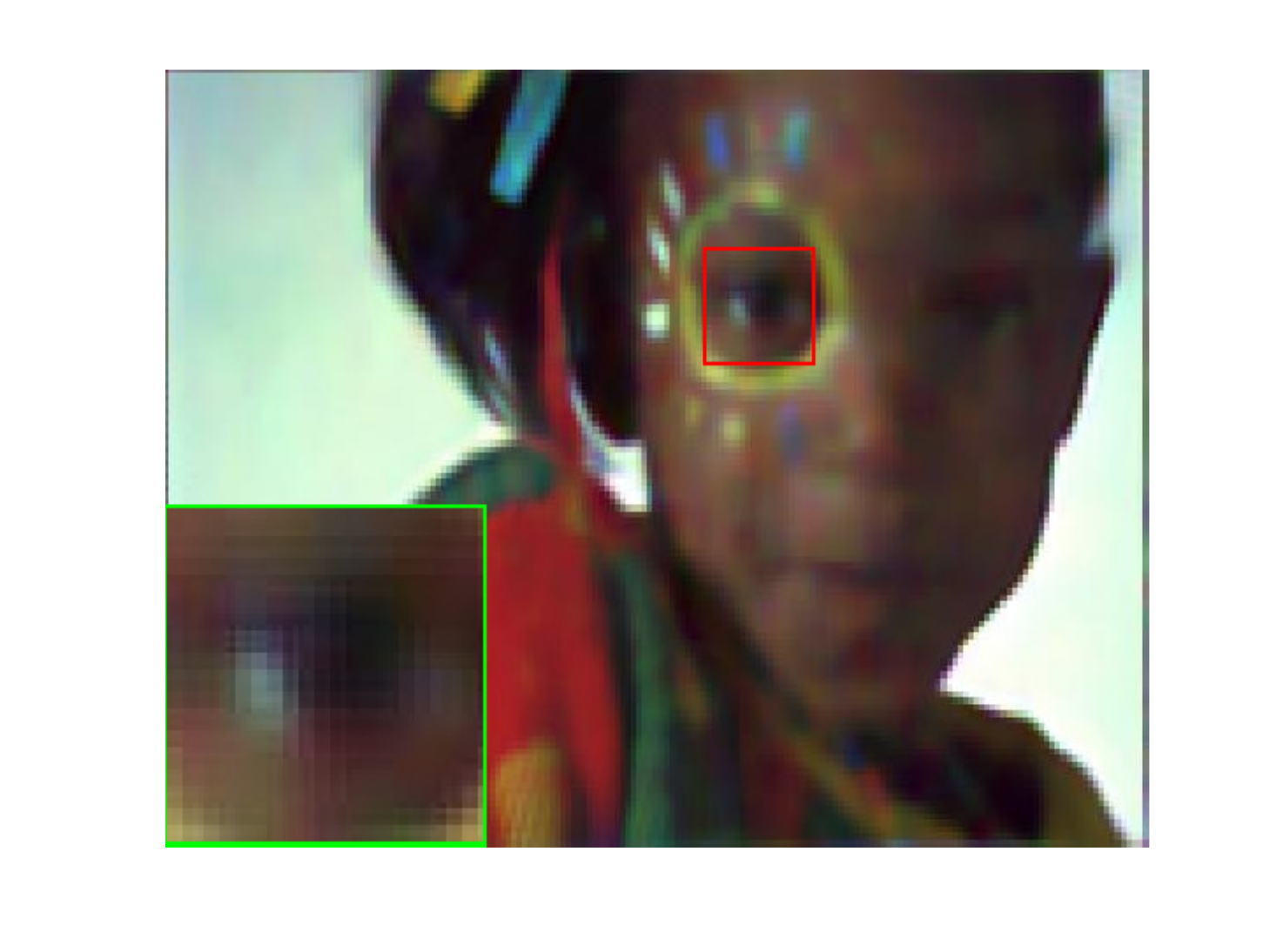}
	}
	\hspace{-0.3in}
	\subfigure[\tiny{QHOSVD}]{
		\includegraphics[width=2.7cm,height=2.5cm]{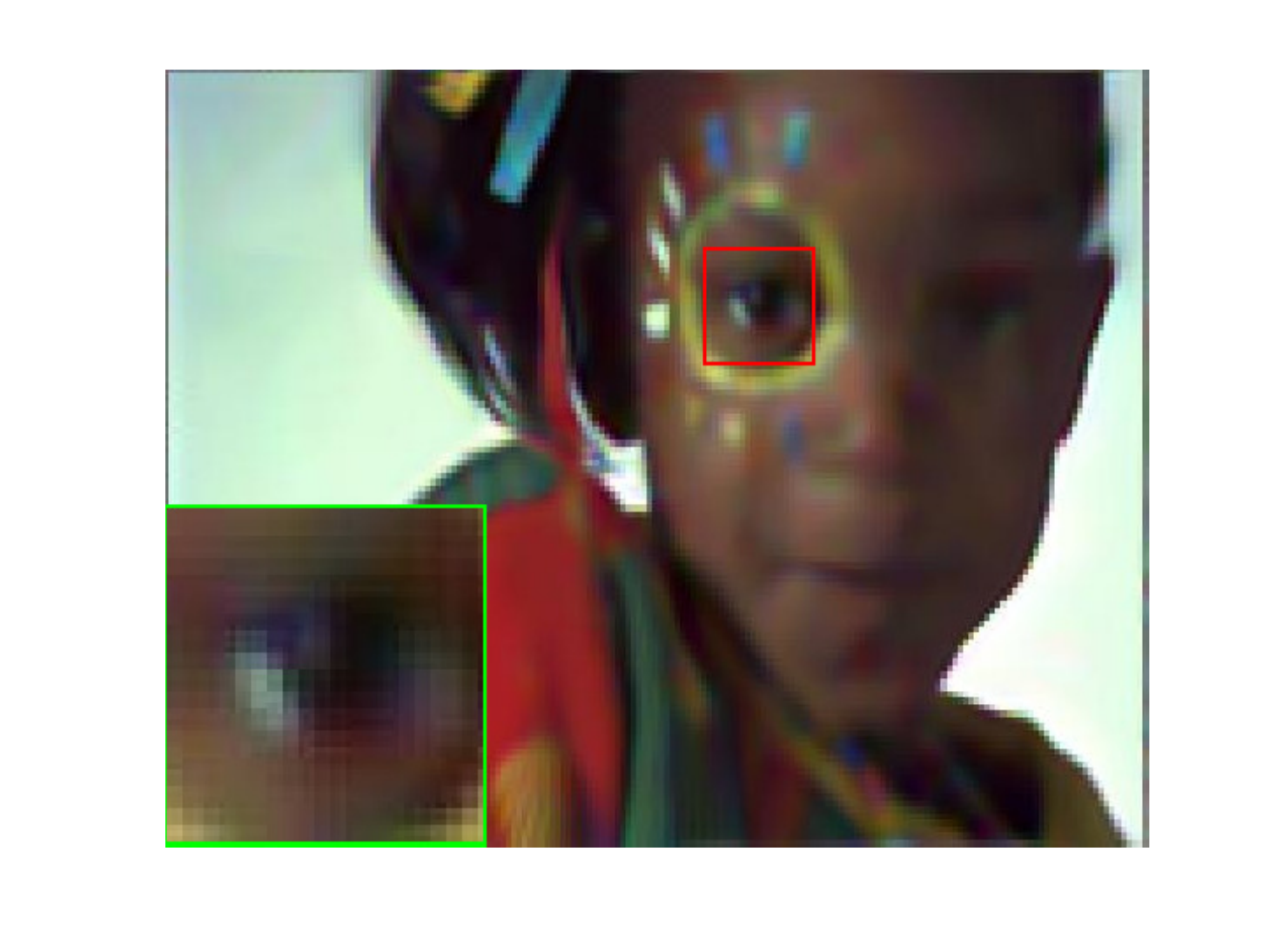}
	}	
	\caption{Color image denoising results on the Image(9) ($\sigma=20$).}
	\label{fig04}
\end{figure}

\begin{figure}[htbp]
	\centering
	\hspace{-0.3in}
	\subfigure[\tiny{Original}]{
		\includegraphics[width=2.7cm,height=2.5cm]{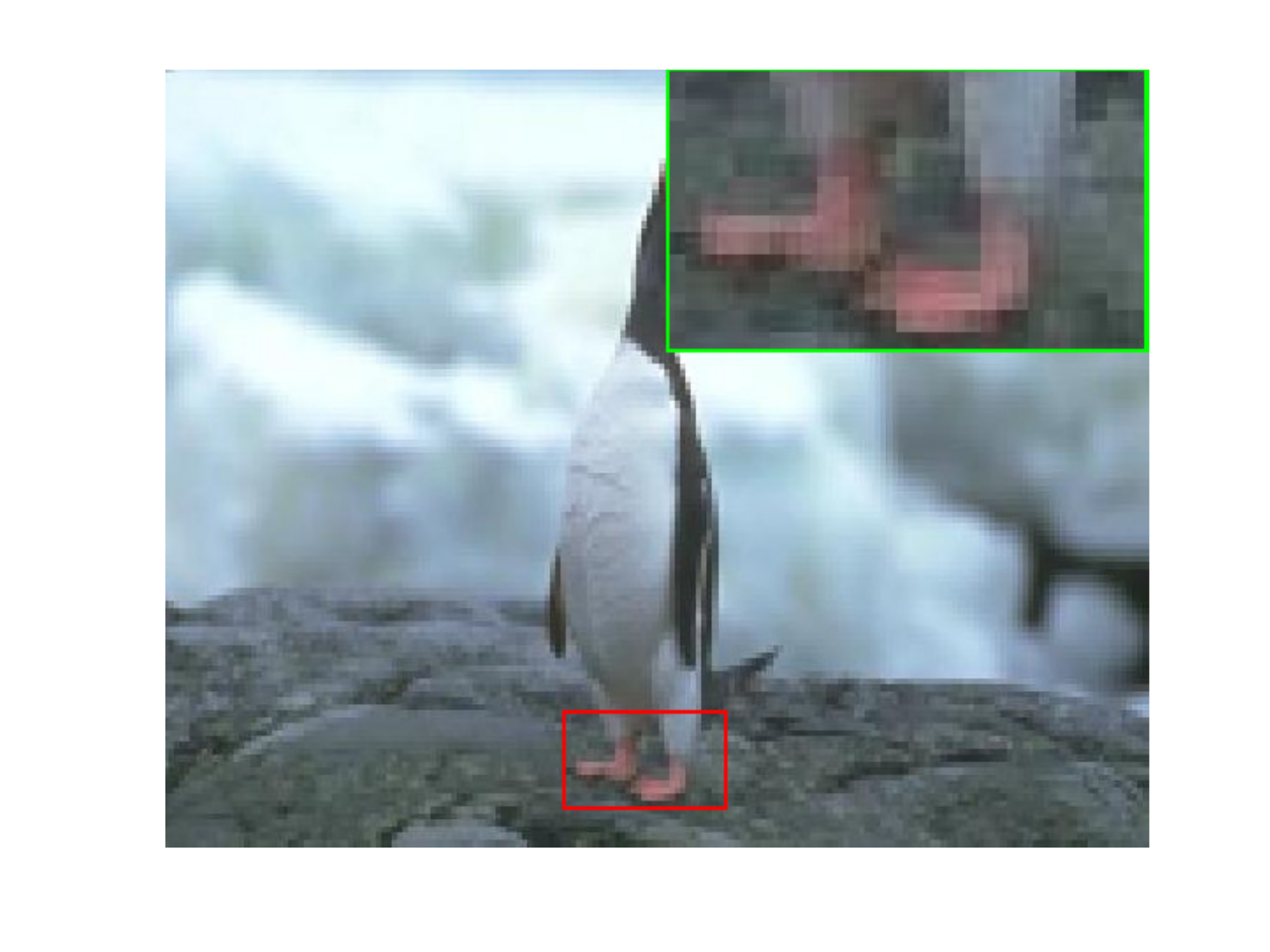}
	}\\
	\hspace{-0.3in}
	\subfigure[\tiny{Observed}]{
		\includegraphics[width=2.7cm,height=2.5cm]{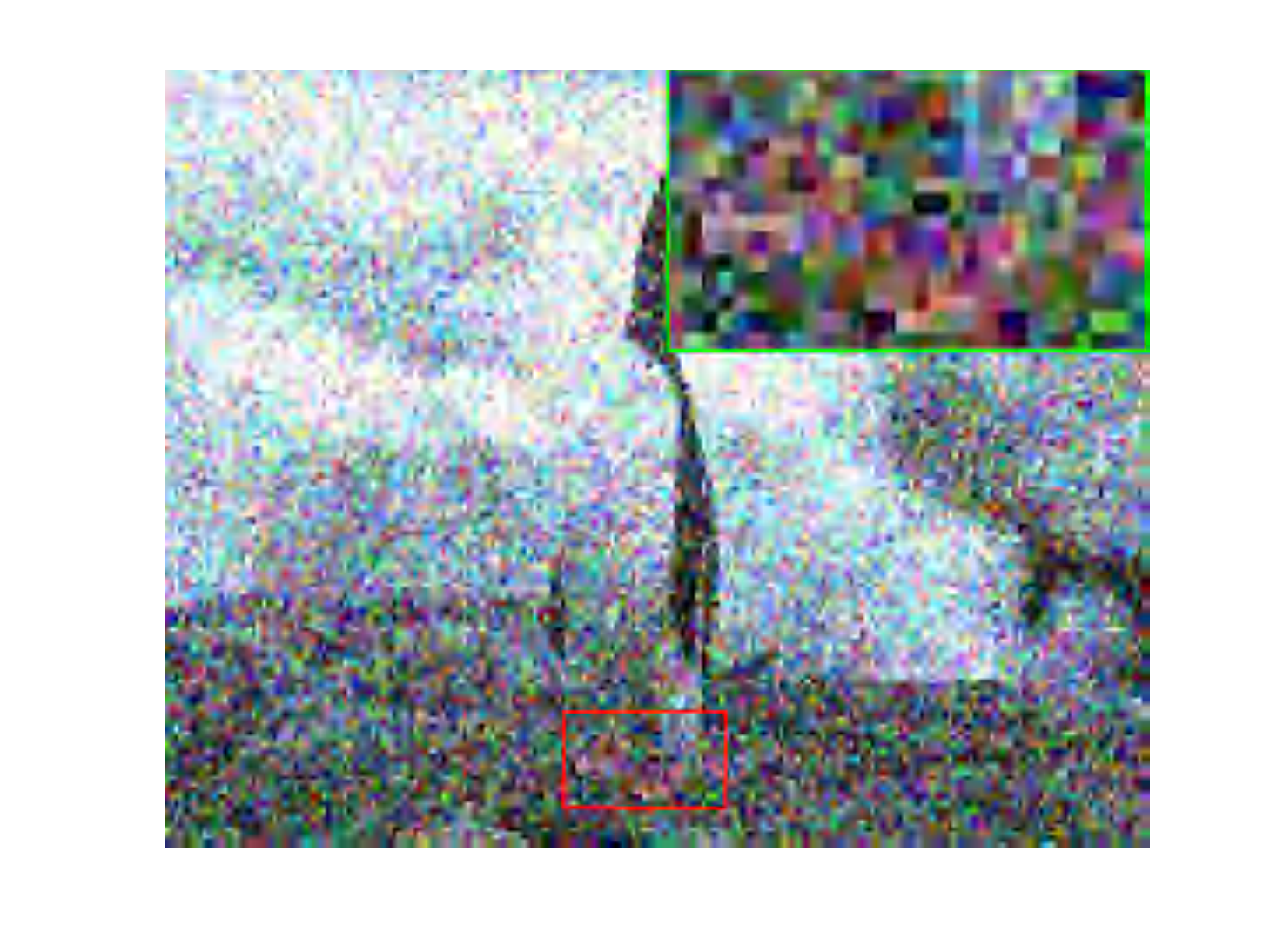}
	}
	\hspace{-0.3in}
	\subfigure[\tiny{QNNM} \cite{DBLP:journals/tip/ChenXZ20}]{
		\includegraphics[width=2.7cm,height=2.5cm]{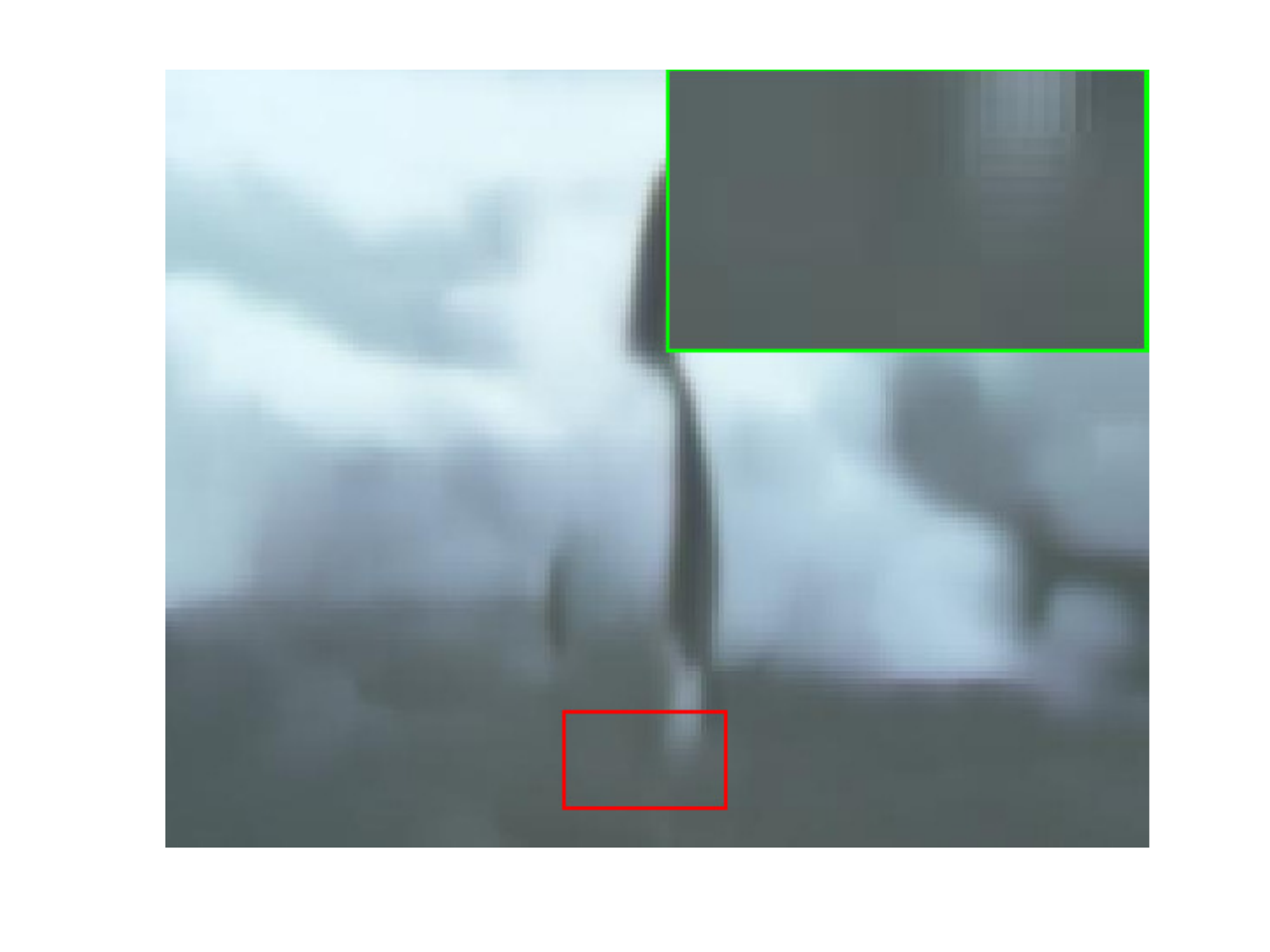}
	}
	\hspace{-0.3in}
	\subfigure[\tiny{QWNNM} \cite{DBLP:journals/ijon/YuZY19}]{
		\includegraphics[width=2.7cm,height=2.5cm]{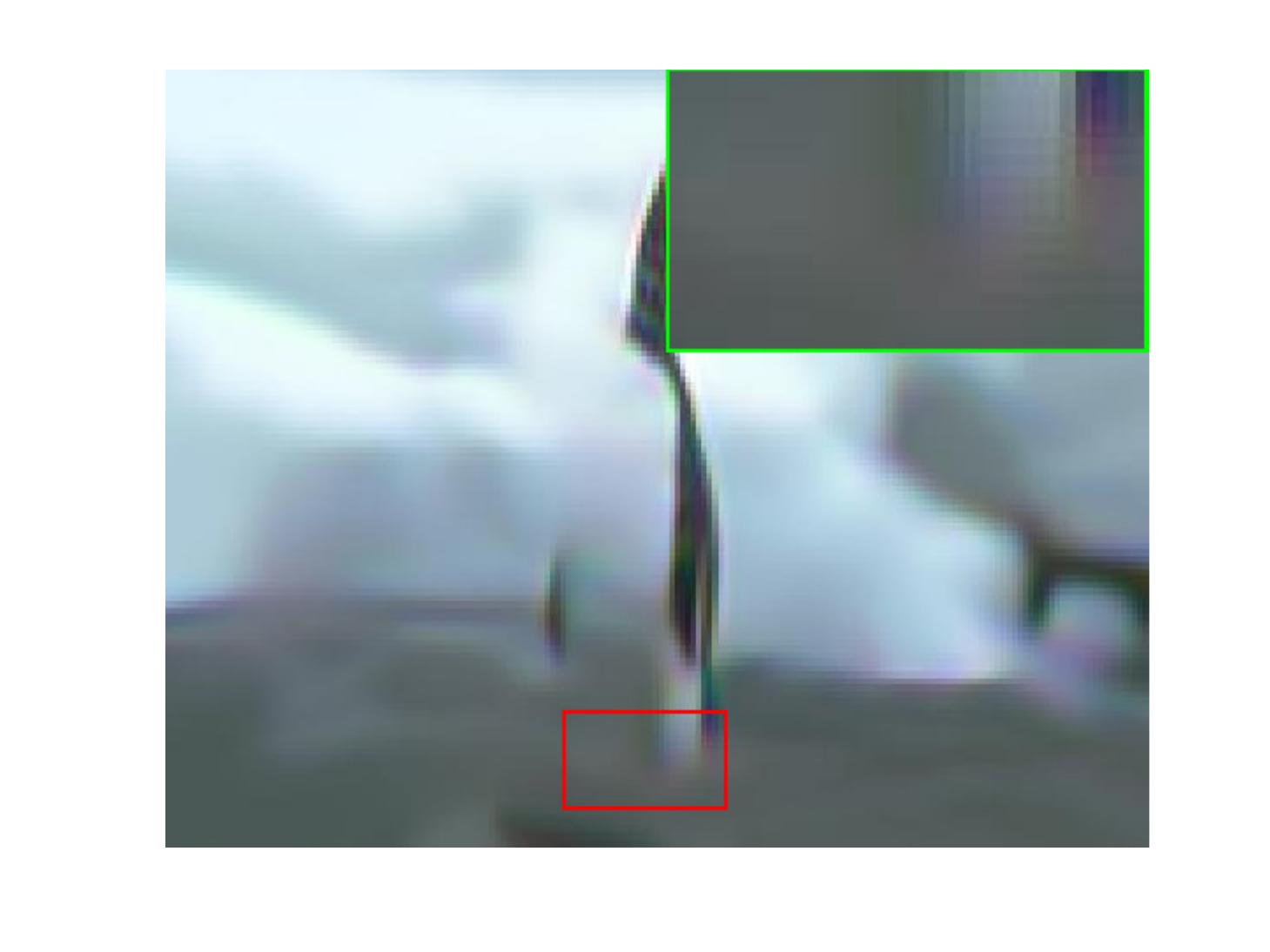}
	}\\
	\hspace{-0.3in}
	\subfigure[\tiny{LRQA-WSNN} \cite{DBLP:journals/tip/ChenXZ20}]{
		\includegraphics[width=2.7cm,height=2.5cm]{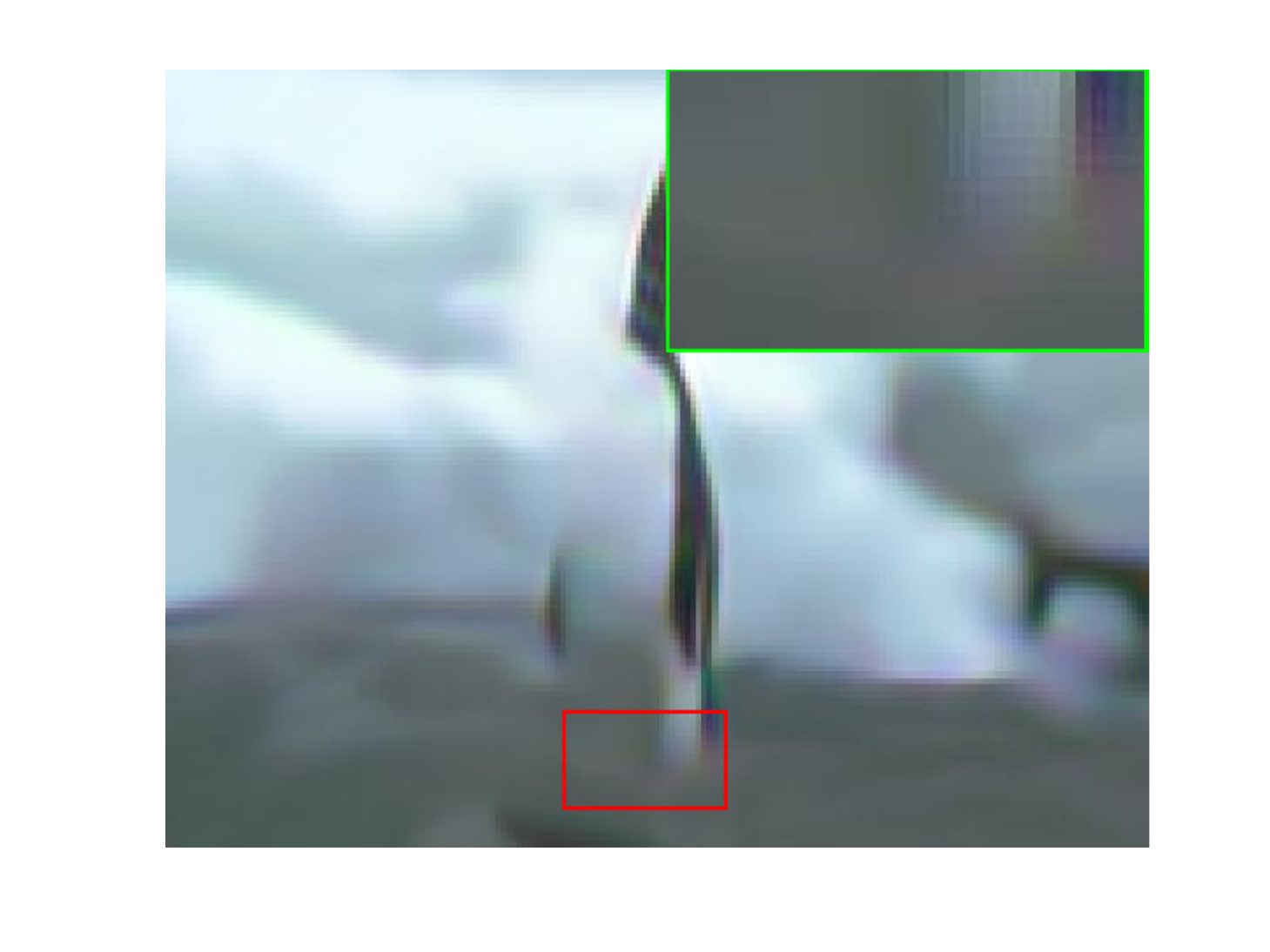}
	}
	\hspace{-0.3in}
	\subfigure[\tiny{HOSVD}  \cite{DBLP:journals/access/GaoGZCZ19}]{
		\includegraphics[width=2.7cm,height=2.5cm]{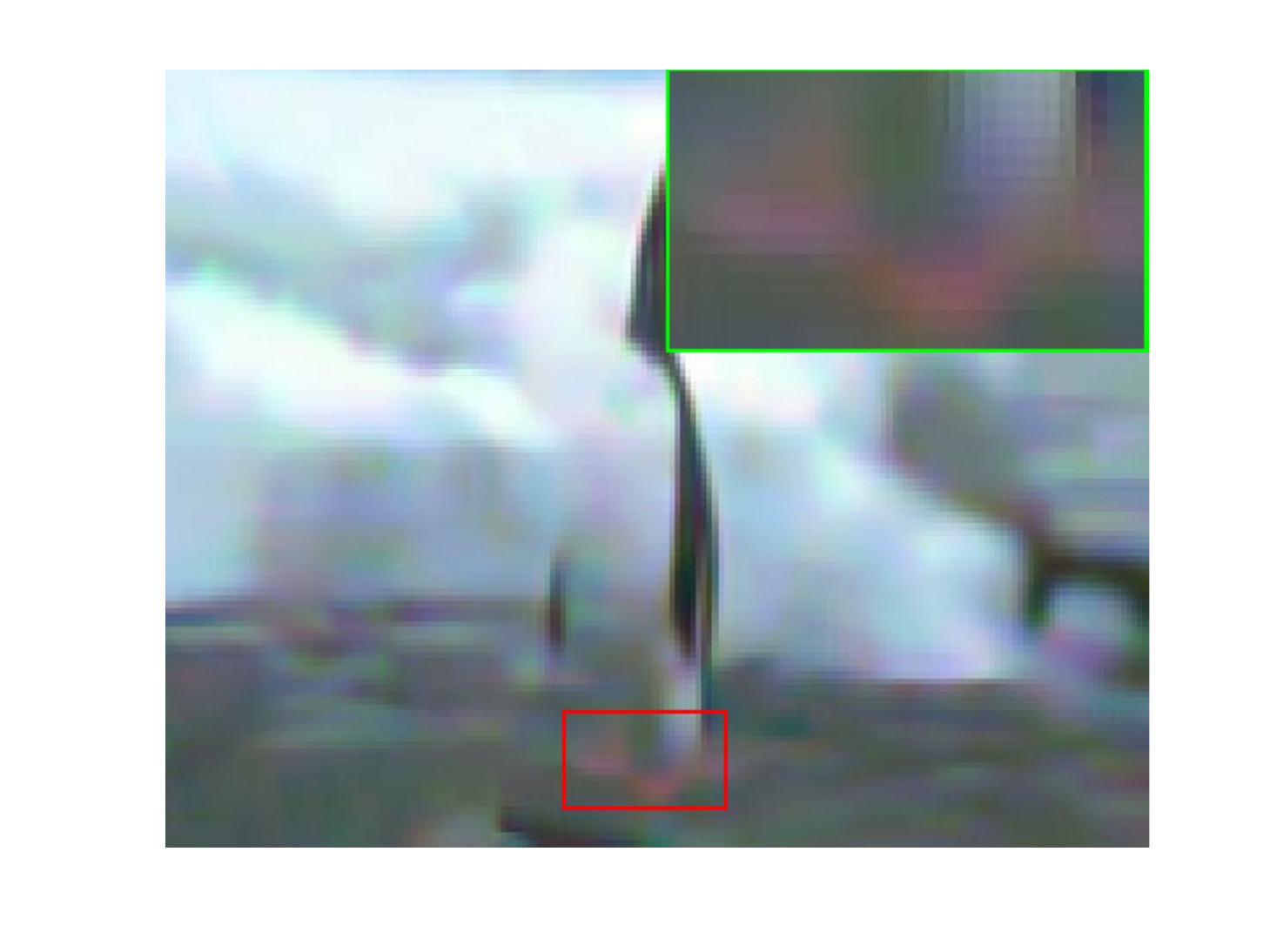}
	}
	\hspace{-0.3in}
	\subfigure[\tiny{QHOSVD}]{
		\includegraphics[width=2.7cm,height=2.5cm]{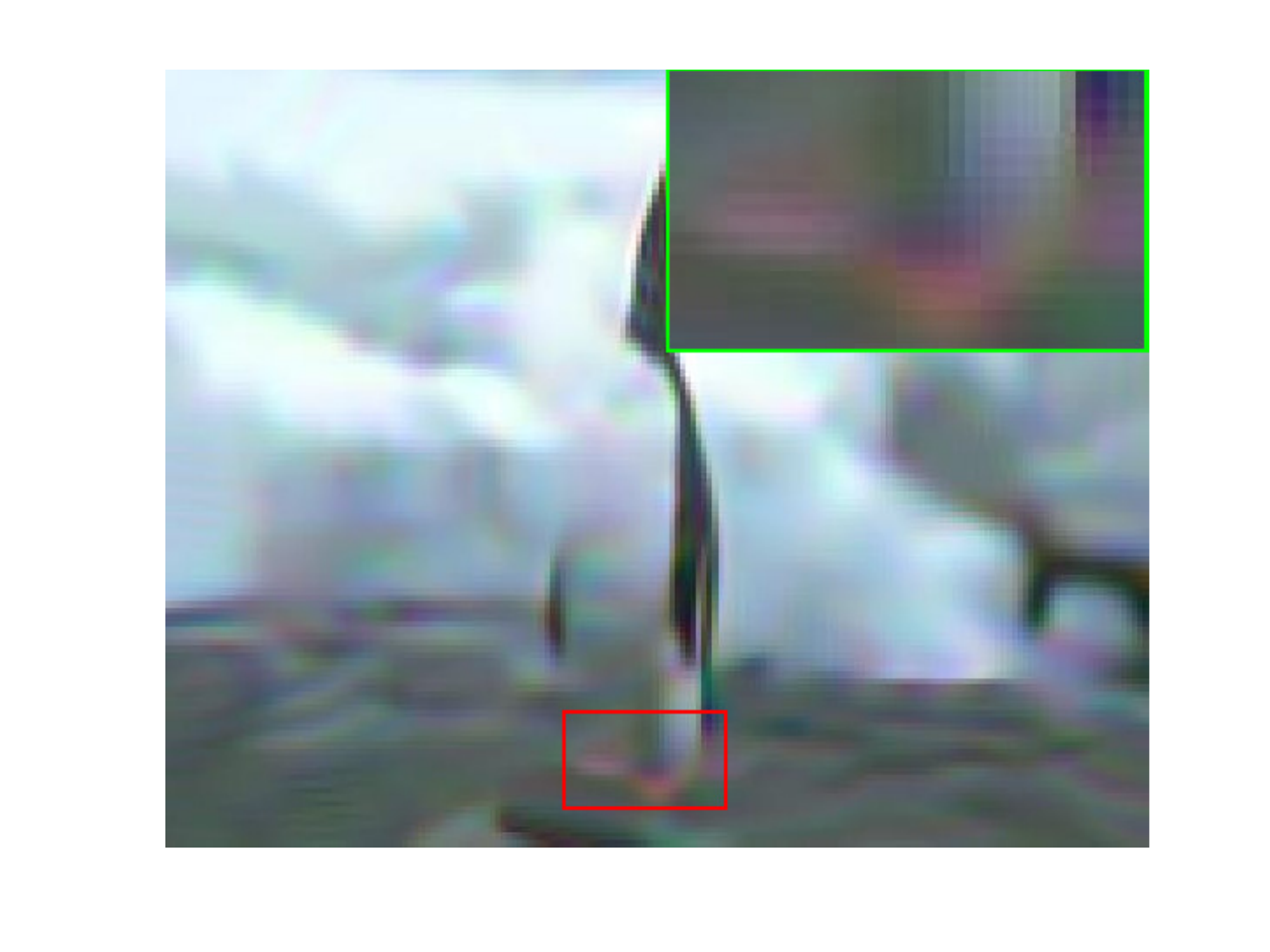}
	}	
	\caption{Color image denoising results on the Image(1) ($\sigma=50$).}
	\label{fig05}
\end{figure}

\begin{figure}[htbp]
	\centering
	\hspace{-0.3in}
	\subfigure[\tiny{Original}]{
		\includegraphics[width=2.7cm,height=2.5cm]{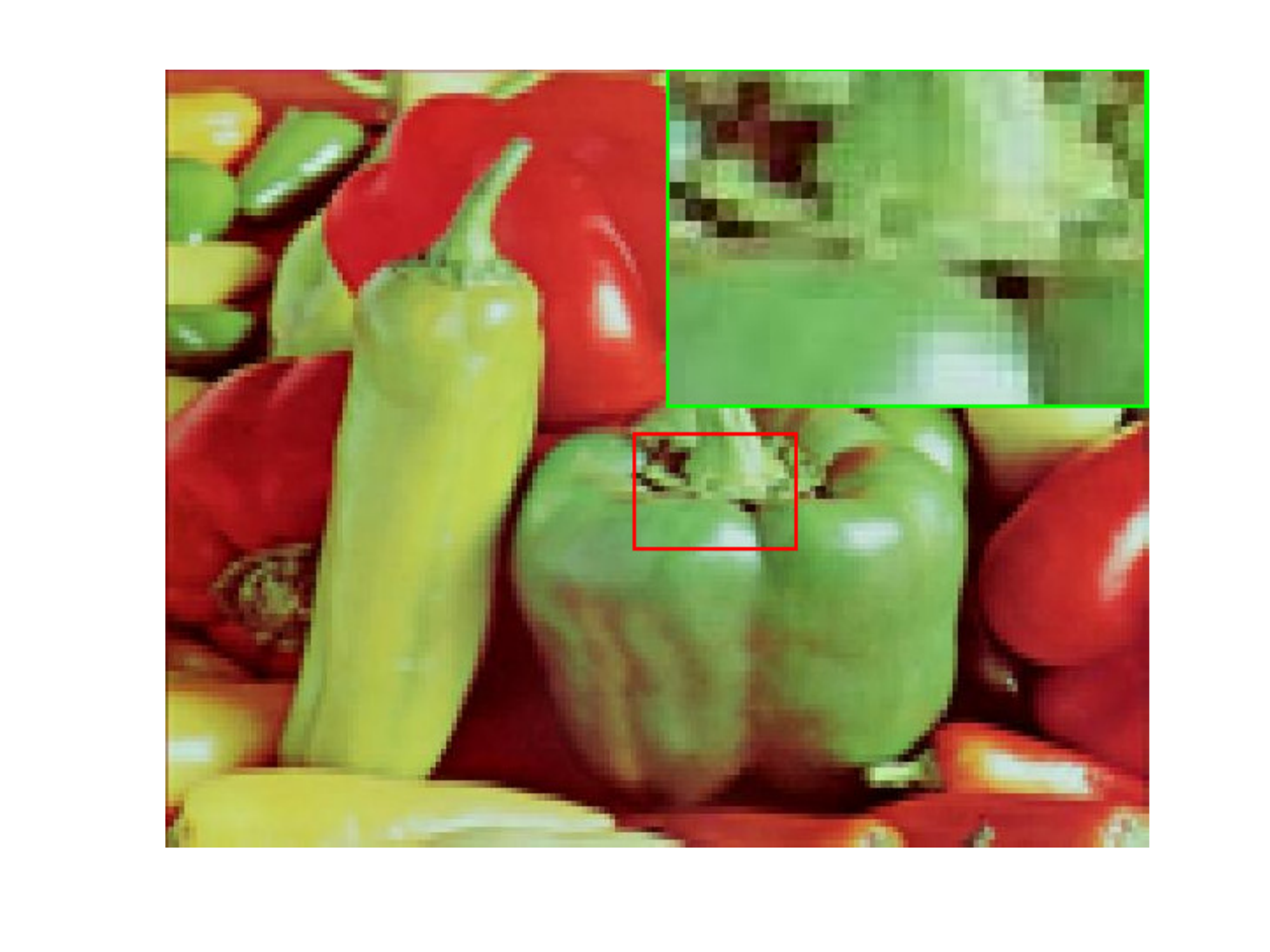}
	}\\
	\hspace{-0.3in}
	\subfigure[\tiny{Observed}]{
		\includegraphics[width=2.7cm,height=2.5cm]{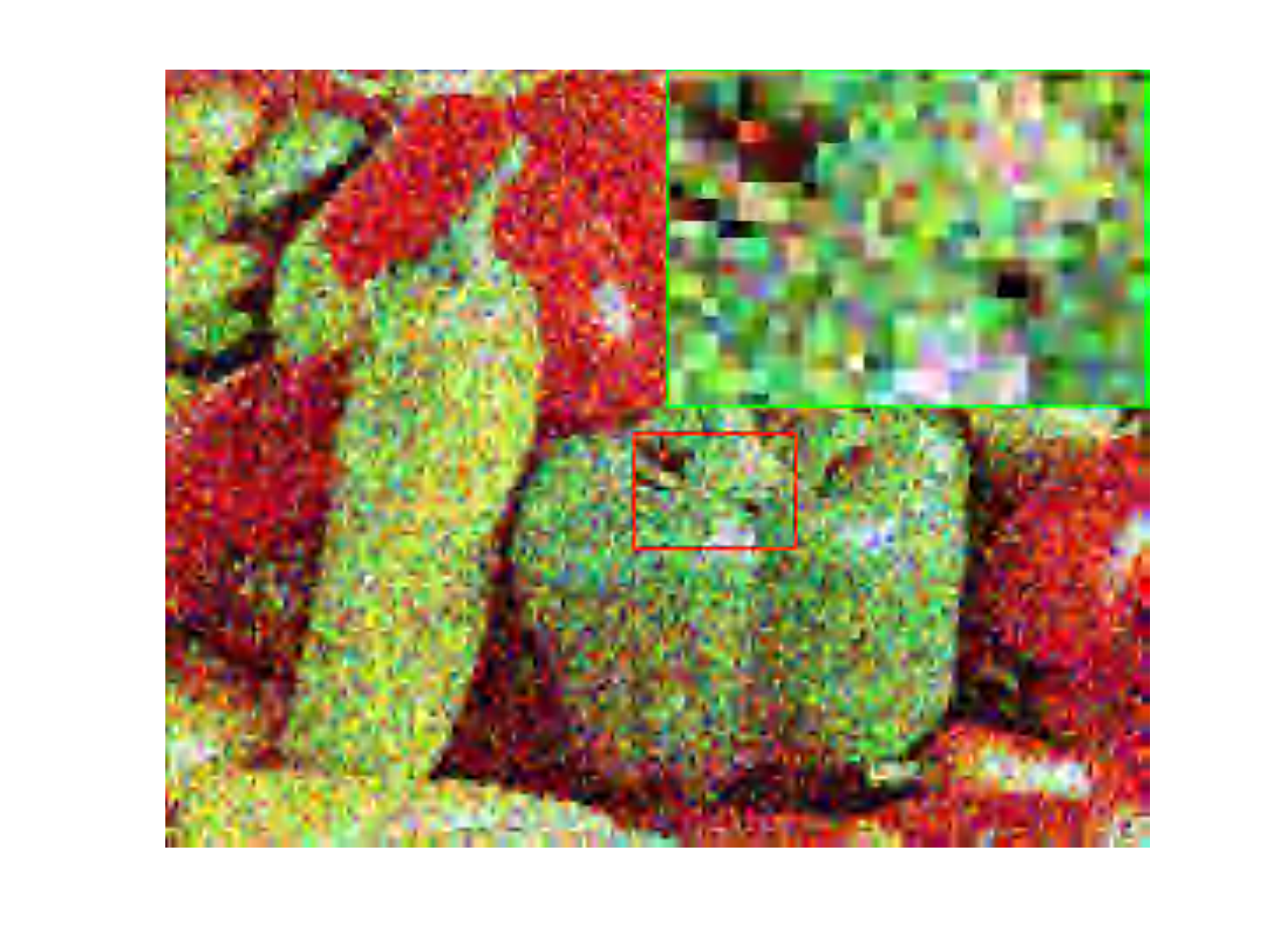}
	}
	\hspace{-0.3in}
	\subfigure[\tiny{QNNM} \cite{DBLP:journals/tip/ChenXZ20}]{
		\includegraphics[width=2.7cm,height=2.5cm]{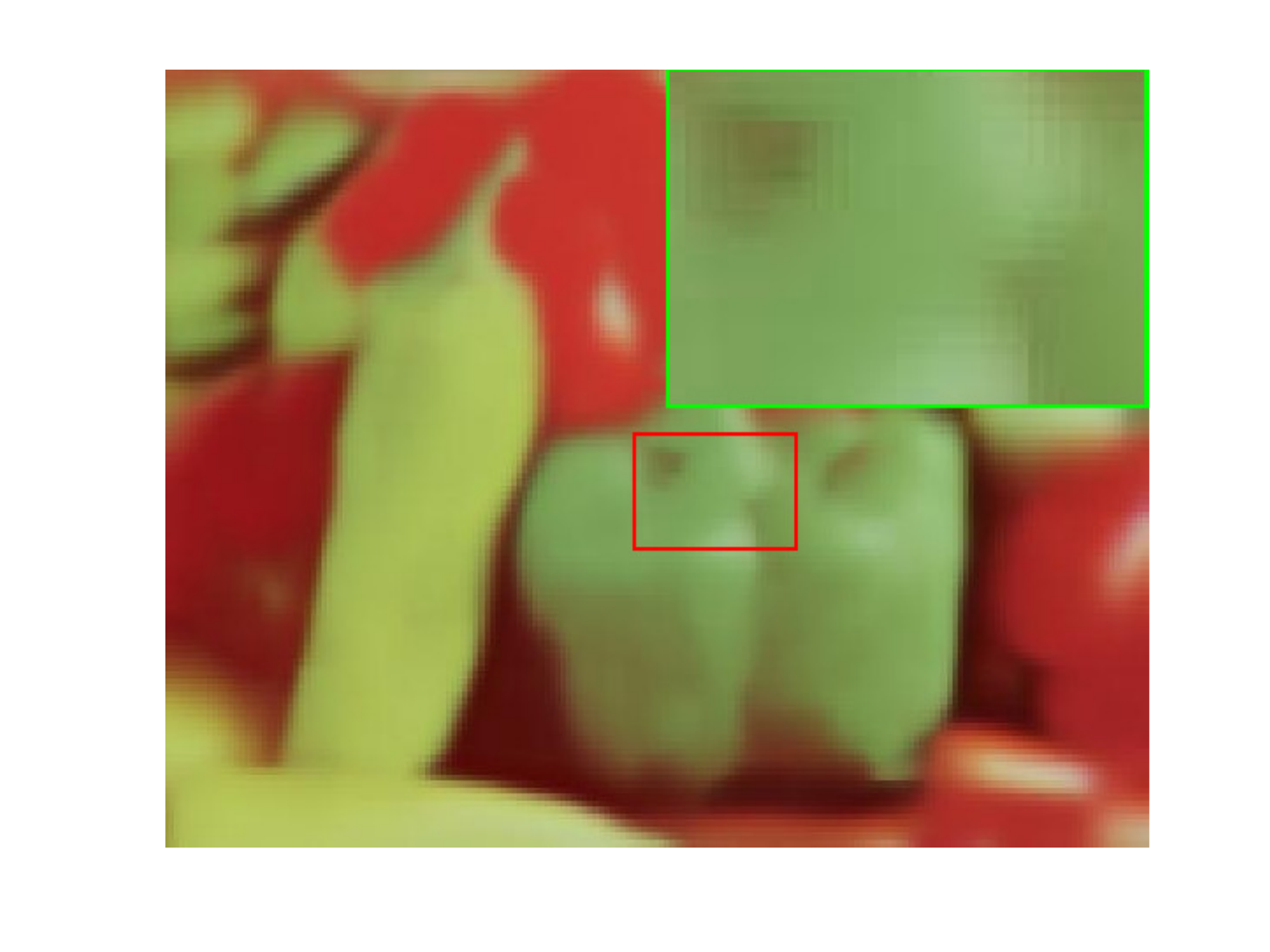}
	}
	\hspace{-0.3in}
	\subfigure[\tiny{QWNNM} \cite{DBLP:journals/ijon/YuZY19}]{
		\includegraphics[width=2.7cm,height=2.5cm]{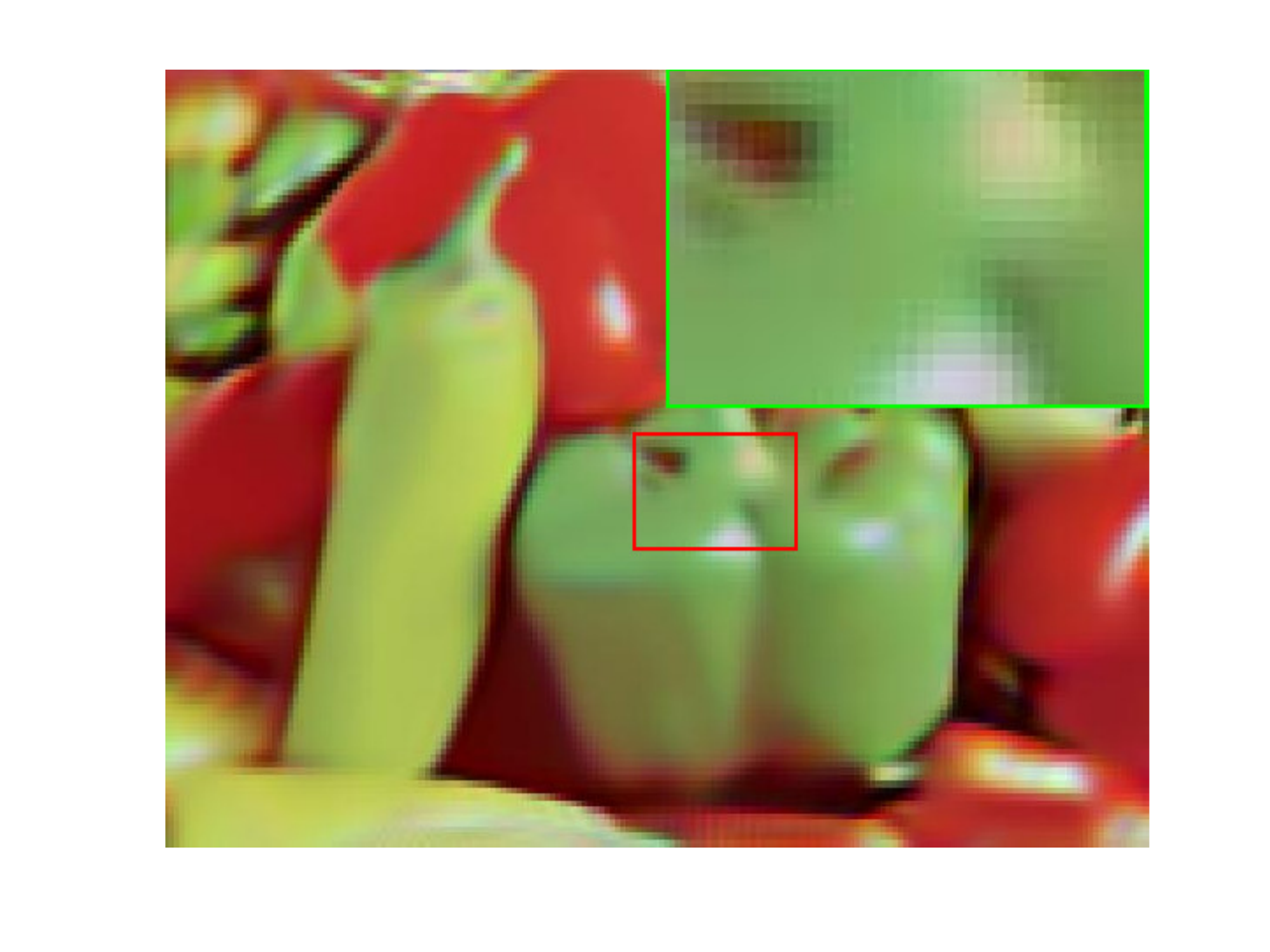}
	}\\
	\hspace{-0.3in}
	\subfigure[\tiny{LRQA-WSNN} \cite{DBLP:journals/tip/ChenXZ20}]{
		\includegraphics[width=2.7cm,height=2.5cm]{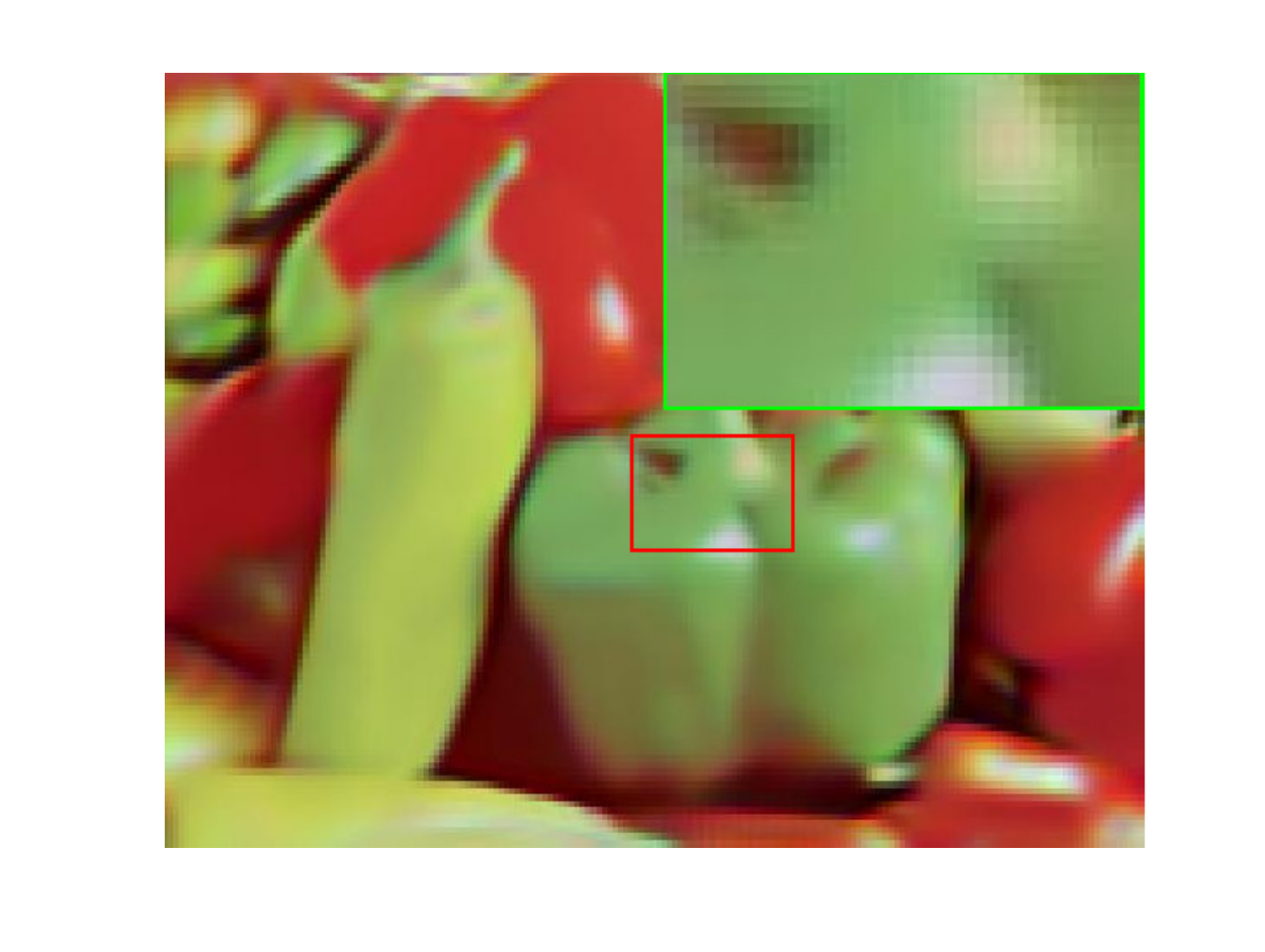}
	}
	\hspace{-0.3in}
	\subfigure[\tiny{HOSVD}  \cite{DBLP:journals/access/GaoGZCZ19}]{
		\includegraphics[width=2.7cm,height=2.5cm]{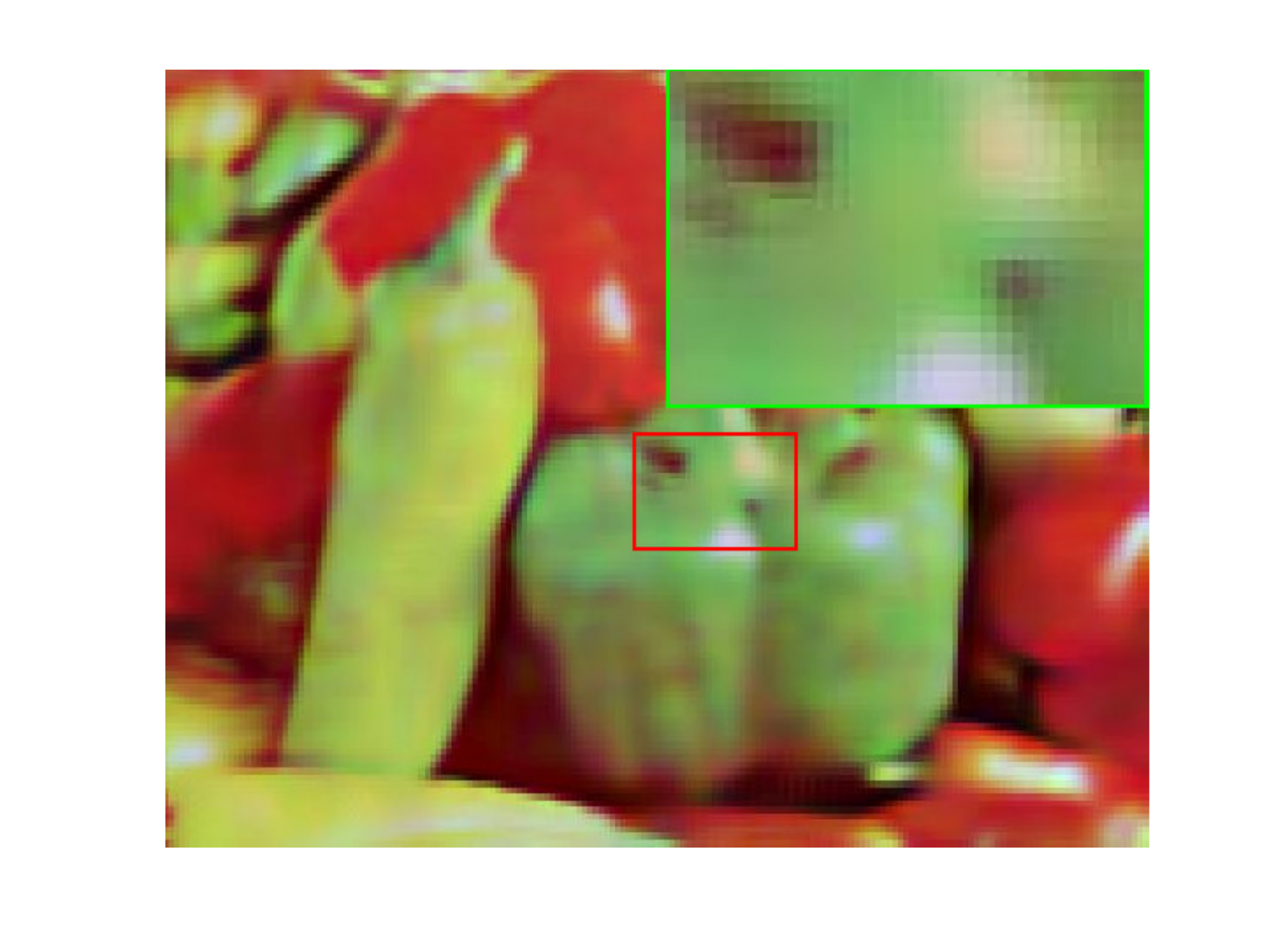}
	}
	\hspace{-0.3in}
	\subfigure[\tiny{QHOSVD}]{
		\includegraphics[width=2.7cm,height=2.5cm]{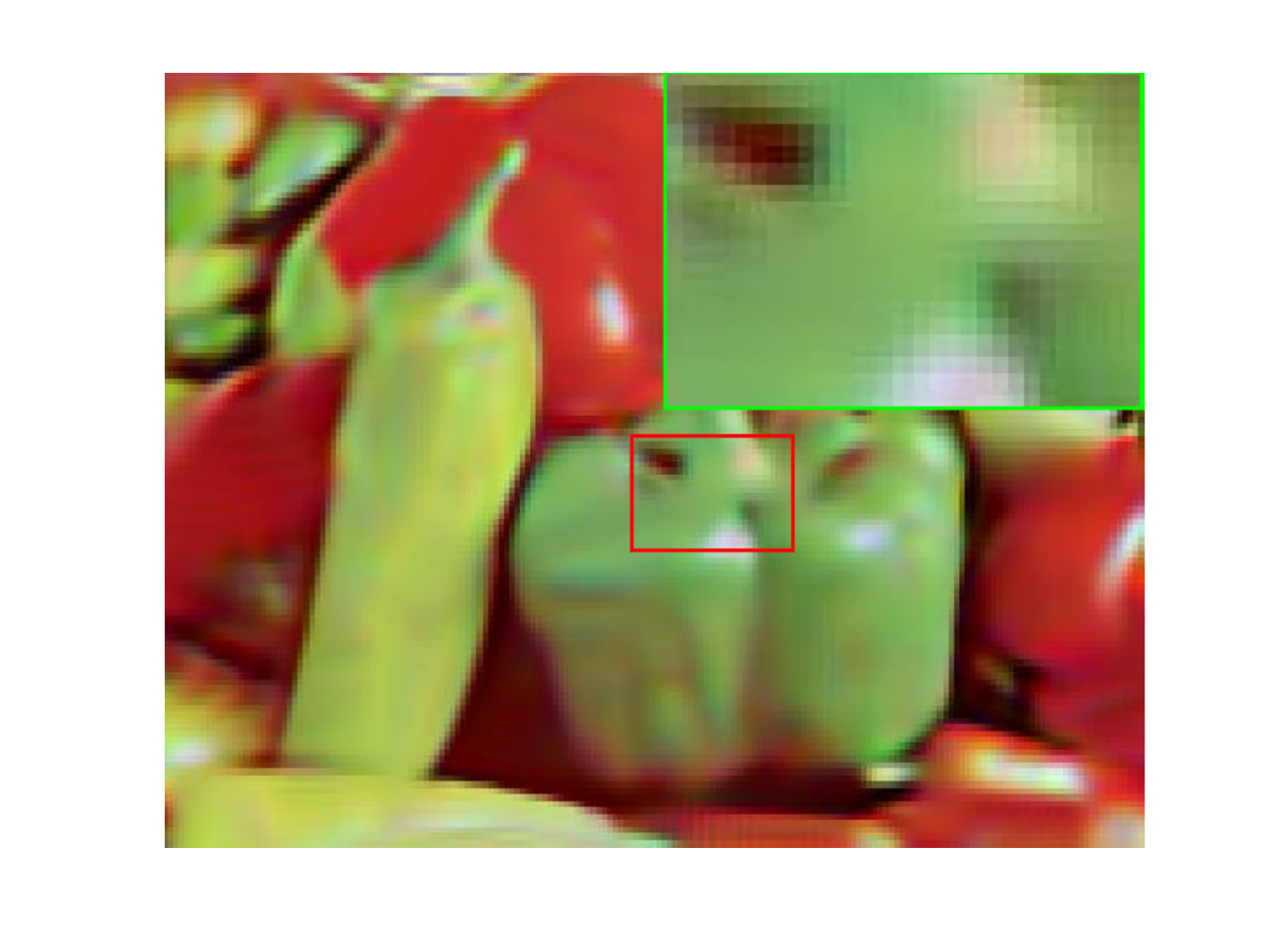}
	}	
	\caption{Color image denoising results on the Image(7) ($\sigma=50$).}
	\label{fig06}
\end{figure}

\textbf{Results and discussions:}
TABLE \ref{Index4SR_2} reports the quantitative PSNR and SSIM values (and the average values of them) of all denoising methods on the ten testing color images with different noise level. Fig. \ref{fig03}-Fig. \ref{fig06}  visually show the denoising results of several color images. From all the experimental results about color image denoising, we can observe
and summarize the following points.
\begin{itemize}
	\item The defined QHOSVD can indeed be applied to color image denoising problems.
	\item  In the overwhelming majority
	of cases, the proposed QHOSVD-based method shows the best performance among the compared latest ones (\emph{see} TABLE \ref{Index4SR_2}). Visually, the QHOSVD-based method preserves the detail of the color images better, which can notably alleviate the oversmooth problem of QWNNM and LRQA-WSNN (\emph{see} Fig. \ref{fig03}-Fig. \ref{fig06}).
	\item The average runtime (when $\sigma=10$) of QNNM, QWNNM, LRQA-WSNN, HOSVD, and QHOSVD are, $1024.623s$, $1036.158s$, $1250.813s$, $406.486s$, and $4065.666s$, respectively.
	The main computation burden of the QHOSVD-based method is performing the QSVD and the NSS procedure, which are generally time-consuming.
\end{itemize}

\section{Conclusion}
\label{sec:Conc}
In this paper, the problem of generalizing the HOSVD to the quaternion domain is investigated. We define QHOSVD and give the calculation procedure. Since the QHOSVD combines the benefits of the quaternion tool and the HOSVD, it can be well applied to many color image processing problems. Moreover, we propose a  multi-focus color image fusion method and a  color image denoising method, as examples of the QHOSVD in color image processing. Experimental results demonstrate the effectiveness of the developed methods.

Since the computation based on quaternion algebra is still time-consuming, such as the calculation of the QSVD, one of the future works is to study the fast algorithms of the QSVD and other quaternion-based operations to make the proposed QHSVD-based methods more efficient. In addition, we would like to extend the defined QHOSVD to other color image processing tasks, such as color face recognition, color image  super-resolution, color image inpainting, and so on.

\appendices
\section{Basic knowledge of quaternion algebras}
\label{a_sec1}
A quaternion $\dot{q}\in\mathbb{H}$ with a real component and three imaginary components is defined as
\begin{equation}
	\label{equ2}
	\dot{q}=q_{0}+q_{1}i+q_{2}j+q_{3}k,
\end{equation}
where $q_{l}\in\mathbb{R}\: (l=0,1,2,3)$, and $i, j, k$ are
imaginary number units and obey the quaternion rules that
\begin{align}
	\left\{
	\begin{array}{lc}
		i^{2}=j^{2}=k^{2}=ijk=-1,\\
		ij=-ji=k, jk=-kj=i, ki=-ik=j.
	\end{array}
	\right.
\end{align}
$\dot{q}$ can be decomposed into a real part $\mathfrak{R}(\dot{q}):=q_{0}$ and an imaginary part $\mathfrak{I}(\dot{q}):=q_{1}i+q_{2}j+q_{3}k$ such that $\dot{q}=\mathfrak{R}(\dot{q})+\mathfrak{I}(\dot{q})$.
If the real part $\mathfrak{R}(\dot{q})=0$, $\dot{q}$ is named a pure quaternion. The addition and multiplication of quaternions are similar to those of complex numbers, except that the multiplication of quaternions does not satisfy the commutative law, \emph{i.e.}, in general $\dot{p}\dot{q}\neq\dot{q}\dot{p}$.
The conjugate and the modulus of a quaternion $\dot{q}$ are,
respectively, defined as 
\begin{align*}
	\dot{q}^{\ast}&=q_{0}-q_{1}i-q_{2}j-q_{3}k,\\
	|\dot{q}|&=\sqrt{\dot{q}\dot{q}^{\ast}}=\sqrt{q_{0}^{2}+q_{1}^{2}+q_{2}^{2}+q_{3}^{2}}.
\end{align*}

Analogously, a quaternion matrix $\dot{\mathbf{Q}}=(\dot{q}_{mn})\in\mathbb{H}^{M\times N}$ is written
as $\dot{\mathbf{Q}}=\mathbf{Q}_{0}+\mathbf{Q}_{1}i+\mathbf{Q}_{2}j+\mathbf{Q}_{3}k$, where $\mathbf{Q}_{l}\in\mathbb{R}^{M\times N}\: (l=0,1,2,3)$, $\dot{\mathbf{Q}}$ is named a pure quaternion matrix when $\mathfrak{R}(\dot{\mathbf{Q}})=\mathbf{Q}_{0}=\mathbf{0}$. The quaternion matrix Frobenius norm and $L_{1}$-norm are respectively defined as\small{ $\|\dot{\mathbf{Q}}\|_{F}=\sqrt{\sum_{m=1}^{M}\sum_{n=1}^{N}|\dot{q}_{mn}|^{2}}$ and \small{$\|\dot{\mathbf{Q}}\|_{L_{1}}=\sum_{m=1}^{M}\sum_{n=1}^{N}|\dot{q}_{mn}|$}. The quaternion singular value decomposition (QSVD) of $\dot{\mathbf{Q}}$ is defined by \cite{10029950538}: $\dot{\mathbf{Q}}=\dot{\mathbf{U}}\mathbf{\Sigma}\dot{\mathbf{V}}^{H}$, where $\dot{\mathbf{U}}\in\mathbb{H}^{M\times M}$ and $\dot{\mathbf{V}}\in\mathbb{H}^{N\times N}$ are unitary quaternion matrices, $\mathbf{\Sigma}$ contains all the singular values of $\dot{\mathbf{Q}}$.

Readers can find more details on quaternion algebra in \cite{10029950538, Girard2007Quaternions, Altmann1986Rotations}.

\section*{Acknowledgment}
This work was supported by The Science and Technology Development Fund, Macau SAR (File no. FDCT/085/2018/A2) and University of Macau (File no. MYRG2019-00039-FST).

\ifCLASSOPTIONcaptionsoff
\newpage
\fi

\bibliographystyle{IEEEtran}
\bibliography{Myreference}
\end{document}